\documentclass[preprintnumbers,superscriptaddress,nofootinbib,aps,prd,floatfix]{revtex4}

\pdfoutput=1
\usepackage{graphicx}
\usepackage{latexsym}
\usepackage{amsfonts}
\usepackage{amsmath,amssymb}
\usepackage{slashed}
\usepackage{feynmp}
\usepackage{hyperref}
\usepackage{url}
\usepackage{color}
\usepackage{cancel}
\usepackage{multirow,array}
 \usepackage{float}
 \usepackage{subfigure}
 
\begin{document}

\title{Wide or Narrow? The Phenomenology of 750 GeV Diphotons}

\author{Matthew R.~Buckley}
\affiliation{Department of Physics and Astronomy, Rutgers University, Piscataway, NJ 08854, USA}

\begin{abstract}
I perform a combined analysis of the ATLAS and CMS diphoton data, using both Run-I and Run-II results, including those released at the 2016 Moriond conference. I find combining the ATLAS and CMS results from Run-II increases the statistical significance of the reported 750~GeV anomaly, assuming a spin-0 mediator coupling to gluons or heavy quarks with a width much smaller than the detector resolution. This significance does not decrease when the 8~TeV data is included.  A spin-2 mediator is disfavored compared to the spin-0 case. The cross section required to fit the ATLAS anomaly is in tension with the aggregate data, all of which prefers a smaller value. The  best fit for all models I consider is a $4.0\sigma$ local significance for a 750~GeV spin-0 mediator coupling to gluons with a cross section of 4~fb at 13~TeV (assuming narrow width) or 10~fb (assuming $\Gamma=45$~GeV).
\end{abstract}

\maketitle

\section{Introduction \label{sec:intro}}

The ATLAS announcement \cite{atlas13} of a $3.6\sigma$ local excess in diphotons with invariant masses near $m_{\gamma\gamma} \sim 750$~GeV in the first batch of LHC Run-II data, combined with the CMS Collaboration announcing \cite{cms13} a $2.6\sigma$ local excess in the same channel and invariant mass, has sent the theoretical physics community into a frenzy of model-building, with 150 papers on the topic appearing on arXiv within a month. The ``$\gamma\gamma$ Resonance that Stole Christmas'' \cite{Craig:2015lra} is of course tremendously exciting. It is the LHC's most  statistically significant deviation from the Standard Model of particle physics made public since the discovery of the 125~GeV Higgs boson. The addition of 0.6~fb$^{-1}$ of CMS data gathered without the presence of the solenoid field \cite{CMS_0}, as well a reanalysis of the 13 and 8~TeV data by the ATLAS Collaboration \cite{ATLAS_redo} has increased the statistical preference for new physics, with CMS now reporting a $3.4\sigma$ local significance in their combined data.

At this stage, the information on which theorists can build models is somewhat minimal, assuming that these results are not just a somewhat unlikely statistical fluctuation and there is actually new physics to be begin with. A $3.6\sigma$ signal is not very large, and additional slicing of the data in order to ask more detailed questions will typically yield only small statistical preference for any possible result. The goal of this paper is to investigate what information can be extracted from the combination of ATLAS and CMS data in the diphoton channel, including the most recent analyses released at the 2016 Moriond conference \cite{CMS_0,ATLAS_redo}. In particular, I am interested in the statistical significance of the combined 13~TeV Run-II data, and in the significance of combinations with the Run-I 8~TeV data. In addition, the ATLAS Collaboration claims a better fit ($3.9\sigma$ local) to a resonance at 750~GeV with a width of 45~GeV, as opposed to a narrow resonance with width much less than the experimental resolution (as might be naively expected). Fitting this wide resonance into theoretical models has been the focus of many of the 150 papers on the topic. In this paper, I consider the combined statistical preference for this width. 

This paper should be viewed as a follow-up to the work of Refs.~\cite{Knapen:2015dap,Gupta:2015zzs,Falkowski:2015swt}, which are some of the earliest theoretical papers on the 750~GeV anomaly, and have largely set the parameters which later papers have adopted. Where our results overlap, my results broadly agree with those found in Refs.~\cite{Knapen:2015dap,Gupta:2015zzs,Falkowski:2015swt}; in particular in terms of the best-fit cross sections for the anomaly. This paper goes into somewhat more detail in fitting the signal in the ATLAS and CMS data, by floating the fit to the background functional forms as signal is added. I also perform fits to spin-2 resonances, and investigate the preference of the combined data for a wide or narrow resonance. This paper does not attempt to address the implications of this anomaly for other search channels at the LHC, new physics models beyond the resonance itself, or derive any limits on fundamental couplings. Such questions have been well addressed in the literature. I restrict this paper solely to the {\it input} parameters for  theoretical studies: the cross sections, widths, and statistical significance of the anomaly itself.

The next section describes my combination fits to the ATLAS and CMS diphoton data, using both 13 and 8~TeV data. For simplicity (and due to the computational limitations of the event generation), I consider only local significance in my combination fits, rather than including the look-elsewhere-effect. The statistical question at hand here is the significance of the anomaly at 750~GeV specifically, thus somewhat justifying this choice. I parameterize the possible couplings of the mediator to gluons and light quarks separately, and consider both spin-0 and spin-2 resonances. I present additional results for second-generation and $b$-quark couplings in Appendix~\ref{sec:appendix}. In Section~\ref{sec:width}, I then turn to the question of the statistical preference for a wide or narrow resonance; using the best-fit parameters from Section~\ref{sec:pheno}. 

I will discuss my conclusions in Section~\ref{sec:conclusions}, but I summarize some of the more salient points here:
\begin{enumerate}
\item The required cross section to fit the anomaly reported by ATLAS is in tension with the 8~TeV results, as well as the required cross section to fit the CMS anomaly.

\item Combining all data sets yields a local significance of $\sim 4.0\sigma$ for a 750~GeV spin-0 resonance produced through couplings to gluons or heavy quarks. While quoted statistical significances must be taken with a grain of salt, as they are obtained using binned data without inclusion of systematic errors, I find the combination yields a net {\it increase} in the statistical significance as compared to the ATLAS data alone. 

\item The spin-2 interpretation is mildly disfavored compared to a spin-0 mediator. This is due to correlations in the photon momenta which results in a relative decrease in the ATLAS acceptance compared to CMS.

\item The combination of ATLAS and CMS 13~TeV data has a slight statistical preference for a spin-0 mediator with a natural width much smaller than the experimental resolution, as compared to the $\Gamma = 45$~GeV preferred by ATLAS alone. When the 8~TeV data is added, there is a slight statistical preference for a wide resonance over the narrow option, as it is easier to hide a wide resonance in the 8~TeV background. 

\end{enumerate}

\section{Combination Fits \label{sec:pheno}}

In this section, I consider the ATLAS and CMS searches for diphoton resonances. There are six data sets to consider: the ATLAS \cite{atlas13} and CMS \cite{cms13} Run-II searches at 13~TeV -- which contain the anomalous excess that have caused so much excitement of late -- and the previous 8~TeV Run-I diphoton resonance searches from the two experiments \cite{Aad:2014ioa,Khachatryan:2015qba}. To this, CMS has a smaller data set, gathered when their solenoid magnetic field was off during the 13~TeV Run-II \cite{CMS_0}, released at the Moriond 2016 conference. Note that ATLAS had two possible Run-I diphoton searches at the time of the announcement of the Run-II anomaly: one from the Higgs working group \cite{Aad:2014ioa}, and one from the exotica working group \cite{Aad:2015mna}. These have slightly different event selection criteria; the exotica search was developed to target spin-2 graviton searches, while the former is designed to search for Higgs-like scalar resonances. After the reanalysis of both Run-I and Run-II data released by ATLAS at the Moriond conference, a choice of two event selection criteria was applied to both the ATLAS 13 and 8 TeV data sets, one targeted for spin-0 resonances, and the other for spin-2 \cite{ATLAS_redo}. In this paper, I use the appropriate ATLAS event selection as I consider the two spin options. For the remainder of this paper, I will refer to these six data sets by the names \textsc{Atlas13}~\cite{atlas13,ATLAS_redo}, \textsc{Cms13}~\cite{cms13}, \textsc{Cms13/0T}~\cite{CMS_0} (for the data collected when the magnetic field was 0~T), \textsc{Atlas8}~\cite{ATLAS_redo}, and \textsc{Cms8}~\cite{Khachatryan:2015qba}. 

The selection criteria for each of the data sets varies slightly. In Table~\ref{tab:selection}, I list the requirements for diphoton events to end up in the analysis region for each experimental search. In addition, each search implements isolation criteria on the photon candidates. \textsc{Cms13} has two search regions, one where both photons are central (denoted ``BB'' here, for ``barrel-barrel'' events), and one where one photon is forward (``BE,'' for barrel-endcap). The \textsc{Cms13/0T} data uses identical selection criteria. The CMS Collaboration analysis of their \textsc{Cms8} results uses four signal regions, depending on whether the leading- and subleading-$p_T$ photons have converted in the detector. For the range of photon $p_T$ of interest here, this level of distinction is less important, and for simplicity I used the combined \textsc{Cms8} data set. Both the \textsc{Atlas13} and \textsc{Atlas8} data uses separate selection criteria for their spin-0 and spin-2 searches.

\begin{table}[ht]
\begin{tabular}{|c||c||c|c|c|c|c|c|}
\hline
 & ${\cal L}$~(fb$^{-1}$) &  $E_{T1}$~(GeV) & $E_{T2}$~(GeV) & $|\eta| < $ & $ < |\eta|< $ & $E_{T1}/m_{\gamma\gamma}$  & $E_{T2}/m_{\gamma\gamma}$ \\ \hline \hline
\textsc{Atlas13} spin-0 & 3.2 & 40 & 30 & 2.47 & 1.37--1.52 & 0.4 & 0.3 \\ 
\textsc{Atlas13} spin-2 & 3.2 & 55 & 55 & 2.47 & 1.37--1.52 & -- & -- \\ \hline
\textsc{Cms13}(\textsc{Cms13/0T})-BB & 2.7(0.6) & 75 & 75 & 1.44 & -- & -- & -- \\ \hline
\textsc{Cms13}(\textsc{Cms13/0T})-BE & 2.7(0.6) & 75 & 75 & 1.44/2.5 & 1.44--1.57 & -- & -- \\ \hline
\textsc{Atlas8} spin-0 & 20.3 & 22 & 22 & 2.37 & 1.37--1.56 & 0.4 & 0.3 \\ 
\textsc{Atlas8} spin-2 & 20.3 & 50 & 50 & 2.37 & 1.37--1.56 & -- & -- \\  \hline
\textsc{Cms8} & 19.7 & 33 & 25 & 2.5 & 1.44--1.57 & 0.33 & 0.25 \\ \hline
\end{tabular}
\caption{Selection criteria for the six experimental searches I consider in this paper. For \textsc{Cms13} and \textsc{Cms13/0T}, events are divided into barrel-barrel (BB) and barrel-endcap (BE) categories. The column labeled ``$<|\eta|<$'' indicates the transition $\eta$ region excluded from the selected events. The dual upper limits on $|\eta|$ for the \textsc{Cms8}-BE category indicate that one photon must be in the barrel ($|\eta| <1.44$) and one must be in the endcap ($|\eta|<2.5$). Two different selection criteria are used by ATLAS for their spin-0 and spin-2 analyses.\label{tab:selection}}
\end{table}

\begin{figure}[ht]
\includegraphics[width=0.3\columnwidth]{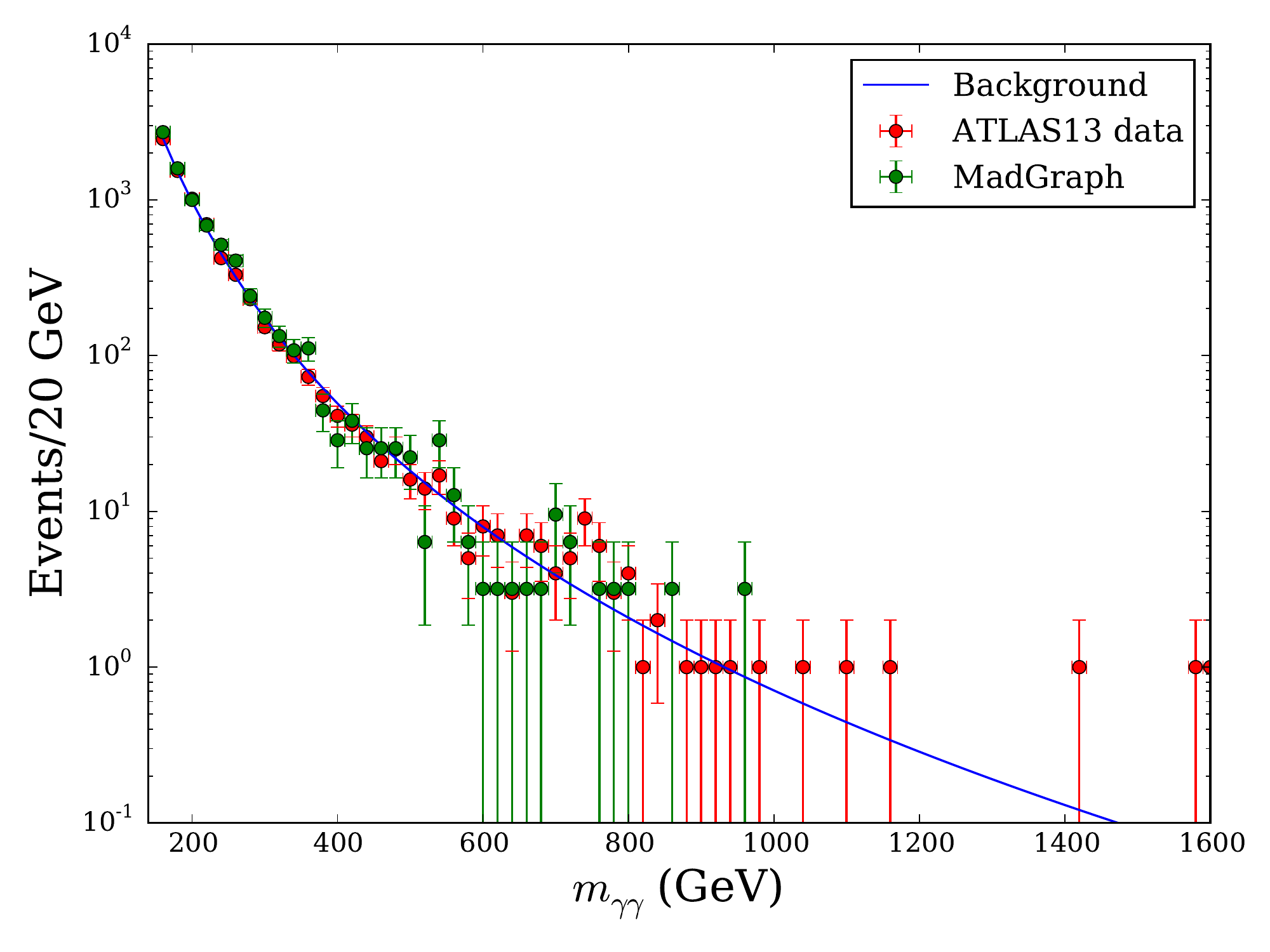} \\
\includegraphics[width=0.3\columnwidth]{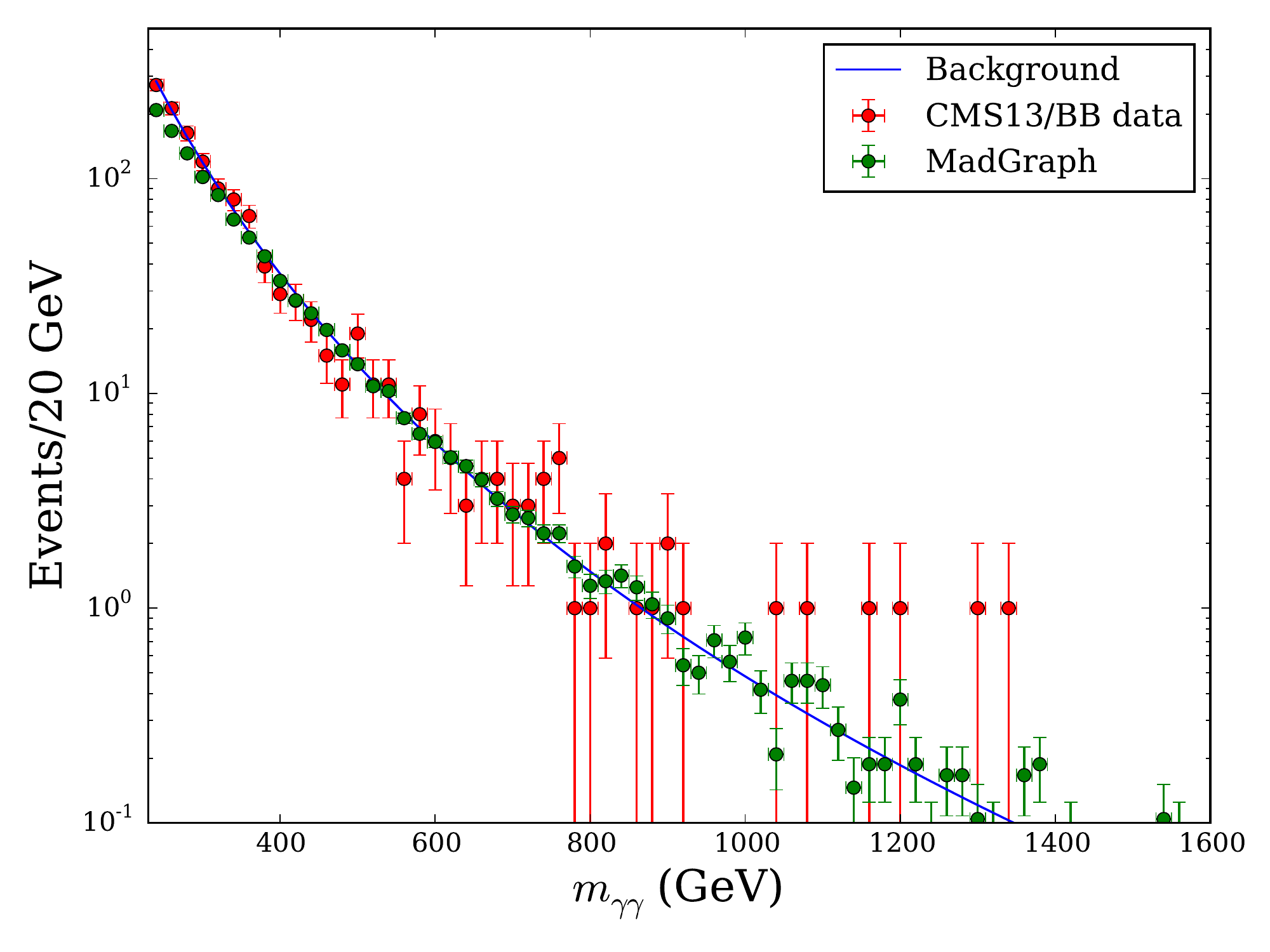}\includegraphics[width=0.3\columnwidth]{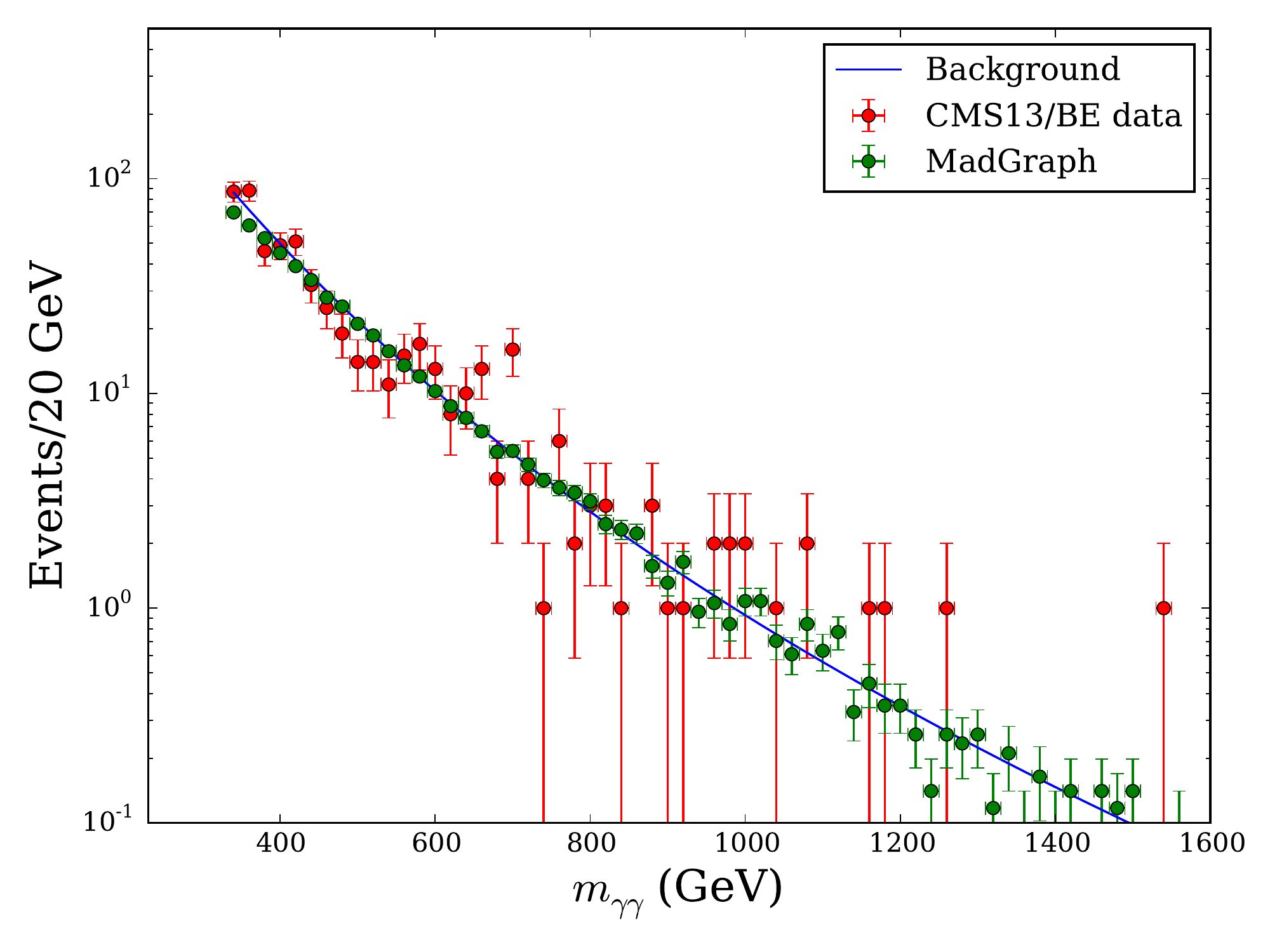}
\includegraphics[width=0.3\columnwidth]{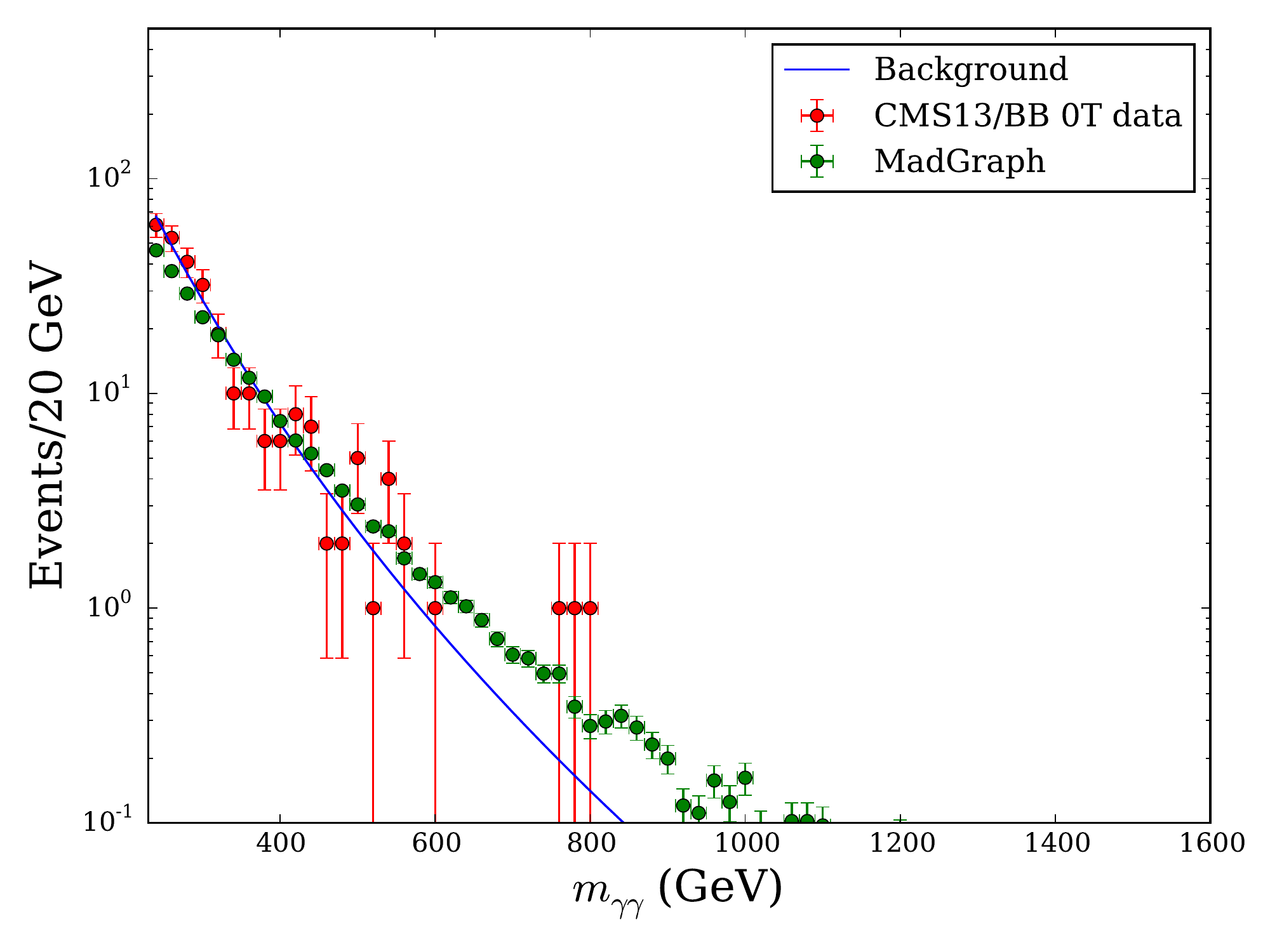}\includegraphics[width=0.3\columnwidth]{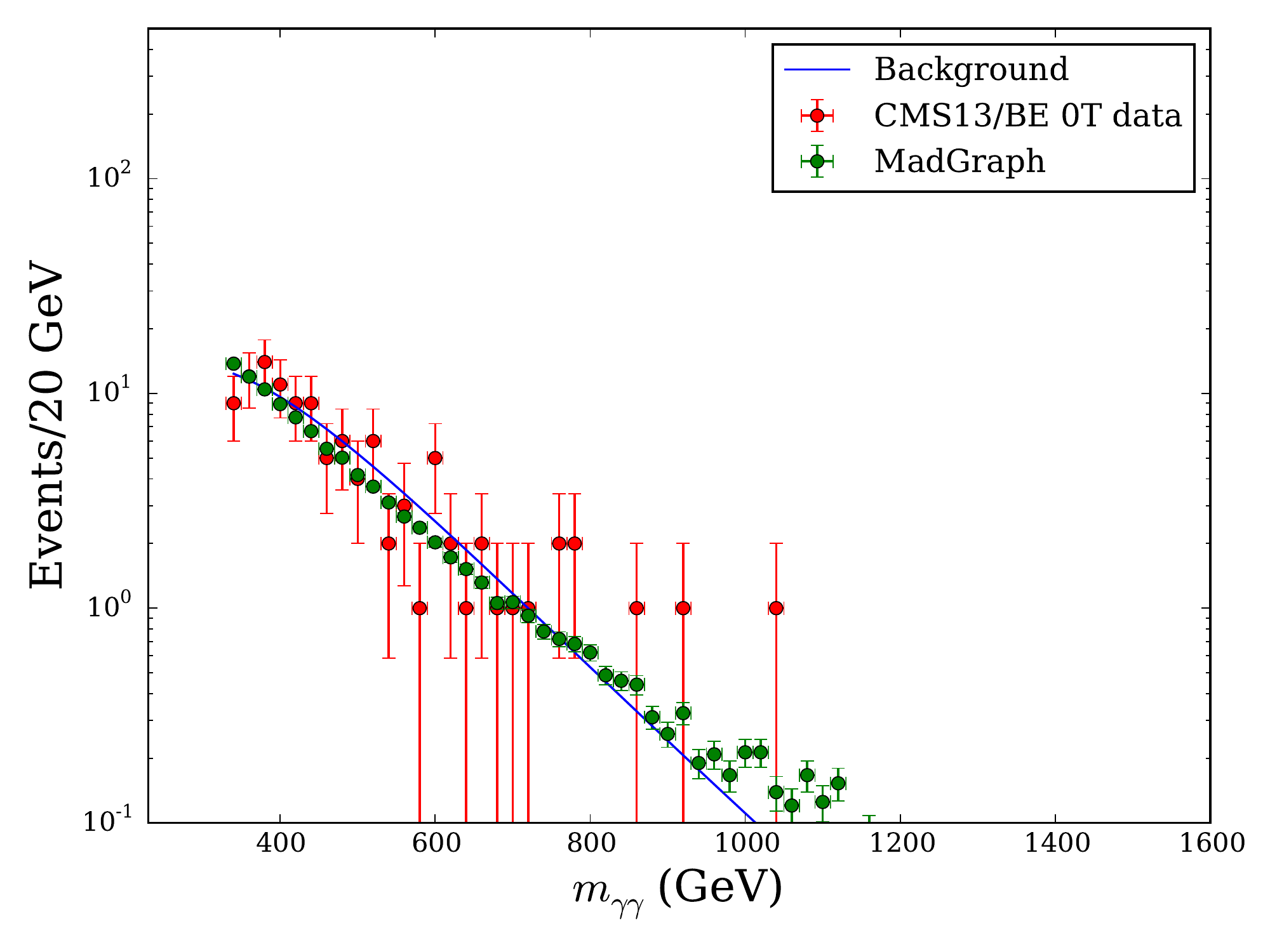}
\includegraphics[width=0.3\columnwidth]{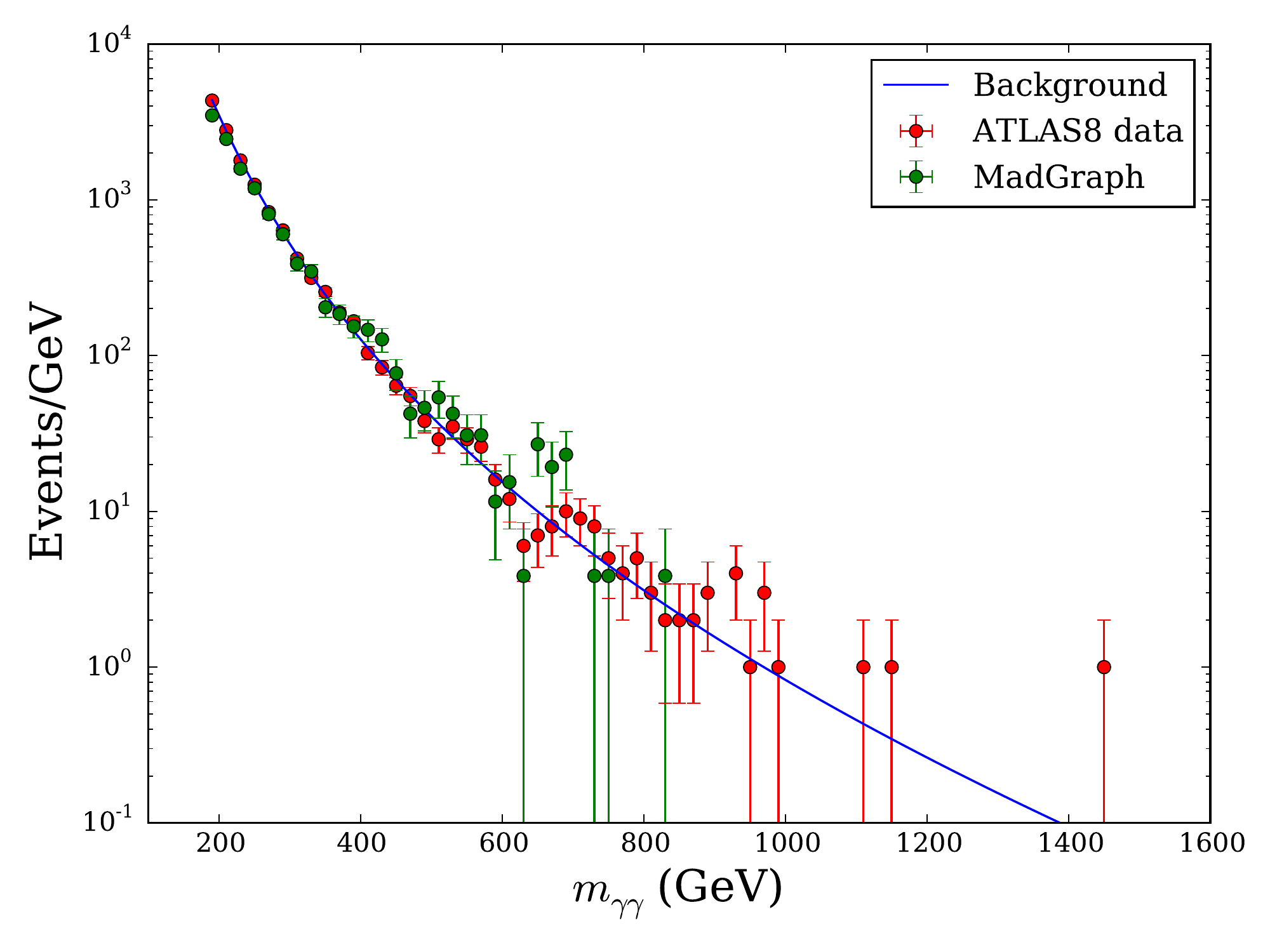}\includegraphics[width=0.3\columnwidth]{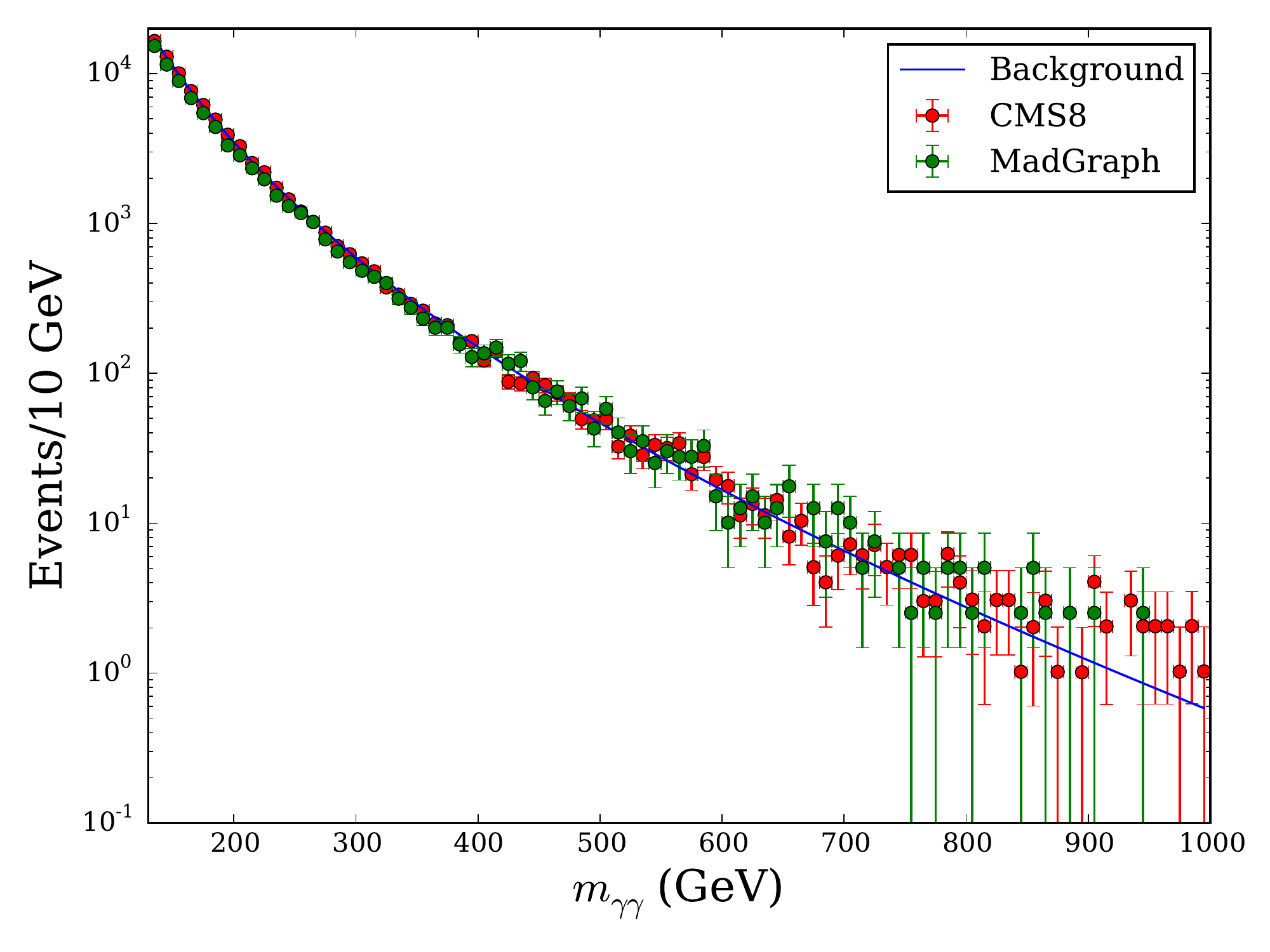}
\caption{Digitized data (red points) from spin-0 \textsc{Atlas13}~\cite{atlas13} (top), \textsc{Cms13}~\cite{cms13} (\textsc{Cms13/0T}~\cite{CMS_0}) in the barrel-barrel (2$^{\rm nd}$(3$^{\rm rd}$) row left) and barrel-endcap (2$^{\rm nd}$(3$^{\rm rd}$)) categories, spin-0 \textsc{Atlas8}~\cite{Aad:2014ioa} (lower left), and \textsc{Cms8}~\cite{Khachatryan:2015qba} (lower right). Best-fit background functions Eqs.~\eqref{eq:atlas_bg}--\eqref{eq:cms8_bg} are shown in blue. \textsc{MadGraph5} simulated background events are shown in green. \label{fig:background_only}}
\end{figure}

As a resonance search, we are interested in peak-like structures in the differential distribution of the diphoton invariant mass $m_{\gamma\gamma}$. As a first step, I verified my ability to reproduce the reported fits to the background functional forms, assuming no signal injected. In each analysis, the backgrounds are fit to a data-driven two-parameter function, assuming Poisson statistics. These functions are
\begin{eqnarray}
\textsc{Atlas8}(\textsc{Atlas13}) : f(m_{\gamma\gamma};a_0,b) & = & \left(1-\left(\frac{m_{\gamma\gamma}}{\sqrt{s}}\right)^{1/3}\right)^b\left(\frac{m_{\gamma\gamma}}{\sqrt{s}}\right)^{a_0}, \label{eq:atlas_bg} \\
\textsc{Cms13}(\textsc{Cms13/0T}): f(m_{\gamma\gamma};a,b) & = & m_{\gamma\gamma}^{a+b\log m_{\gamma\gamma}}, \label{eq:cms13_bg} \\
\textsc{Cms8}: f(m_{\gamma\gamma};p_1,p_2) & = & e^{-p_1m_{\gamma\gamma}}m_{\gamma\gamma}^{-p_2}. \label{eq:cms8_bg}
\end{eqnarray}
After digitizing the binned $m_{\gamma\gamma}$ data from each experiment, I fit to these functional forms, marginalizing over the two free parameters assuming Poisson statistics. While the experiments themselves obviously have access to much more information of the unbinned diphoton events, I am restricted to the public data, which is of course binned. This loss of information will result in some degradation of statistical power, as will be seen, but the difference is not large. My resulting best-fit backgrounds are shown in Figure~\ref{fig:background_only} (using the spin-0 selection for the ATLAS data) overlaid with the experimental data for the six experimental searches. In all cases, I can successfully reproduce the best-fit backgrounds found by the experimental collaborations

It should be noted that these functional forms are data-driven, and out of six diphoton analyses, three different functional forms were chosen. It has been noted that changing the functional forms to increase support at high invariant mass could possibly reduce the significance of the observed excess \cite{Davis:2016hlw}. This is made possible  by the low statistics of diphoton counts at large $m_{\gamma\gamma}$. Further, the 750~GeV diphoton excess sits near the tail of the 8~TeV ATLAS and CMS analyses. Thus, it is possible to ``hide'' the 13~TeV excess in the 8~TeV by lowering the background function in this region and absorbing the excess into the signal. This is especially notable when the signal is assumed to be a wide resonance, covering much of the high $m_{\gamma\gamma}$ range.

After fitting the background functions to the digitized data, I then use these background-only fits to validate my simulation pipeline. I simulate the primary irreducible background of $p p \to \gamma \gamma + X$ using \textsc{MadGraph5}~\cite{Alwall:2014hca}, matched up to two jets at $p_T = 10$~GeV using \textsc{Pythia6}~\cite{Sjostrand:2006za}. Detector simulation is performed using \textsc{Delphes3}~\cite{deFavereau:2013fsa}, with the default ATLAS and CMS detector cards. A $K$-factor of between $1.4-1.8$ was needed to match the experimental yields. The resulting distributions, normalized using the $K$-factors, are also shown in Figure~\ref{fig:background_only}. While the simulated $m_{\gamma\gamma}$ distribution is largely in good agreement, some deviation is observed at low invariant masses. This deviation is due to the lack of box diagrams in the \textsc{MadGraph5} simulation. Fortunately this occurs far from the signal region. Therefore, this simulation technique should be acceptable for the generation of signal events. 

I now turn to the excess at 750~GeV in the 13~TeV data. I fit the data to two possibilities: either a spin-0 or spin-2 particle decaying to two photons with a mass near 750~GeV.\footnote{Spin-1 mediators decaying to diphotons are ruled out by the Landau-Yang theorem, though it may be possible to find gauge bosons mediator solutions through sufficient theoretical model-building efforts \cite{Chala:2015cev}.} In particular, I will discuss the agreement of the four data sets, and the preference in the data (if any) for a wide or narrow resonance. 

\subsection{Spin-0 Resonance \label{sec:spin0}}

Of the avalanche of theory papers discussing the \textsc{Atlas13} and \textsc{Cms13} diphoton anomaly, the majority have considered the spin-0 scenario. Here, I take a model-independent approach, though I do specialize to the CP-even scalar option. The CP-odd pseudoscalar is also possible (see {\it e.g.}~Ref.~\cite{Low:2015qep}), but will result in a very similar analysis. Using \textsc{FeynRules2.3}~\cite{Alloul:2013bka}, I constructed \textsc{MadGraph5} model files for a new scalar spin-0 particle with couplings to {\it one} of the following sets of partons:
\begin{enumerate}
\item Gluons (presumably mediated through a loop) via the interaction
\begin{equation}
{\cal L}_{\Phi-g} \supseteq c_{g} \Phi_g G^{\mu\nu,a}G_{\mu\nu}^a.
\end{equation}
\item Valance quark-antiquark ($u/d$) pairs, through the interaction
\begin{equation}
{\cal L}_{\Phi-q} \supseteq c_{q} \Phi_q \left(\frac{m_{u}}{v} \bar{u}u+\frac{m_{d}}{v}  \bar{d}d\right).
\end{equation}
\item Charm and strange quark-antiquark pairs, through the interaction
\begin{equation}
{\cal L}_{\Phi-Q} \supseteq c_{Q} \Phi_Q \left(\frac{m_{c}}{v}  \bar{c}c+\frac{m_{c}}{v}  \bar{s}s\right).
\end{equation}
\item Bottom quark-antiquark pairs, through 
\begin{equation}
{\cal L}_{\Phi-b} \supseteq c_{b} \Phi_b \frac{m_{b}}{v}  \bar{b}b.
\end{equation}
\end{enumerate}
Here $v$ is the Standard Model Higgs vacuum expectation value $v = 246$~GeV, and the $c_i$ are couplings which can be floated to fit to the observed cross section. Assuming proportionality of interactions to the quark masses is motivated from Minimal Flavor Violation~\cite{D'Ambrosio:2002ex}, but what follows is not too sensitive to this assumption. In all cases, decays to photons are the result of the interaction
\begin{equation}
{\cal L}_{\Phi-\gamma} \supseteq c_{\gamma} \Phi F^{\mu\nu}F_{\mu\nu}.
\end{equation}

The interactions are separated in this way to allow for more fine-grained investigation of the agreement of the 13 and 8~TeV data. If the anomaly is the result of a new particle at $\sim 750$~GeV, the expected signal strength in each experiment should be related by the ratios of the relevant parton distribution functions (p.d.f.s). Any ``realistic'' theory for the anomaly could have couplings to more than one of these sets of partons; in such cases one can reweight the results of this paper. I also note that full, realistic simulation of the gluon-coupling may require resolving the heavy colored particles running in the loop; this is relevant at values of the mediator $p_T$ comparable to the mass of the particles in the loop~\cite{Buckley:2014fba}. Here I assume infinite masses, which presumably is a reasonable assumption for most models, as typically the mediator will be produced nearly at rest and the new colored mediators running the loop must be heavy. I also stress that my analysis assumes that the mediator producing the diphoton excess at $m_{\gamma\gamma} \sim 750$~GeV is indeed a particle with mass near 750~GeV. It is possible that some heavier particle is produced, followed by a cascade decay resulting in the observed excess (see Ref.~\cite{Knapen:2015dap}). By increasing the mass of the mediator, the constraints from the 8~TeV data can be weakened, and only the direct comparison of \textsc{Atlas13} and \textsc{Cms13} would be relevant.

For each model, I generated $p p \to (\Phi \to \gamma\gamma) + X$ events, matched to two jets with a matching scale of 10~GeV, using the \textsc{MadGraph5}/\textsc{Pythia6}/\textsc{Delphes3} simulation chain described previously. No cuts were placed on the $\Phi$ particles at the generator level. I scanned over mediator masses from 700 to 800~GeV, under two assumptions of the width: narrow and wide. The narrow width mediator has a width set by the Lagrangian terms above (typically $\Gamma \lesssim 50~$MeV), which much less than the diphoton invariant mass resolution. The ``wide'' resonance has a width of $\Gamma = 45$~GeV, as this is reported best-fit width in the \textsc{Atlas13} results. Scanning over the widths would be preferable to considering just these two assumptions. However, as I am considering only the binned data, the effective resolution of this scan is poor and extremely computationally intensive.

I find nearly flat signal acceptance for the range of mediator masses considered, for both the narrow and wide hypotheses. The \textsc{Atlas13} analysis has an efficiency of $\sim 50\%$ for spin-0 mediators -- this is somewhat lower than the value quoted in Ref.~\cite{Falkowski:2015swt}. The combined barrel-barrel and barrel-endcap \textsc{Cms13} analysis has an efficiency of $\sim 60\%$, with 40\% of events ending up in the barrel-barrel category, and 20\% in the barrel-endcap. The \textsc{Cms8} search also has a 60\% efficiency, while the \textsc{Atlas8} Higgs search is slightly less than this.

Using the $m_{\gamma\gamma}$ distributions constructed from the simulated $\Phi$ production and decay, I then fit the signal plus background to the data provided from each experiment, floating the normalization of the signal in terms of the production cross section times branching ratio into photons $\sigma \times \mbox{BR}_{\gamma\gamma}$. In each case, I refit the background distributions using the appropriate functional forms Eqs.~\eqref{eq:atlas_bg}--\eqref{eq:cms8_bg}, marginalizing over the background function parameters, and maximizing the log likelihood assuming Poissonian statistics.

Production cross sections for the 8~TeV data are then reweighted to the 13~TeV results using \textsc{MadGraph5} simulation to obtain the necessary p.d.f.~factors. All cross sections quoted in this paper are in terms of the 13~TeV data, and are thus directly comparable. The ratio of 13~TeV cross sections to 8~TeV cross sections (for both wide and narrow resonances), are nearly independent of resonance mass in the 700-800~GeV range considered here. For scalar mediators coupling to gluons ($\Phi_g$), this ratio is $\sim 4.5$, for couplings to valance quarks ($\Phi_q$) it is $\sim 3.1$, for couplings to $s/c$ quarks ($\Phi_Q$) it is $\sim 4.2$, and $\sim 4.0$ for bottom quarks ($\Phi_b$). These cross section ratios come from two-jet matching, and so include initial states other than those that couple directly to the mediator in question. Due to the similarity of the $\Phi_g$, $\Phi_b$, and $\Phi_Q$ p.d.f.~ratios, I will show only $\Phi_g$ in this section, and relegate the $\Phi_b$ and $\Phi_Q$ results to Appendix~\ref{sec:appendix}.

\begin{figure}[th]
\includegraphics[width=0.4\columnwidth]{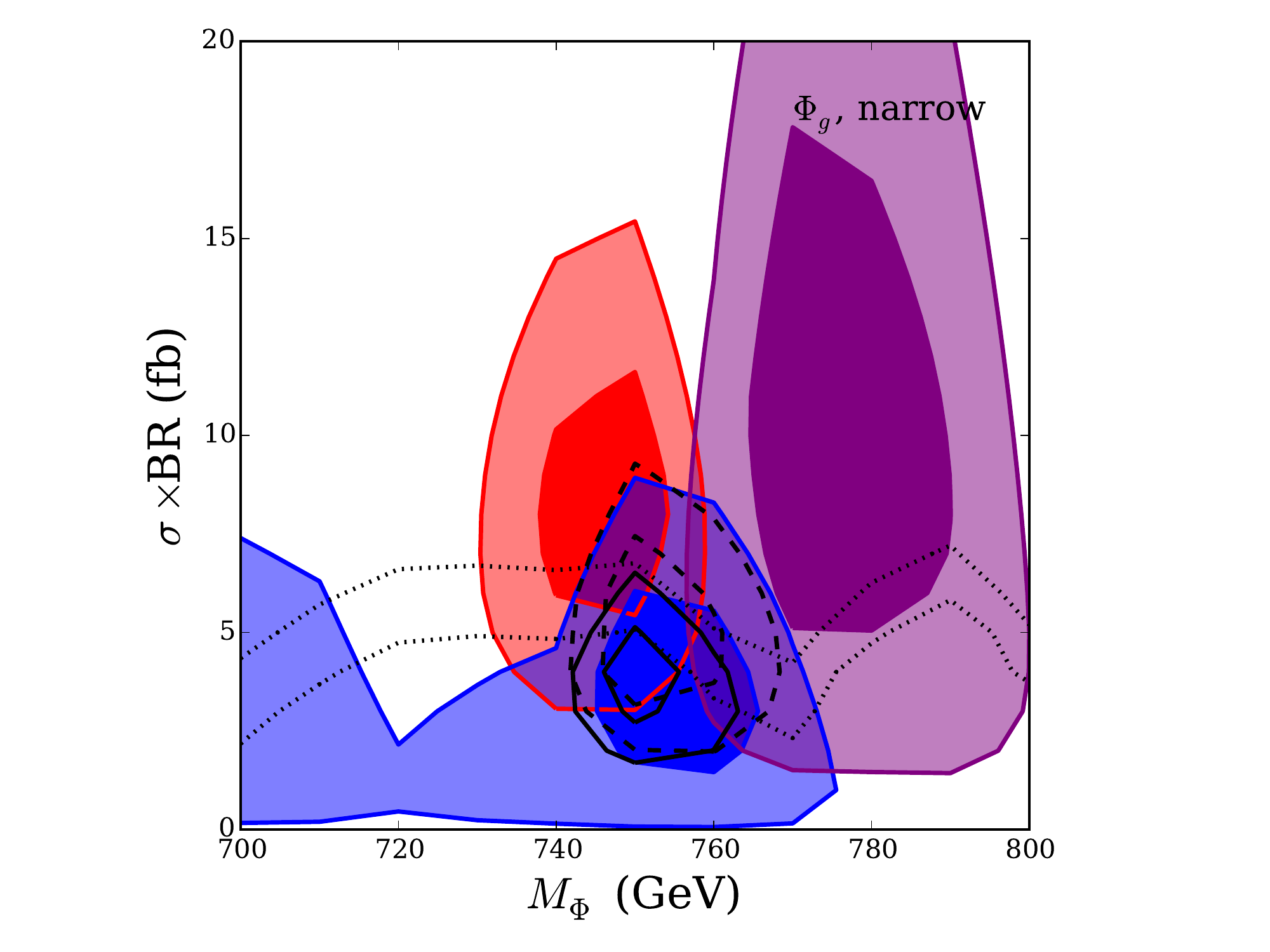}\includegraphics[width=0.4\columnwidth]{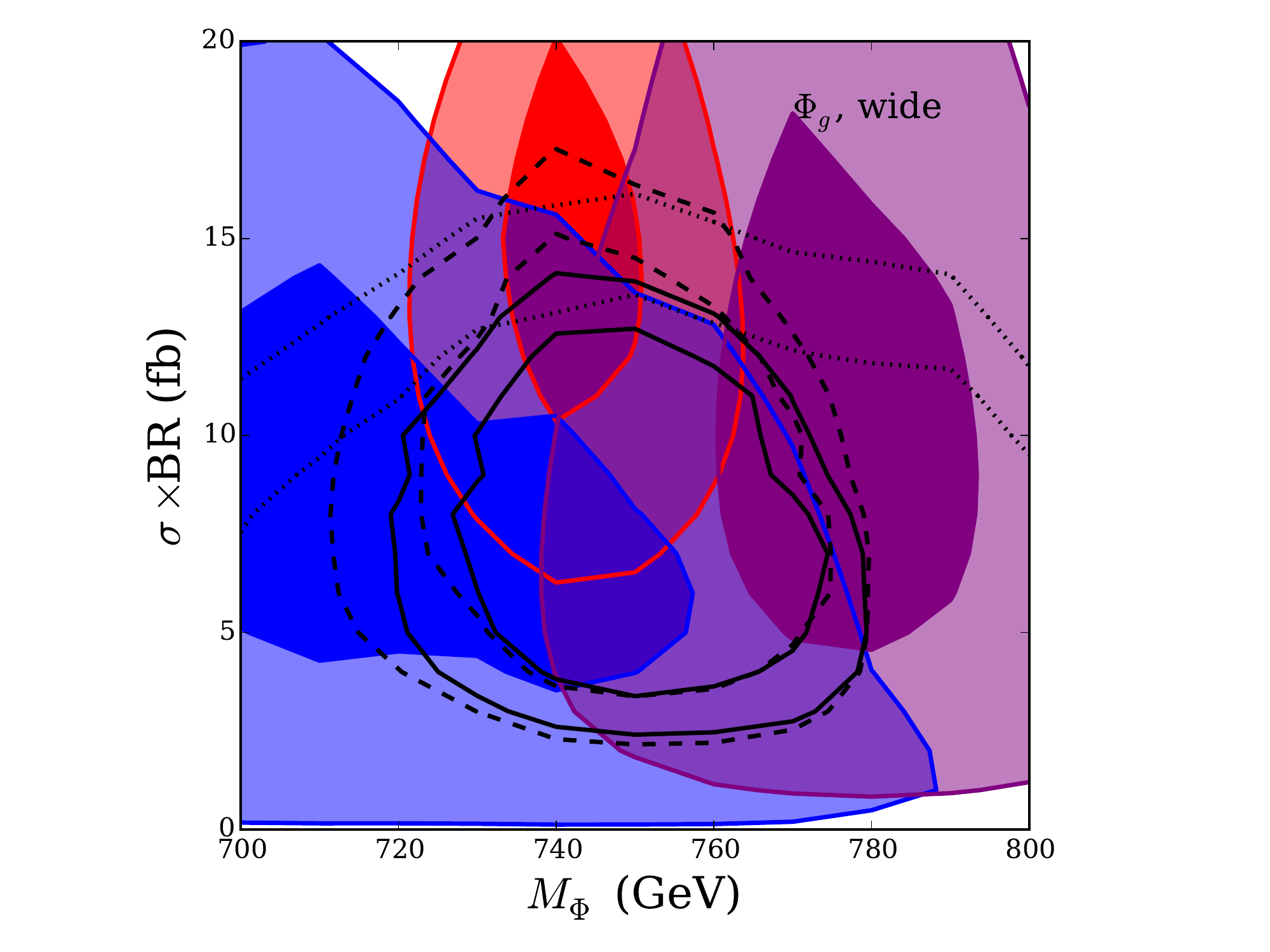}
\includegraphics[width=0.4\columnwidth]{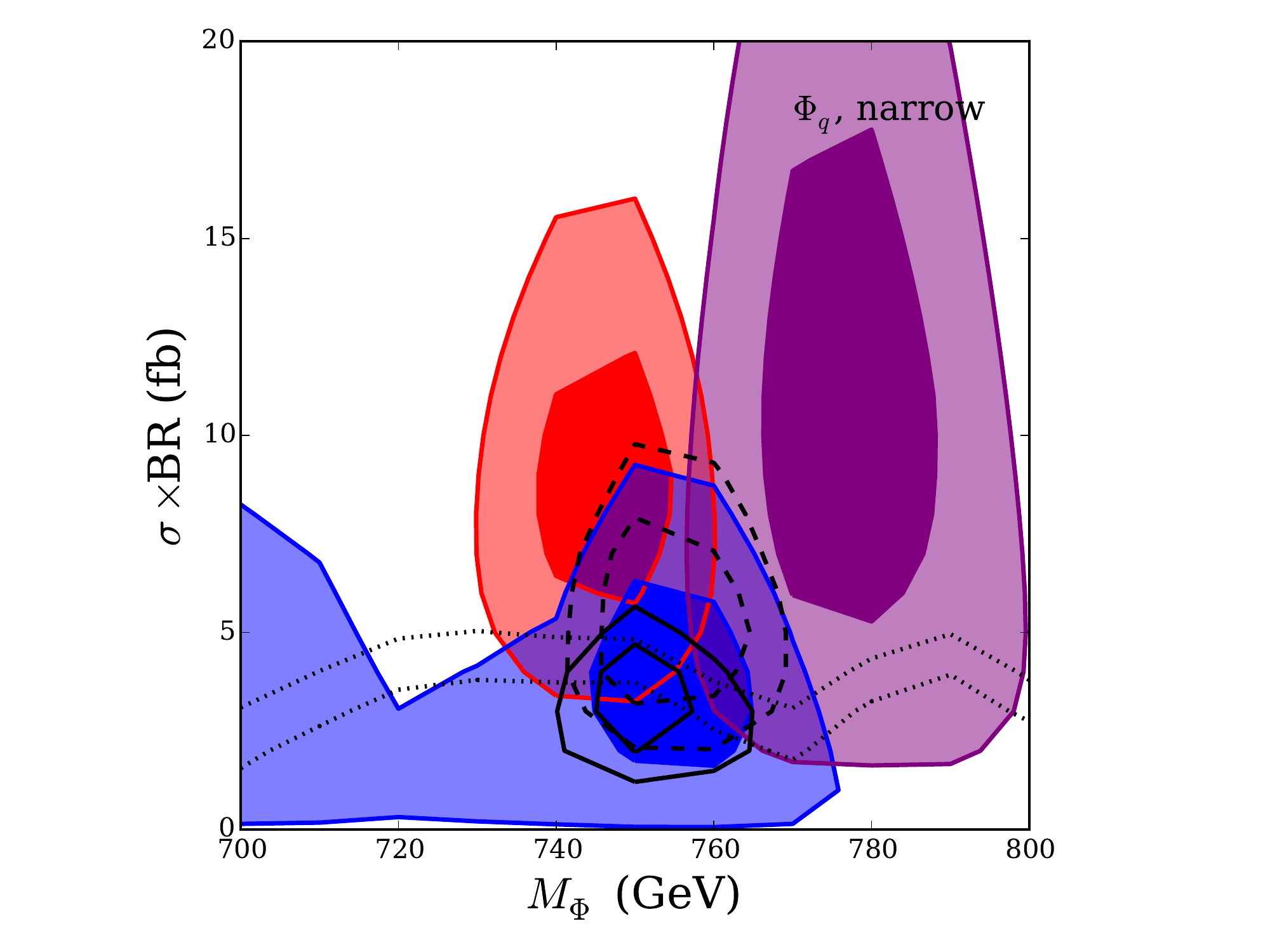}\includegraphics[width=0.4\columnwidth]{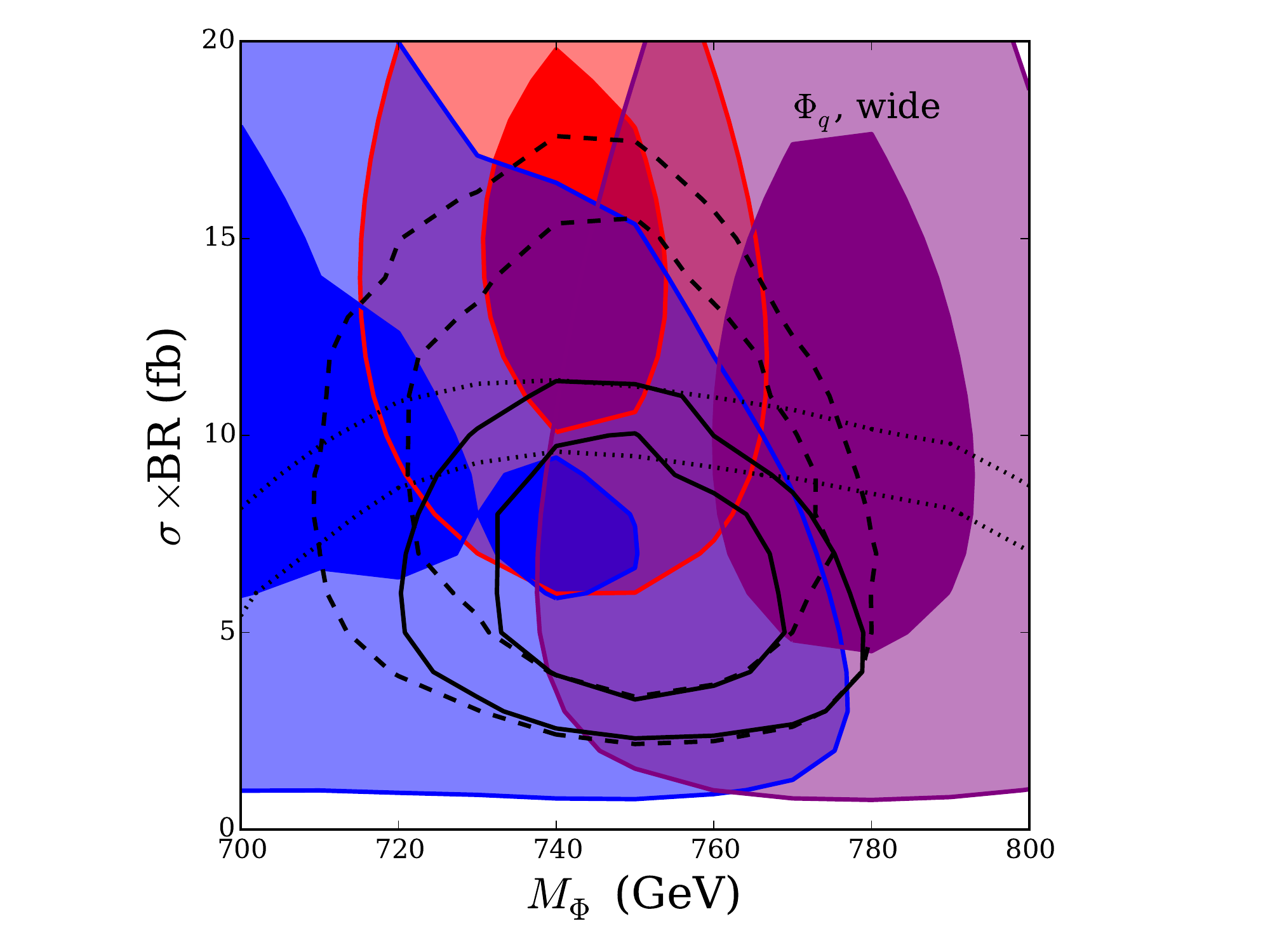}
\caption{Best fit regions (1 and $2\sigma$) of a spin-0 mediator decaying to diphotons, as a function of mediator mass and 13~TeV cross section, assuming the indicated mediator couplings to partons and mediator width. Red regions are the 1 and $2\sigma$ best-fit regions for the \textsc{Atlas13} data, blue is the fit to \textsc{Cms13} data, and purple is the \textsc{Cms13/0T}. The combined best fit for both \textsc{Atlas13}, \textsc{Cms13}, and \textsc{Cms13/0T} (\textsc{Combo13}) are the regions outlined in black dashed lines. The 1 and $2\sigma$ upper limits from the combined 8~TeV data (\textsc{Combo8}) are the black dashed lines (with cross sections converted to 13~TeV-equivalents). The best-fit signal combination of all six data sets (\textsc{Combo}) are the black solid lines.  \label{fig:spin0_xs}}
\end{figure}

\begin{figure}[ht]
\includegraphics[width=0.4\columnwidth]{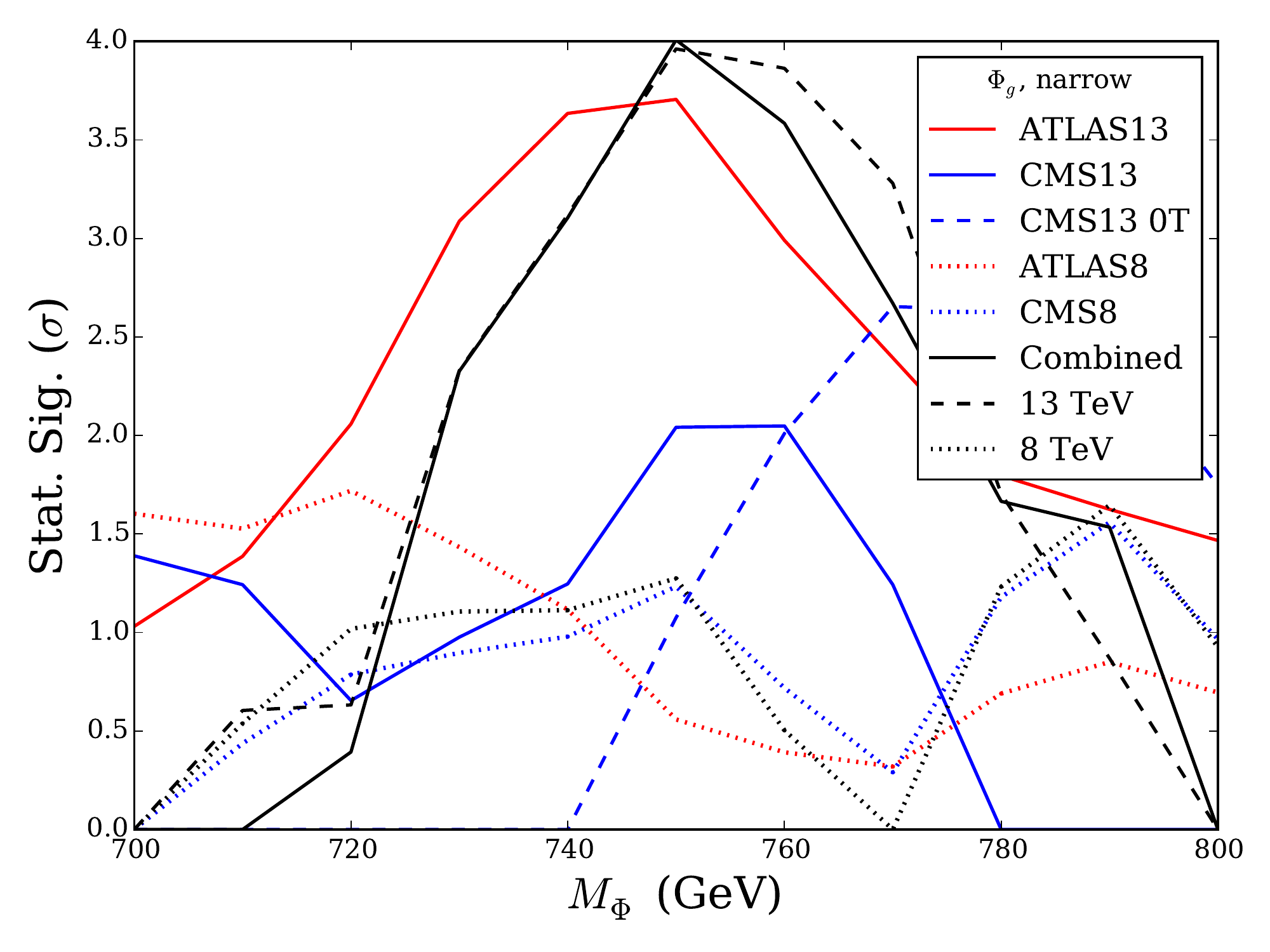}\includegraphics[width=0.4\columnwidth]{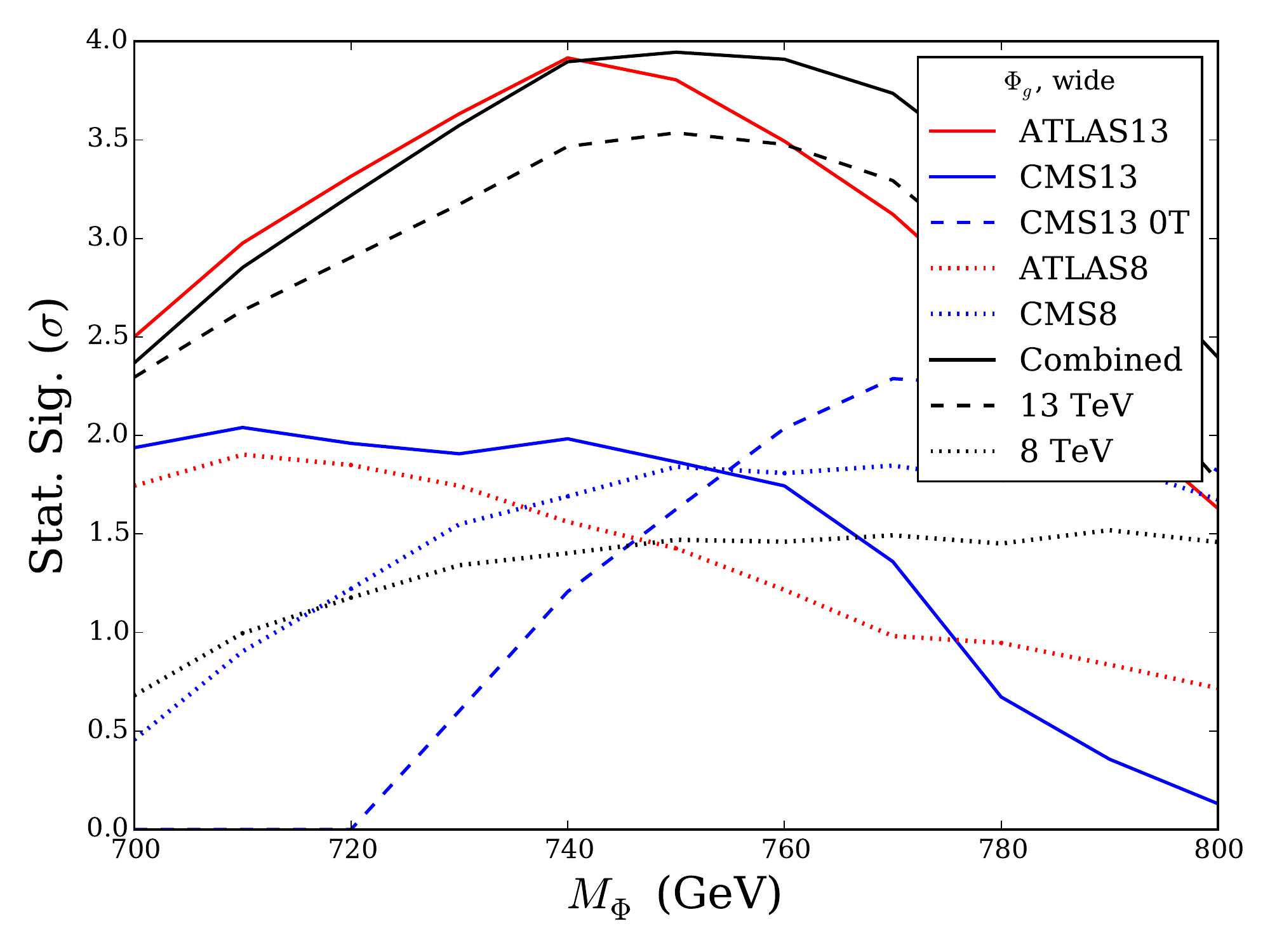}
\includegraphics[width=0.4\columnwidth]{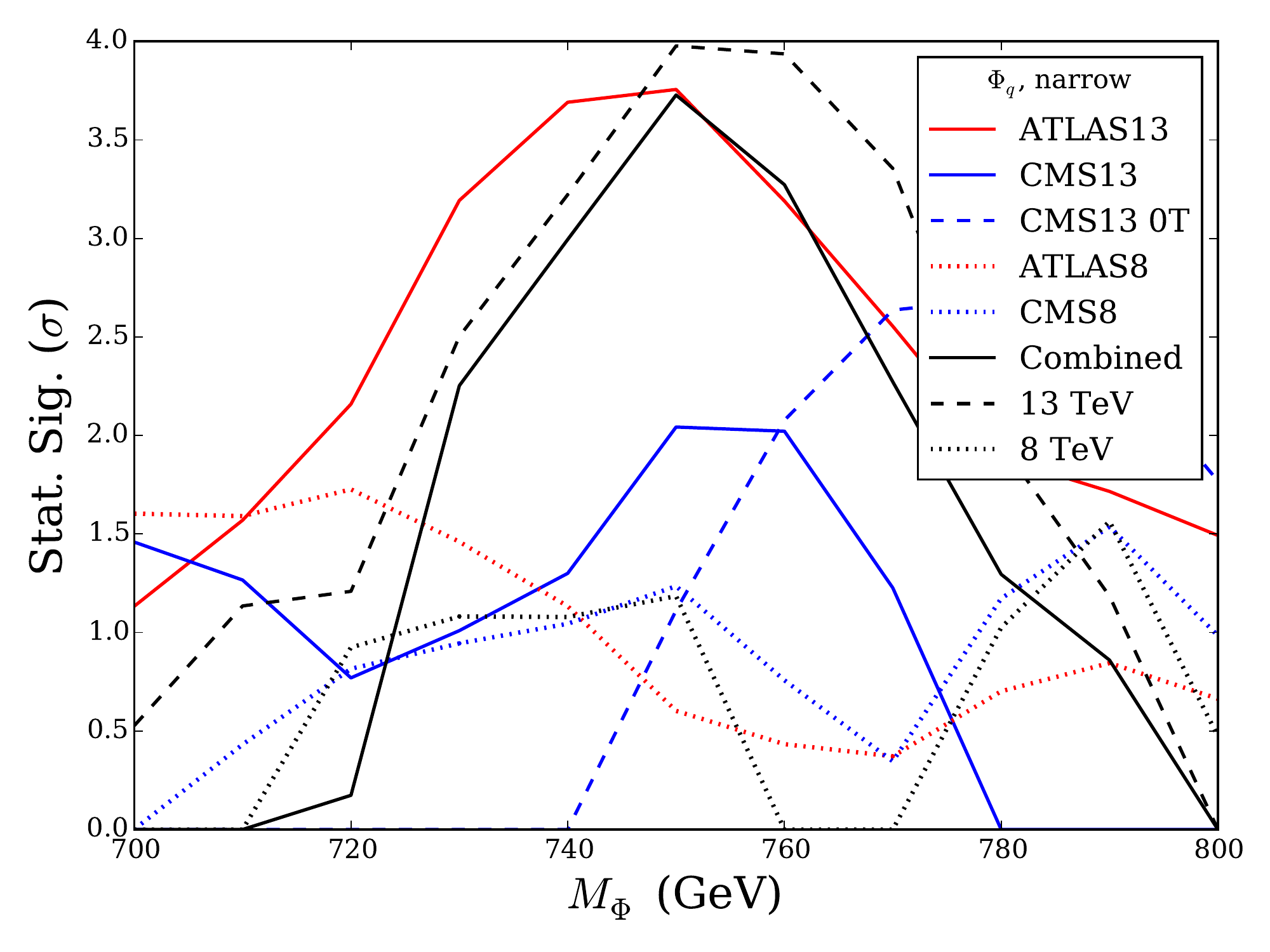}\includegraphics[width=0.4\columnwidth]{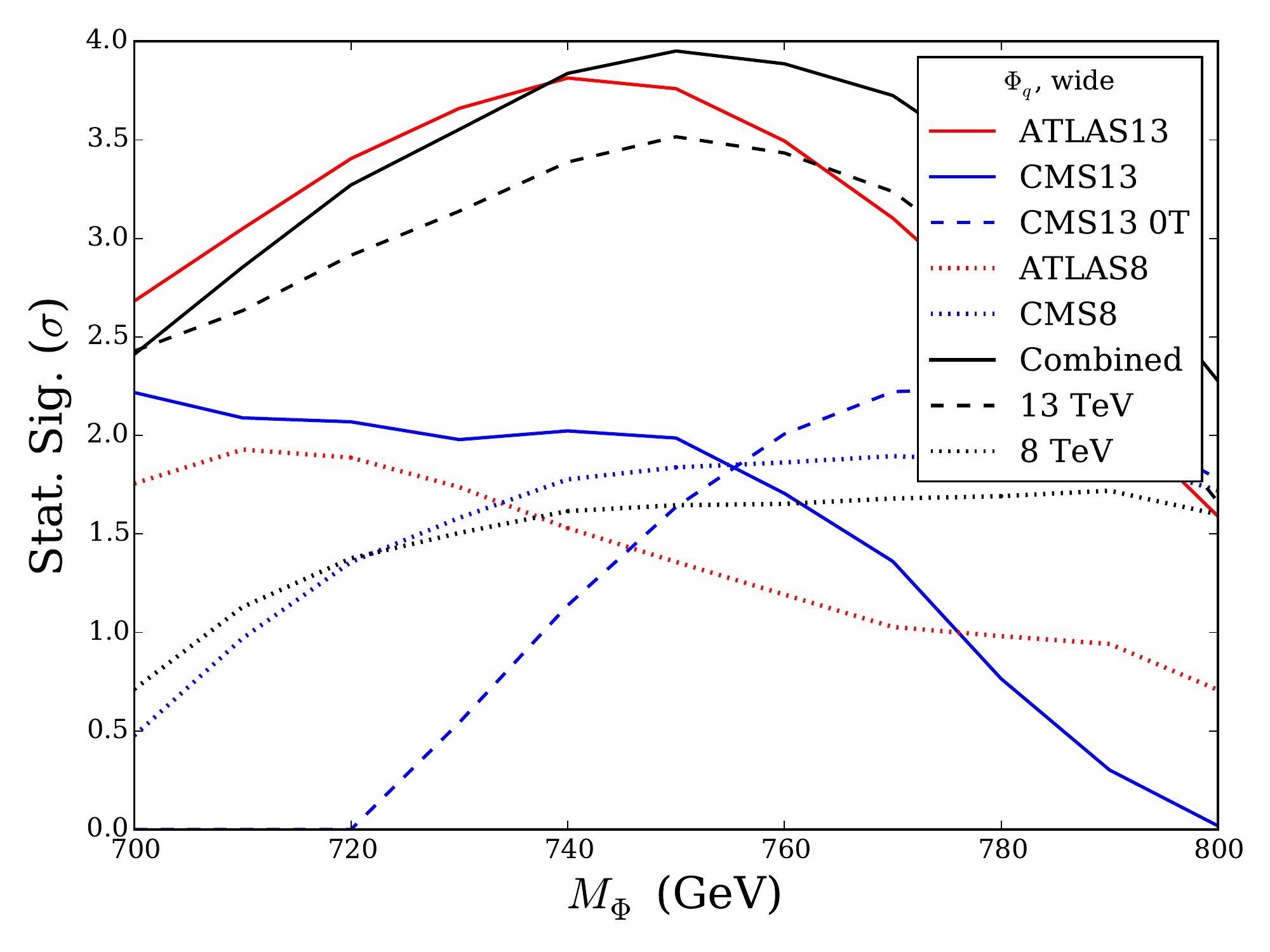}
\caption{Statistical significance for a spin-0 mediator decaying to diphotons, as a function of mediator mass, assuming the indicated mediator couplings to partons and mediator width. At each mass, the cross section is set to the value that maximizes statistical significance for a signal (see Figure~\ref{fig:spin0_xs}). The solid red line is the statistical significance of the \textsc{Atlas13} data alone, solid blue is \textsc{Cms13}, dashed blue is \textsc{Cms13/0T}, and dotted red and blue lines are \textsc{Atlas8} and \textsc{Cms8}, respectively. When comparing across experiments, note that these significances do not correspond to the same value of the cross section. The dashed (dotted) black line is the combination of 13(8)~TeV data, requiring the same cross section in both ATLAS and CMS. The solid black line is the combined significance of all six data sets.  \label{fig:spin0_stat}}
\end{figure}

Using the fitting procedure described, in Figure~\ref{fig:spin0_xs}, I show the best-fit values for the cross-sections time branching ratios into photons, as a function of resonance mass (again, for both choices of overall width). The statistical significance of these best-fit excesses are shown in Figure~\ref{fig:spin0_stat} (for $\Phi_Q$ and $\Phi_b$ interpretations, see Figures~\ref{fig:spin0_xs_app} and \ref{fig:spin0_stat_app} in Appendix~\ref{sec:appendix}). The statistical significance is obtained from the $\Delta \log$ likelihood assuming a single degree of freedom. Best-fit $\sigma \times$BR and statistical significances are shown individually for the \textsc{Atlas13} and \textsc{Cms13} data-sets; as is by now well-understood, these analyses show an excess near 750~GeV. Adding in the \textsc{Cms13/0T} data also shows some preference for a signal slightly about 750~GeV.


My statistical fits must be compared with the quoted values from the ATLAS and CMS Collaborations themselves. For \textsc{Atlas13}, I find a local statistics-only significance for a narrow signal of $\sim 3.6\sigma$ for a particle with a mass of 750~GeV. I further find a marginal improvement to the local significance (up to $\sim 3.9\sigma$) for the $\Gamma = 45$~GeV hypothesis. The full experimental analysis finds $3.6\sigma$ for the narrow width and $3.9\sigma$ for the wider resonance, using the unbinned data and including systematic errors which are not replicable in a theory analysis -- though the exact agreement of my numbers with the experimental results must be seen as coincidental. For the \textsc{Cms13} data, I find a local statistical significance of $2.0\sigma$ for the narrow width hypothesis, while the CMS Collaboration found a local significance of $2.6\sigma$. Thus, my results combining the data sets are also likely to be underestimations of the true significance of various combinations of data sets (despite the fact that I have neglected systematic errors), though of course one cannot be sure barring a full experimental analysis.

Going further, I next consider the evidence for an excess in combinations of the data sets. Looking first at the 13~TeV data, I test the statistical significance of a single resonance with a common $\sigma \times$BR in both the \textsc{Atlas13}, \textsc{Cms13}, and \textsc{Cms13/0T} data. This ``\textsc{Combo13}'' result prefers a signal at 750~GeV with a cross section between the \textsc{Atlas13} and \textsc{Cms13} value, as expected. More interesting perhaps is the change in the statistical significance of this excess: for narrow widths, the \textsc{Combo13} best-fit cross section of $\sim 5$~fb is preferred at $\sim 3.9\sigma$. This is an increase from the \textsc{Atlas13} individual fit, but below the naive expectation one might have from combining the \textsc{Atlas13}, \textsc{Cms13}, and  \textsc{Cms13/0T} significances ($\sim \sqrt{(3.6\sigma)^2+(2.0\sigma)^2+(2.5\sigma)^2} = 4.8\sigma$), as the best-fit cross sections and masses for the three experimental analyses disagree. For wide resonances, combining the \textsc{Atlas13} and \textsc{Cms13} data results in a net decrease in the statistical significance relative to the \textsc{Atlas13} data, to $\sim 3.4\sigma$. This is the first indication that the wide resonance seen by ATLAS is disfavored by the CMS results. 

The preceding set of statements is largely independent of the type of the coupling between the mediator and the proton's partons. However, when adding in the 8~TeV data, I must specify the coupling in order to determine the 8~TeV cross section which is equivalent to the 13~TeV value. For my purposes, it suffices to discuss the coupling to gluons and compare to the coupling to the valence $u/d$ quarks, as couplings to other quark flavors have similar pd.f.s to gluons and thus have very similar conclusions. As can be seen in Figure~\ref{fig:spin0_xs}, the combined 8~TeV data disfavors the \textsc{Atlas13} best-fit cross section for a narrow width at $1\sigma$. 

However, if I instead ask for the best fit region for a single cross section fitting both the 13 and 8~TeV data, I find that there is a good fit for a narrow resonance at a 4~fb cross section, close to the \textsc{Cms13} value, assuming a gluon coupling. This is within the $1\sigma$ region for \textsc{Cms13}, within $2\sigma$ of \textsc{Atlas13}, and is less than $1\sigma$ tension with the 8~TeV data. The statistical preference for this signal is identical to the combined fit to \textsc{Atlas13} and \textsc{Cms13}, at about $4.0\sigma$. 

In the wide resonance interpretation, the 8~TeV data is much less constraining. While the combination of \textsc{Atlas13} and \textsc{Cms13} data reduces the preference for a wide signal, adding the 8~TeV data returns the total statistical significance to $\sim 4.0\sigma$ assuming couplings to gluons -- about the same significance as \textsc{Atlas13} alone. This is driven by the ability for the background models of the 8~TeV data to absorb the signal without resulting in any large excess above the observed smooth distribution. The insensitivity of the 8~TeV data to the broad excess is enough to hide even that larger cross sections required by the light quark coupling.

\subsection{Spin-2 Resonance \label{sec:spin2}}

I now turn to the spin-2 possibility. My general approach is the same as in Section~\ref{sec:spin0}: I investigate individual couplings to the partons one-by-one, in order to make comparisons between the 8 and 13~TeV data. The mediator here is based on a spin-2 Kaluza-Klein graviton $K$, as implemented in \textsc{MadGraph5} by Ref.~\cite{deAquino:2011ix} (see also Refs.~\cite{Giudice:1998ck,Han:1998sg,Hagiwara:2008jb}. I modified the relevant couplings to limit the interactions to gluons, light quarks, second generation quarks, or bottom quarks. In all cases, the coupling to photons is generated through the simplified interaction
\begin{eqnarray}
{\cal L}_{K-\gamma} & \supseteq & c_\gamma K^{\mu\nu}T_{\gamma,\mu\nu}, \\
T^{\mu\nu}_\gamma & = & \eta^{\mu\nu}\left( -\frac{1}{4}F^{\rho\sigma}F_{\rho\sigma}+(\partial_\rho\partial_\sigma A^\sigma)A^\rho +\frac{1}{2} \partial_\rho A^\rho \partial_\sigma A^\sigma \right) \nonumber \\
& & -F^{\mu\rho}F^{\nu}_\rho+(\partial^\mu\partial_\rho A^\rho)A^\nu+(\partial^\nu\partial_\rho A^\rho)A^\mu.
\end{eqnarray}
The terms involving $A$ fields are gauge-fixing terms in the Feynman gauge. The stress-energy tensor for a quark $q$, relevant for the $K$-quarks interaction, is given by
\begin{eqnarray}
T^{\mu\nu} & = & -i\eta^{\mu\nu}\left( \bar{q}\gamma^\rho\partial_\rho q - \frac{1}{2}\partial^\rho(\bar{q}\gamma_\rho q) \right)+\frac{i}{2}\left(\bar{q}\gamma^\mu\partial^\nu q + \bar{q}\gamma^\nu\partial^\mu q-\frac{1}{2} \partial^\mu(\bar{q}\gamma^\nu q)-\frac{1}{2} \partial^\nu(\bar{q}\gamma^\mu q)  \right). \label{eq:Tq}
\end{eqnarray}

The production is through one of the following interactions:
\begin{enumerate}
\item Gluons, through the operator
\begin{eqnarray}
{\cal L}_{K-g} & \supseteq & c_g K_g^{\mu\nu}T_{g,\mu\nu}, \\
T^{\mu\nu}_g & = & \eta^{\mu\nu}\left( -\frac{1}{4}G^{a,\rho\sigma}G^a_{\rho\sigma}+(\partial_\rho\partial_\sigma G^{a,\sigma})G^{a,\rho} +\frac{1}{2} \partial_\rho G^{a,\rho} \partial_\sigma G^{a,\sigma} \right)\nonumber \\
 & & -G^{a,\mu\rho}G^{a,\nu}_\rho+(\partial^\mu\partial_\rho G^{a,\rho})G^{a,\nu}+(\partial^\nu\partial_\rho G^{a,\rho})G^{a,\mu}.
\end{eqnarray}

\item Light valence quarks, $u/d$, through the interaction
\begin{eqnarray}
{\cal L}_{K-q} & \supseteq & c_q K_q^{\mu\nu} T_{q,\mu\nu},
\end{eqnarray}
where $T_q^{\mu\nu}$ is the stress-energy tensor Eq.~\eqref{eq:Tq} with $q=u/d$. 

\item Second generation quarks, $s/c$, through
\begin{eqnarray}
{\cal L}_{K-Q} & \supseteq & c_Q K_Q^{\mu\nu} T_{Q,\mu\nu},
\end{eqnarray}
where $T_Q^{\mu\nu}$ is the stress-energy tensor Eq.~\eqref{eq:Tq} with $q=s/c$. 

\item The bottom quark, through 
\begin{eqnarray}
{\cal L}_{K-b} & \supseteq & c_b K_b^{\mu\nu} T_{b,\mu\nu},
\end{eqnarray}
where $T_b^{\mu\nu}$ is the stress-energy tensor Eq.~\eqref{eq:Tq} with $q=b$. 
\end{enumerate}

The major difference in the analysis when compared to the spin-0 case is change in acceptance of the experiments to diphoton events. These changes are different for each of the six experimental searches I consider. After the ATLAS reanalysis from the 2016 Moriond conference, which uses a separate set of selection criteria for the spin-0 and spin-2 searches, I find that spin-2 signal events have an have an acceptance of $\sim 55\%$ in \textsc{Atlas13} and \textsc{Atlas8} for 750~GeV mediators. This is essentially the same as the acceptance for spin-0 mediators in these experiments. However, there are 75\% more events in the spin-2 analysis. I find the barrel-barrel \textsc{Cms13}(\textsc{Cms13/0T}) signal acceptance is $\sim 35\%$, while the barrel-endcap signal acceptance is 25\%, for a combined efficiency of nearly 60\%, essentially the same as for the spin-0 case. The acceptance of spin-2 signals for \textsc{Cms8} is 45\%, a significant drop from the 60\% acceptance for spin-0 signals.

\begin{figure}[ht]
\includegraphics[width=0.4\columnwidth]{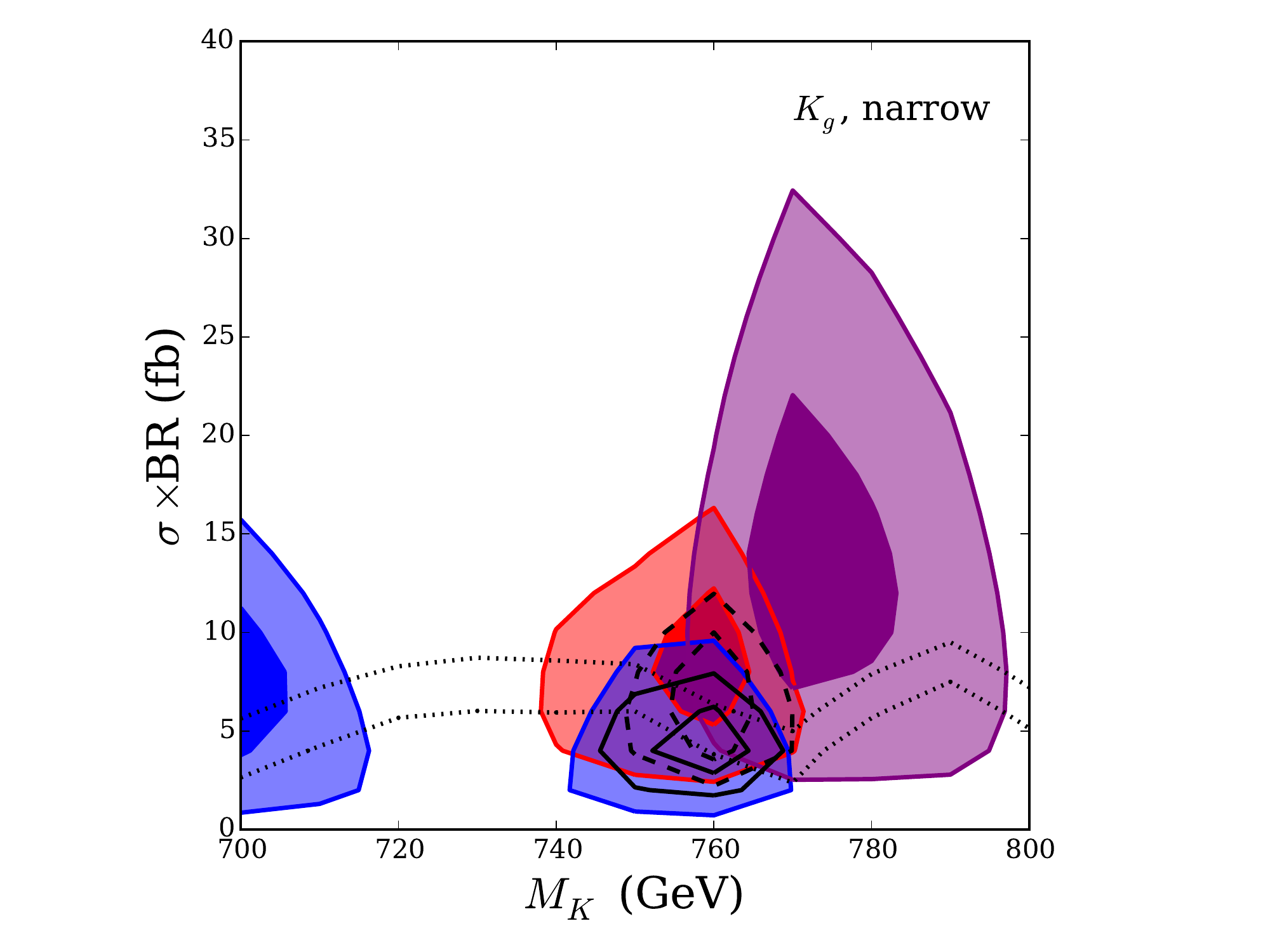}\includegraphics[width=0.4\columnwidth]{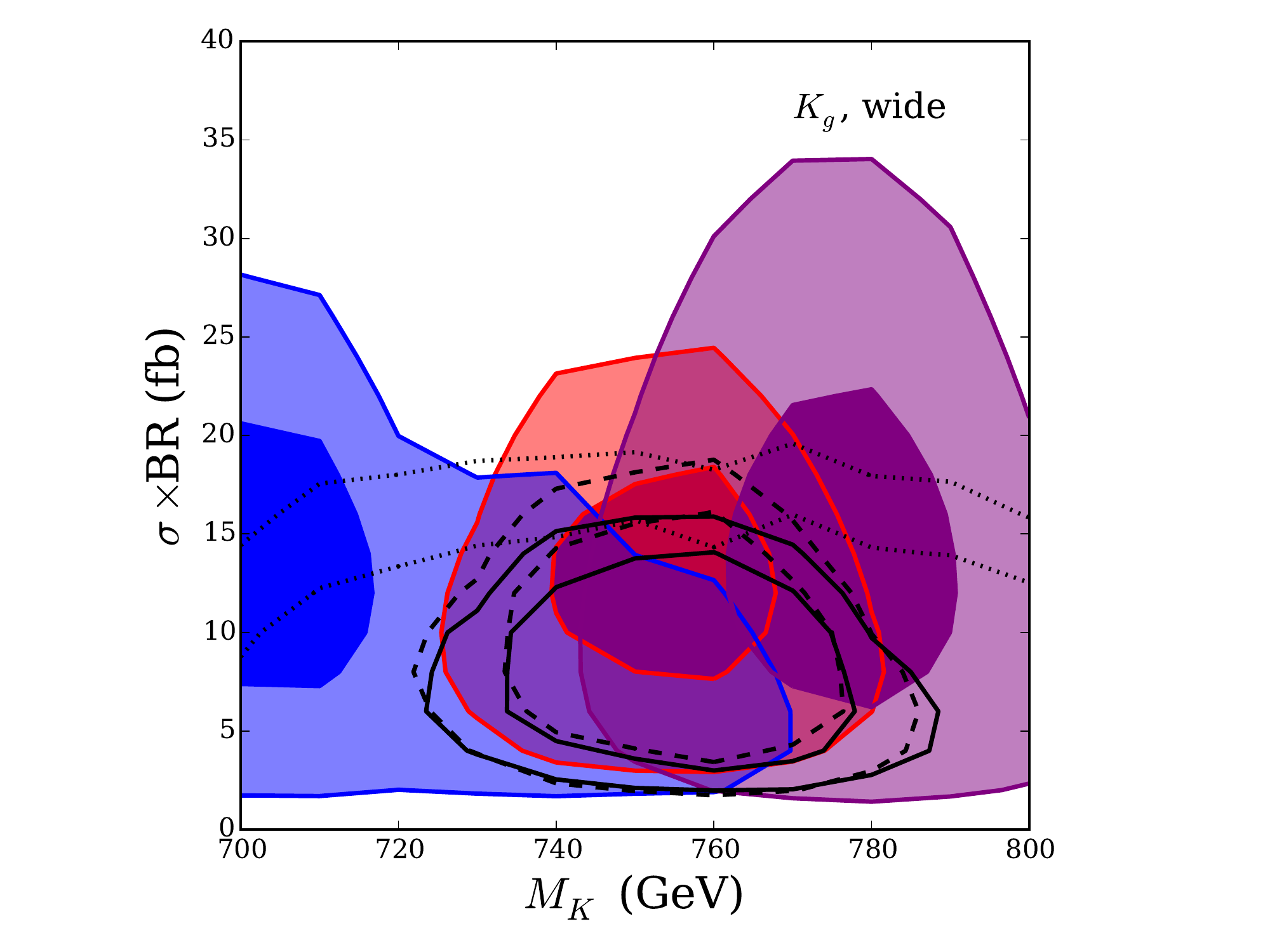}
\includegraphics[width=0.4\columnwidth]{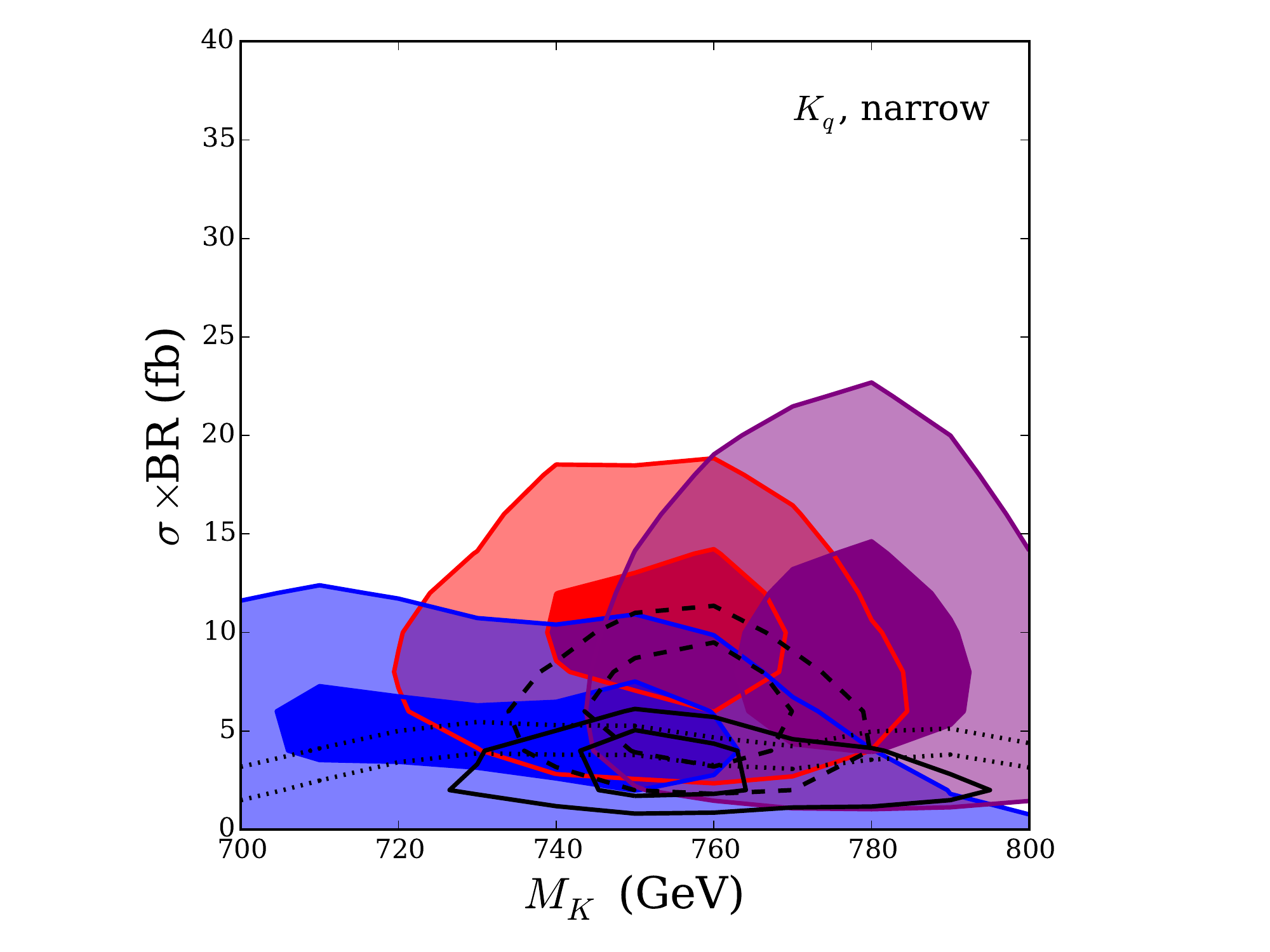}\includegraphics[width=0.4\columnwidth]{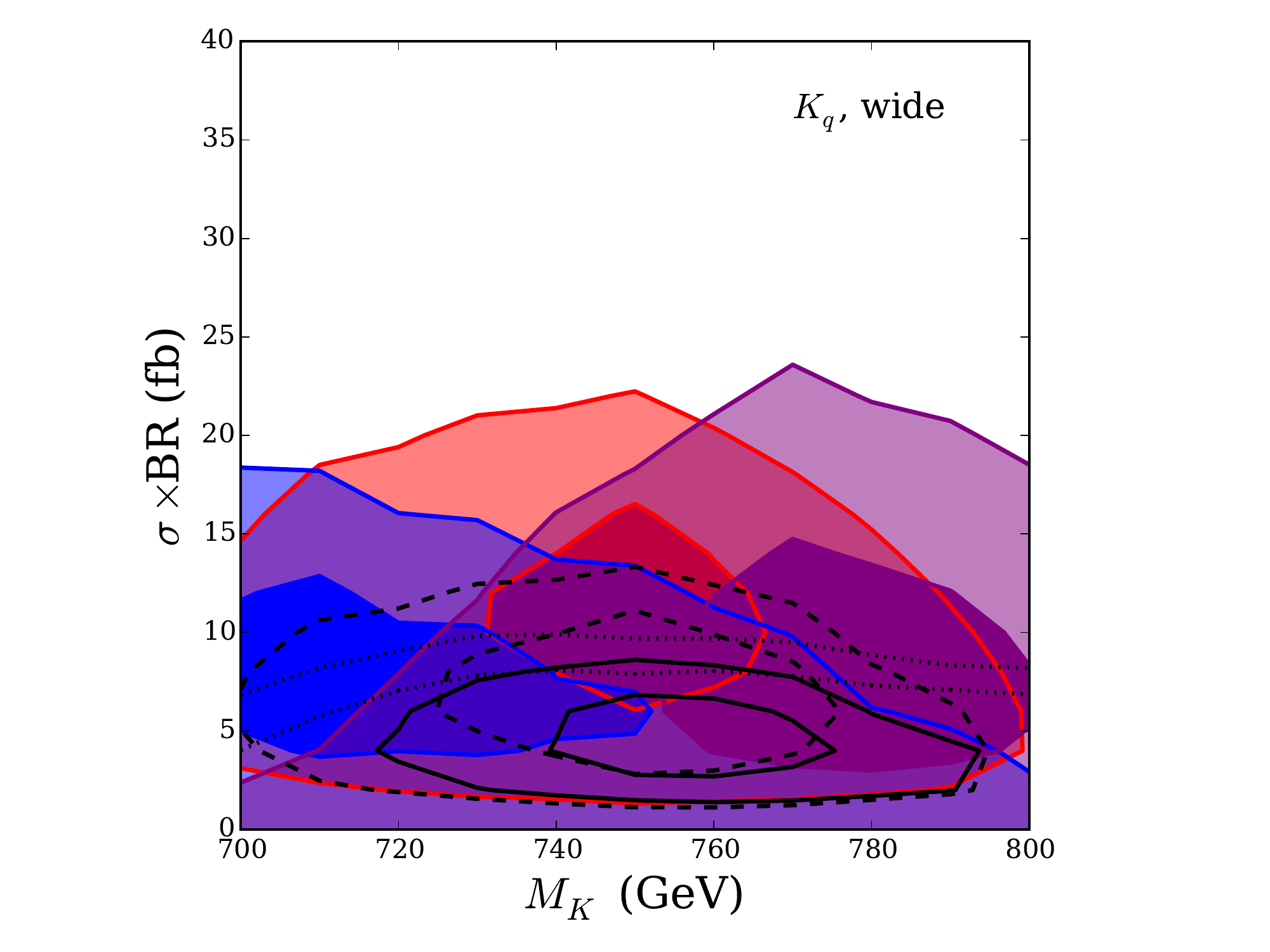}
\caption{Best fit regions (1 and $2\sigma$) of a spin-2 mediator decaying to diphotons, as a function of mediator mass and 13~TeV cross section, assuming the indicated mediator couplings to partons and mediator width. Red regions are the 1 and $2\sigma$ best-fit regions for the \textsc{Atlas13} data, blue is the fit to \textsc{Cms13} data, and purple is the \textsc{Cms13/0T}. The combined best fit for both \textsc{Atlas13}, \textsc{Cms13}, and \textsc{Cms13/0T} (\textsc{Combo13}) are the regions outlined in black dashed lines. The 1 and $2\sigma$ upper limits from the combined 8~TeV data (\textsc{Combo8}) are the black dashed lines (with cross sections converted to 13~TeV-equivalents). The best-fit signal combination of all six data sets (\textsc{Combo}) are the black solid lines. \label{fig:spin2_xs}}
\end{figure}

\begin{figure}[ht]
\includegraphics[width=0.4\columnwidth]{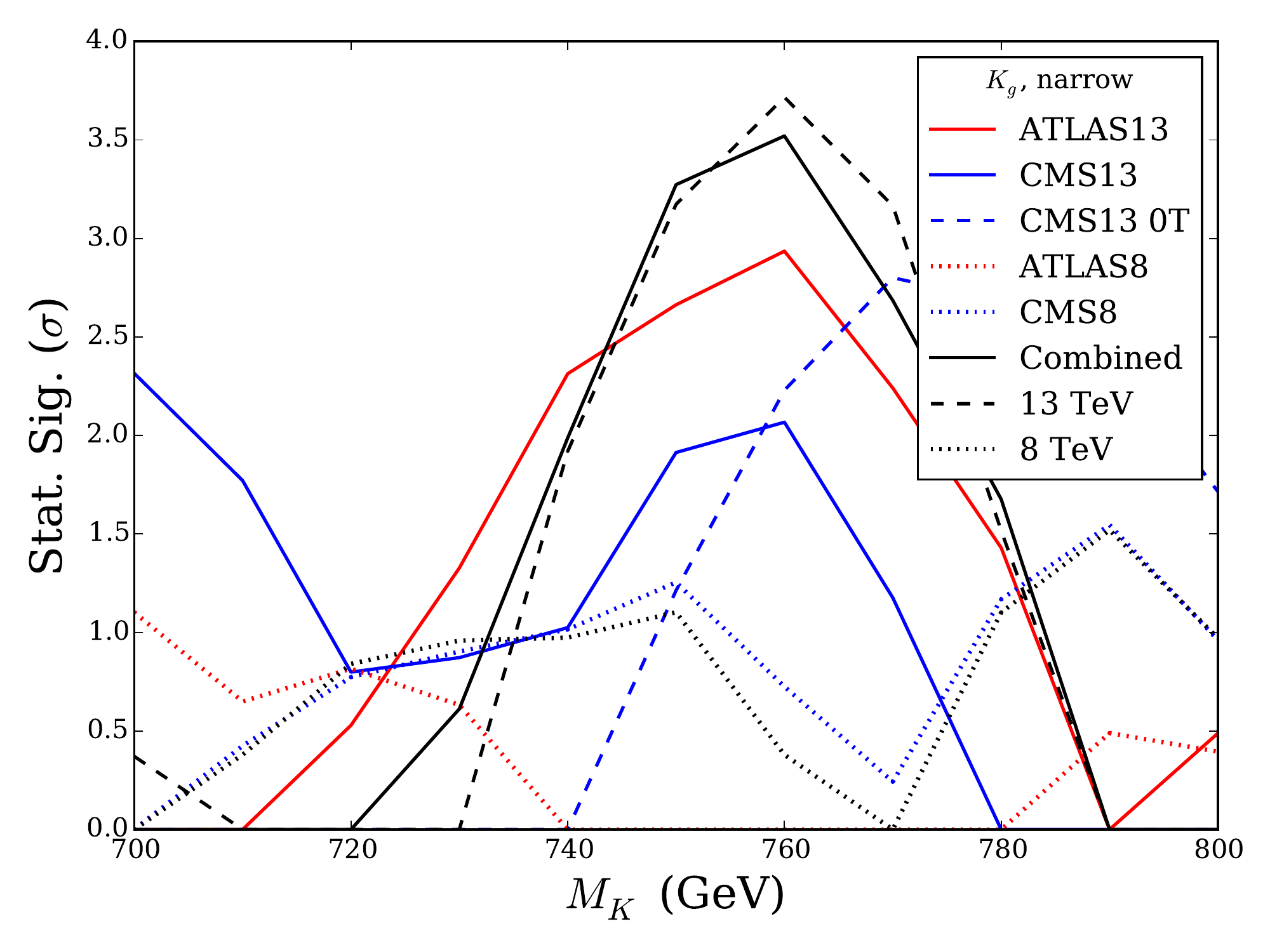}\includegraphics[width=0.4\columnwidth]{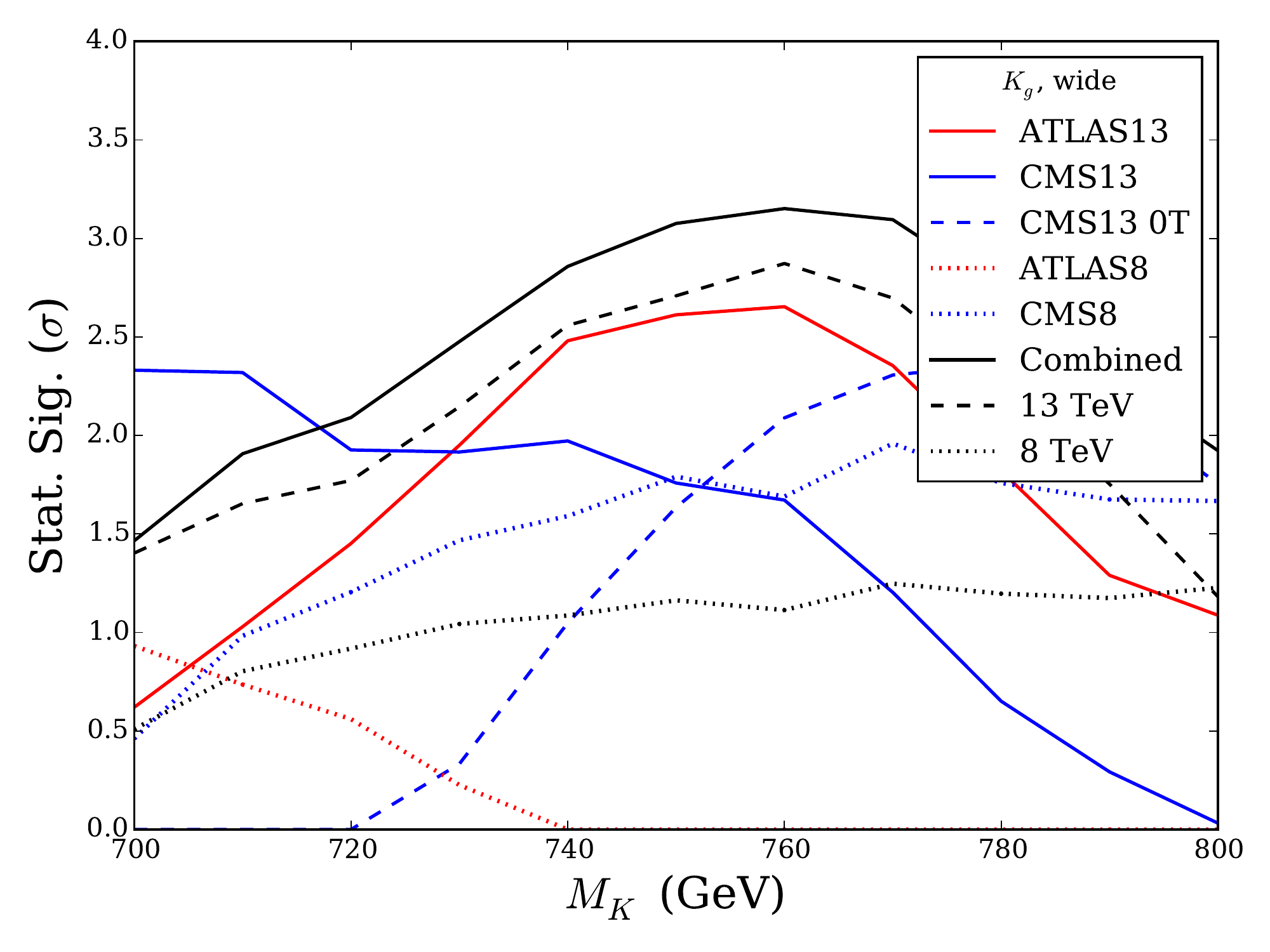}
\includegraphics[width=0.4\columnwidth]{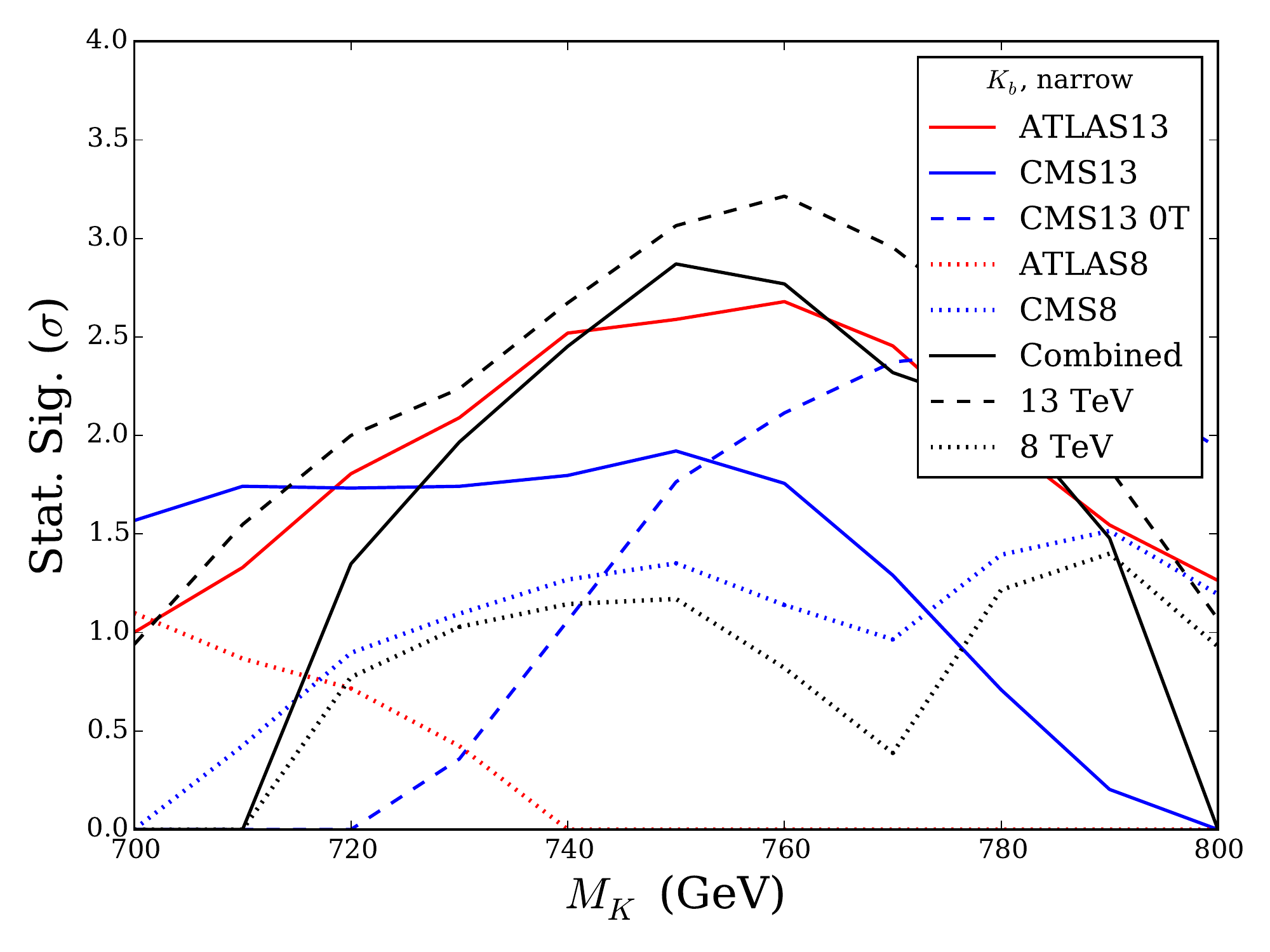}\includegraphics[width=0.4\columnwidth]{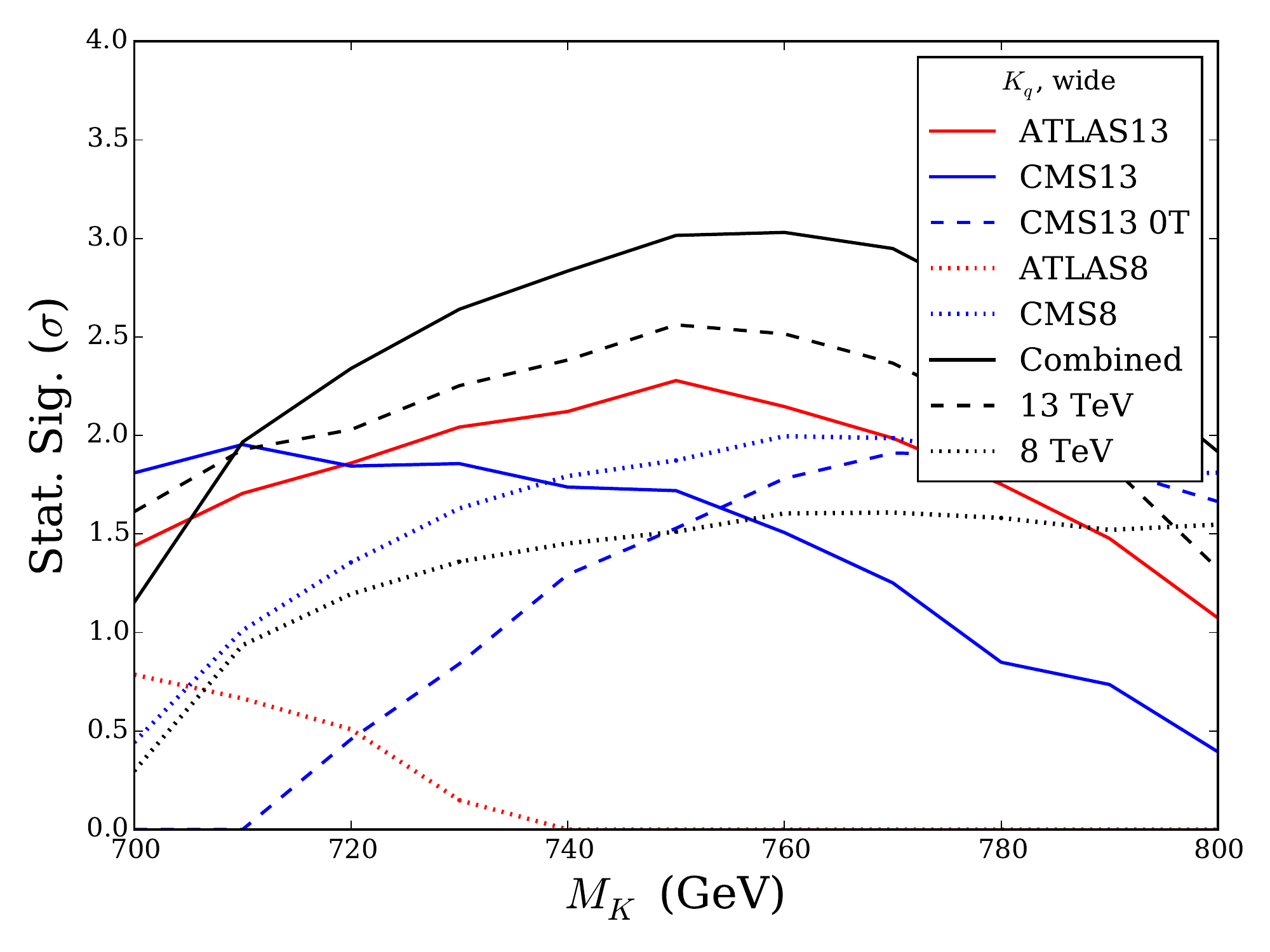}
\caption{Statistical significance for a spin-2 mediator decaying to diphoton, as a function of mediator mass, assuming the indicated mediator couplings to partons and mediator width. At each mass, the cross section is set to the value that maximizes statistical significance for a signal (see Figure~\ref{fig:spin0_xs}). The solid red line is the statistical significance of the \textsc{Atlas13} data alone, solid blue is \textsc{Cms13}, dashed blue is \textsc{Cms13/0T}, and dotted red and blue lines are \textsc{Atlas8} and \textsc{Cms8}, respectively. When comparing across experiments, note that these significances do not correspond to the same value of the cross section. The dashed (dotted) black line is the combination of 13(8)~TeV data, requiring the same cross section in both ATLAS and CMS. The solid black line is the combined significance of all six data sets. \label{fig:spin2_stat}}
\end{figure}

In Figure~\ref{fig:spin2_xs}, I show the best-fit regions for signal cross section as a function of mediator mass for the $K$ particle coupling to gluons or light quarks (couplings to $c/s$ or $b$ quarks are shown in Figure~\ref{fig:spin2_xs_app}, and are very similar to those for gluons). The statistical significance of the best-fit cross sections are shown in Figure~\ref{fig:spin2_stat}. I find that the combined data sets disfavors the spin-2 mediator when compared to the spin-0, with a combined statistical significance of $\sim 3.5\sigma$ for a narrow gluon-initiated resonance. This preference appears to be largely driven by the lower statistical preference for a spin-2 mediator in the \textsc{Atlas13} data. However, it must be stressed that the changes in statistical significance discussed here are less than $1\sigma$, and so are at best examples of mild preferences in the data.

\section{Narrow or Wide? \label{sec:width}}

\begin{figure}[th]
\includegraphics[width=0.3\columnwidth]{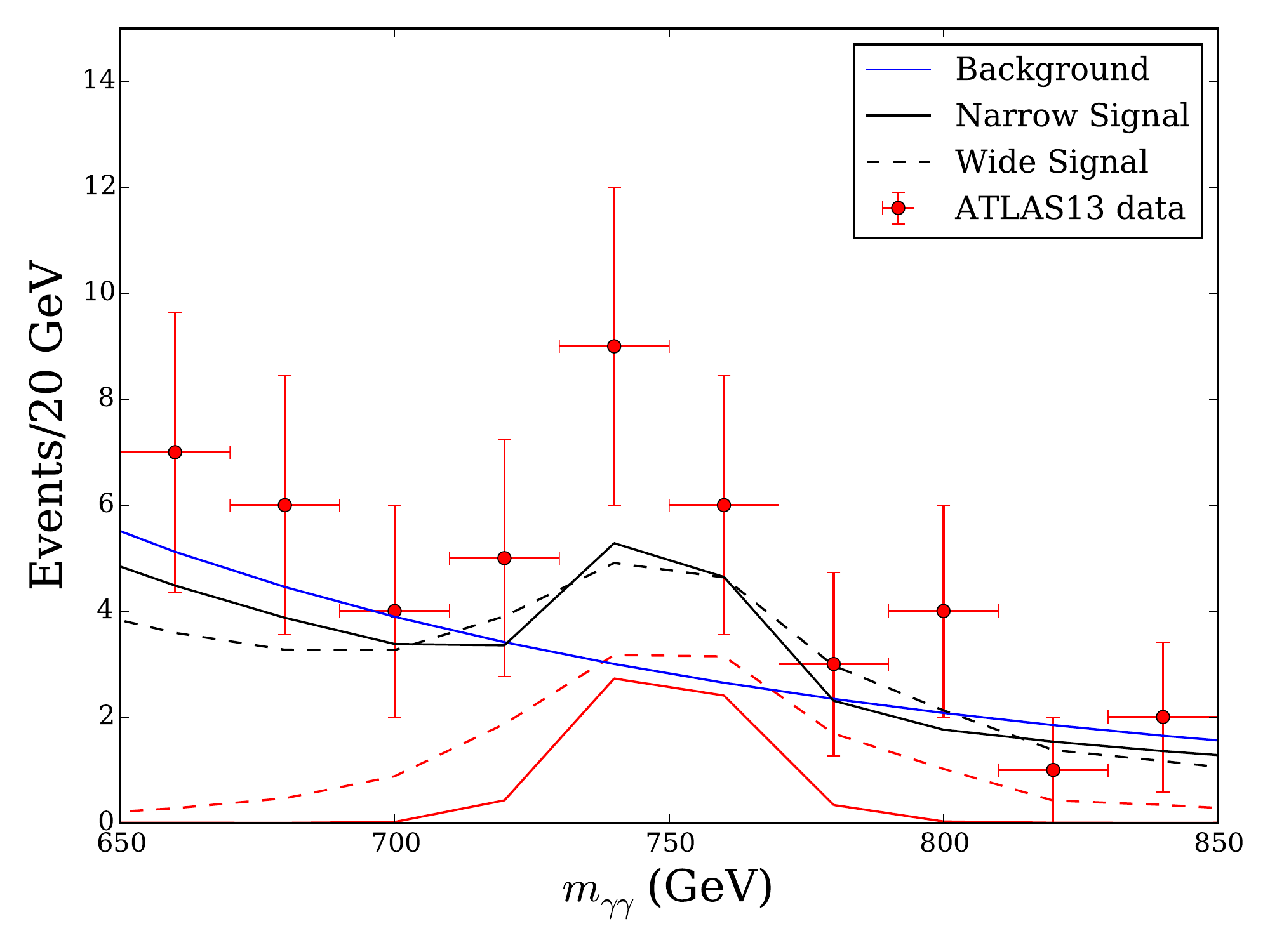} \\
\includegraphics[width=0.3\columnwidth]{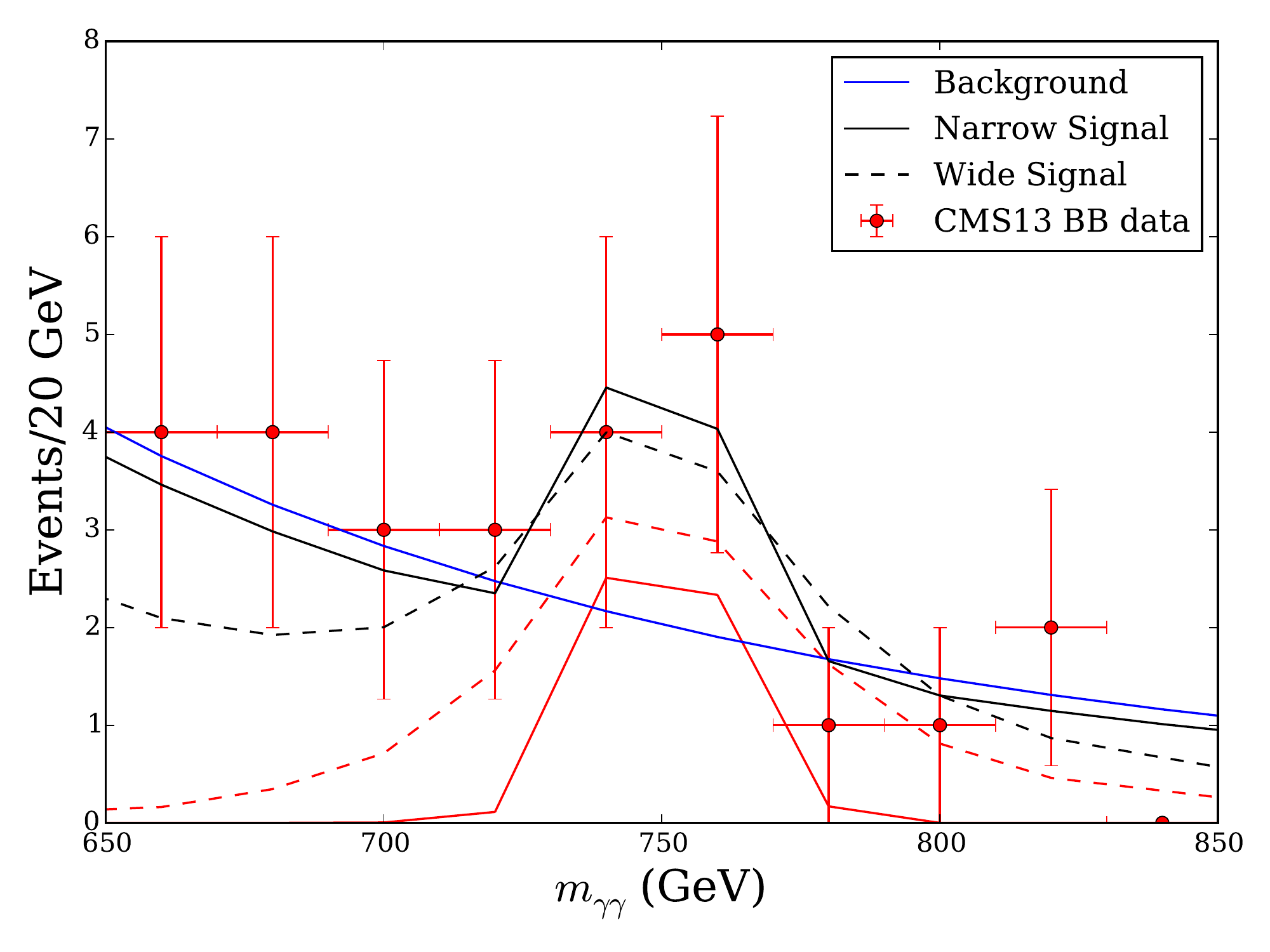}\includegraphics[width=0.3\columnwidth]{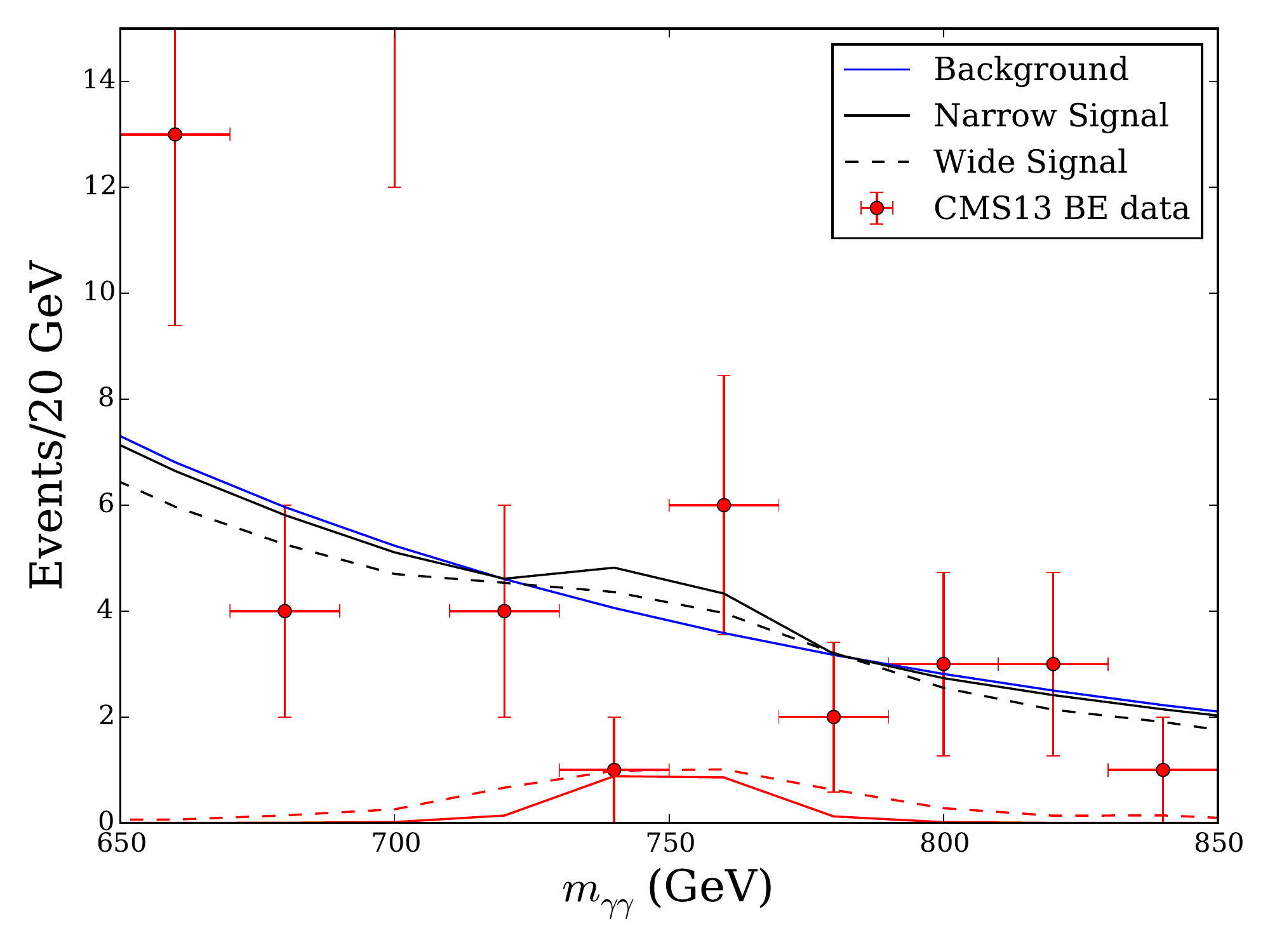}
\includegraphics[width=0.3\columnwidth]{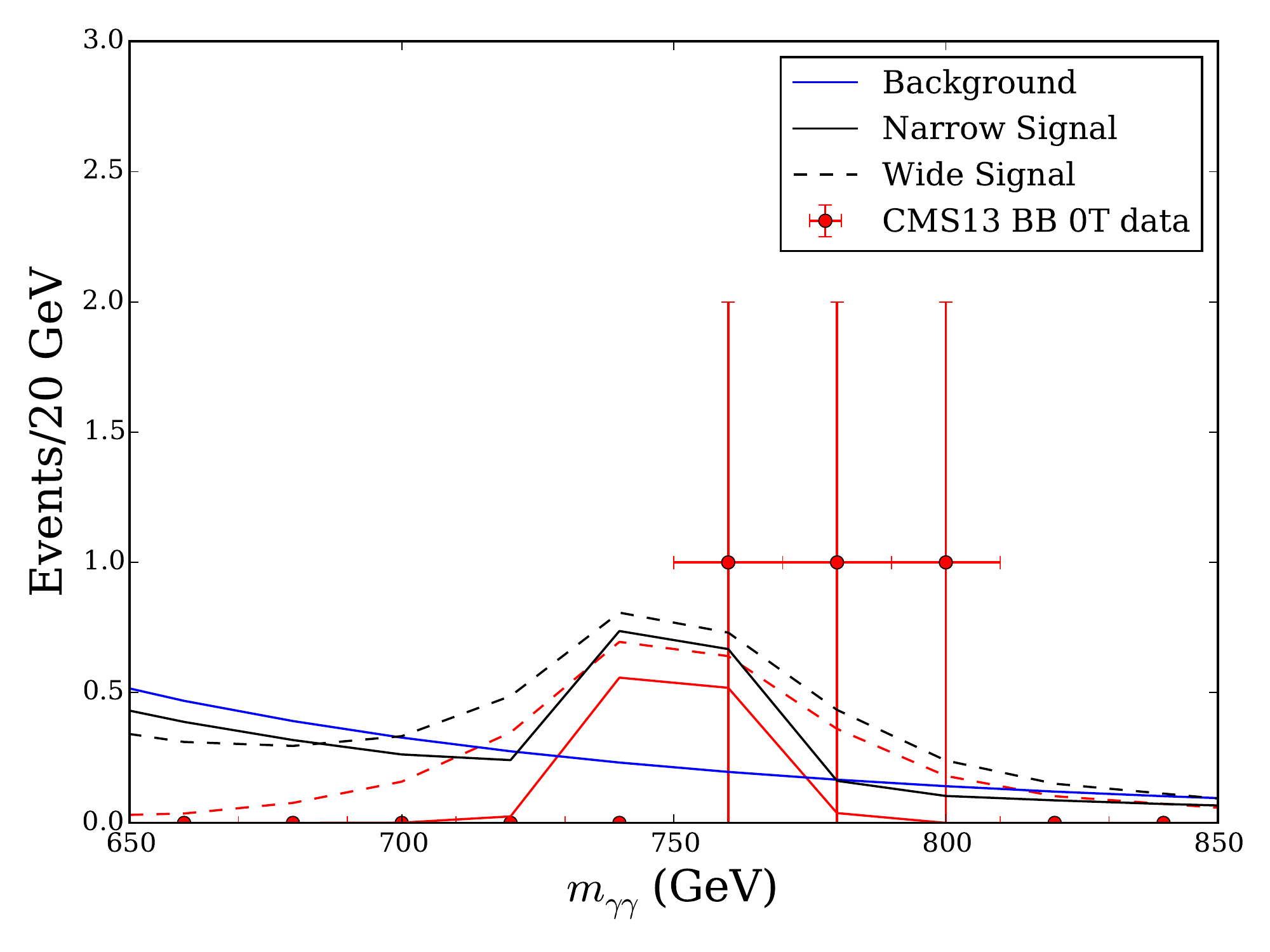}\includegraphics[width=0.3\columnwidth]{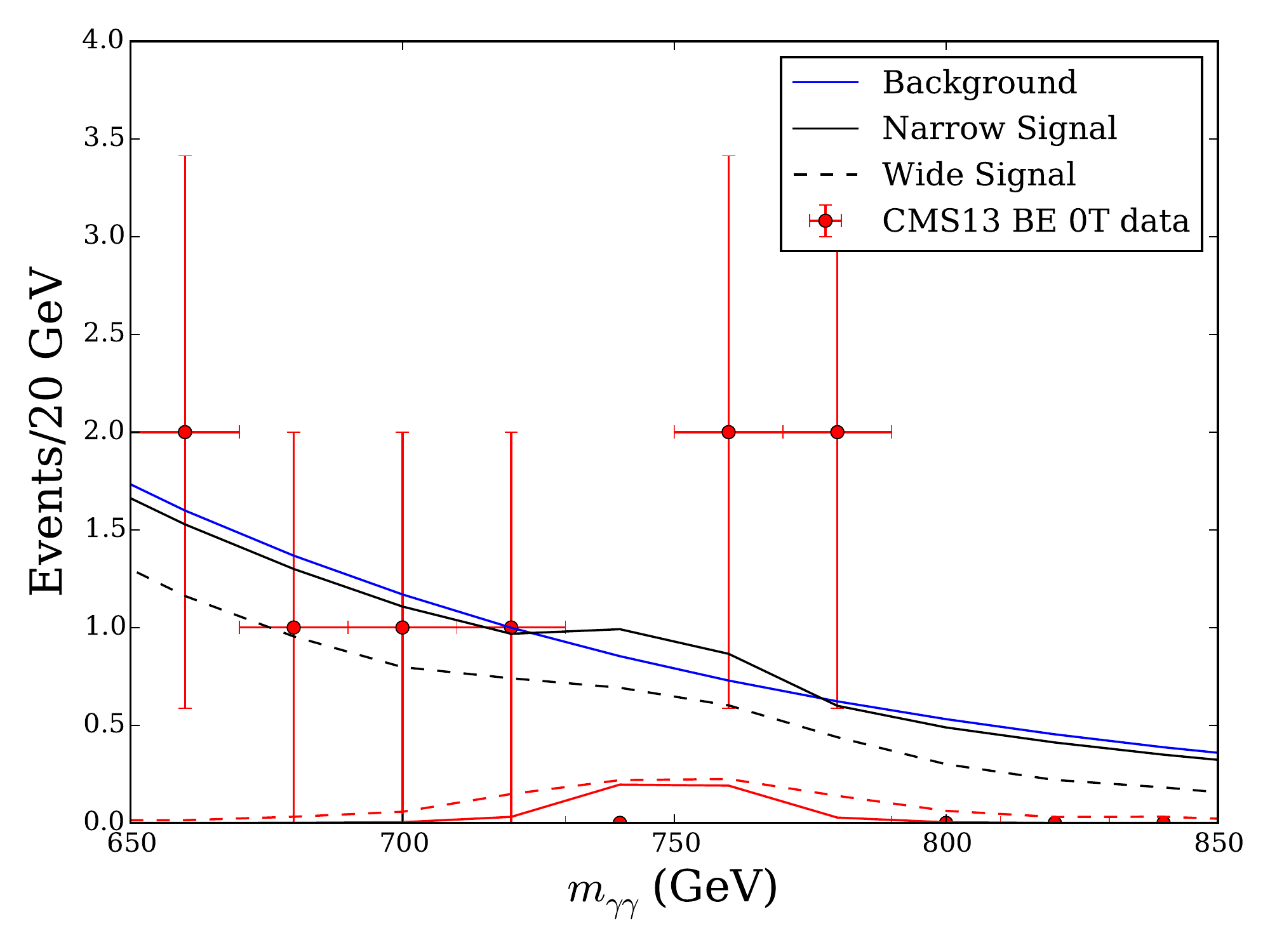}
\includegraphics[width=0.3\columnwidth]{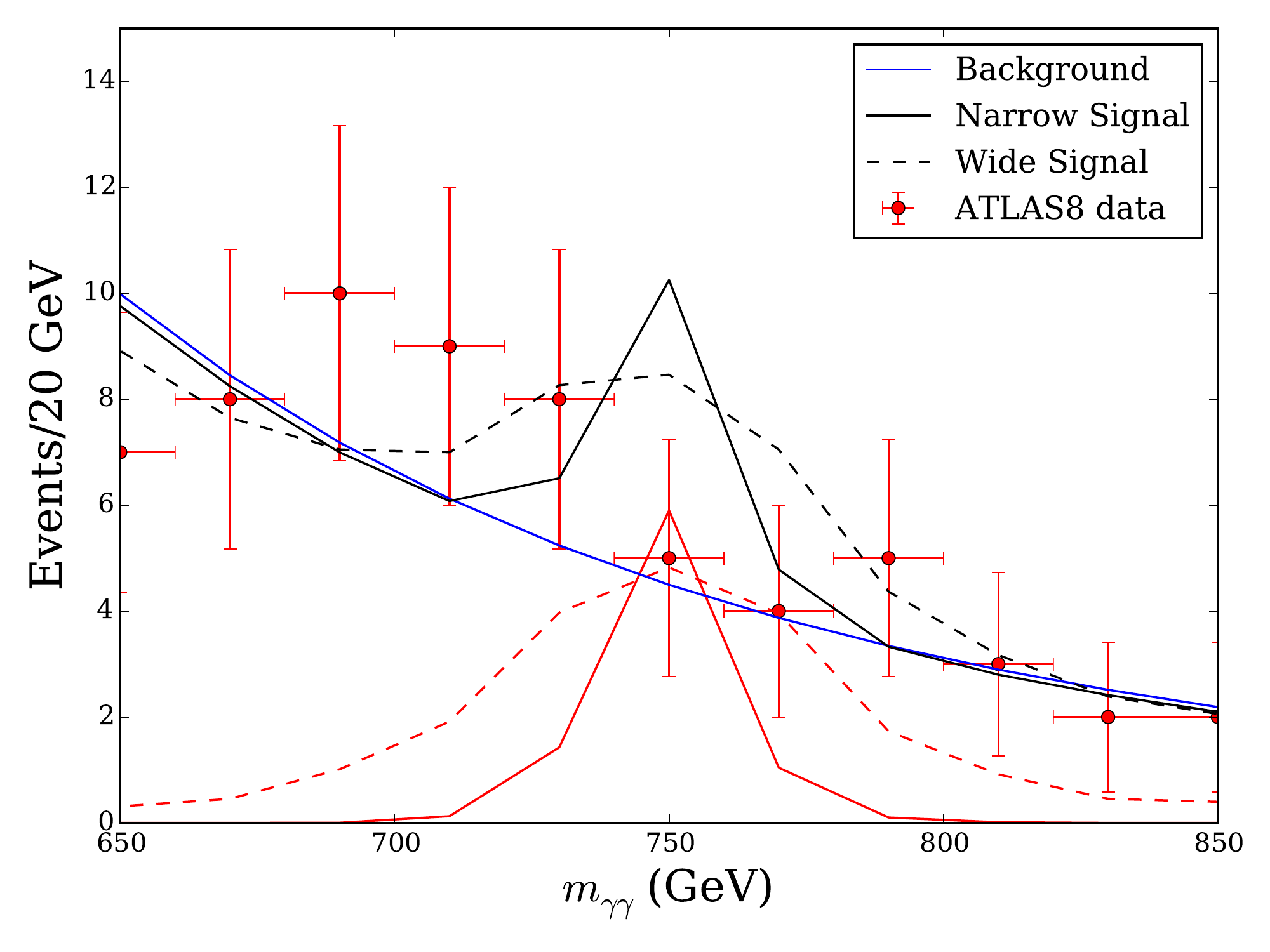}\includegraphics[width=0.3\columnwidth]{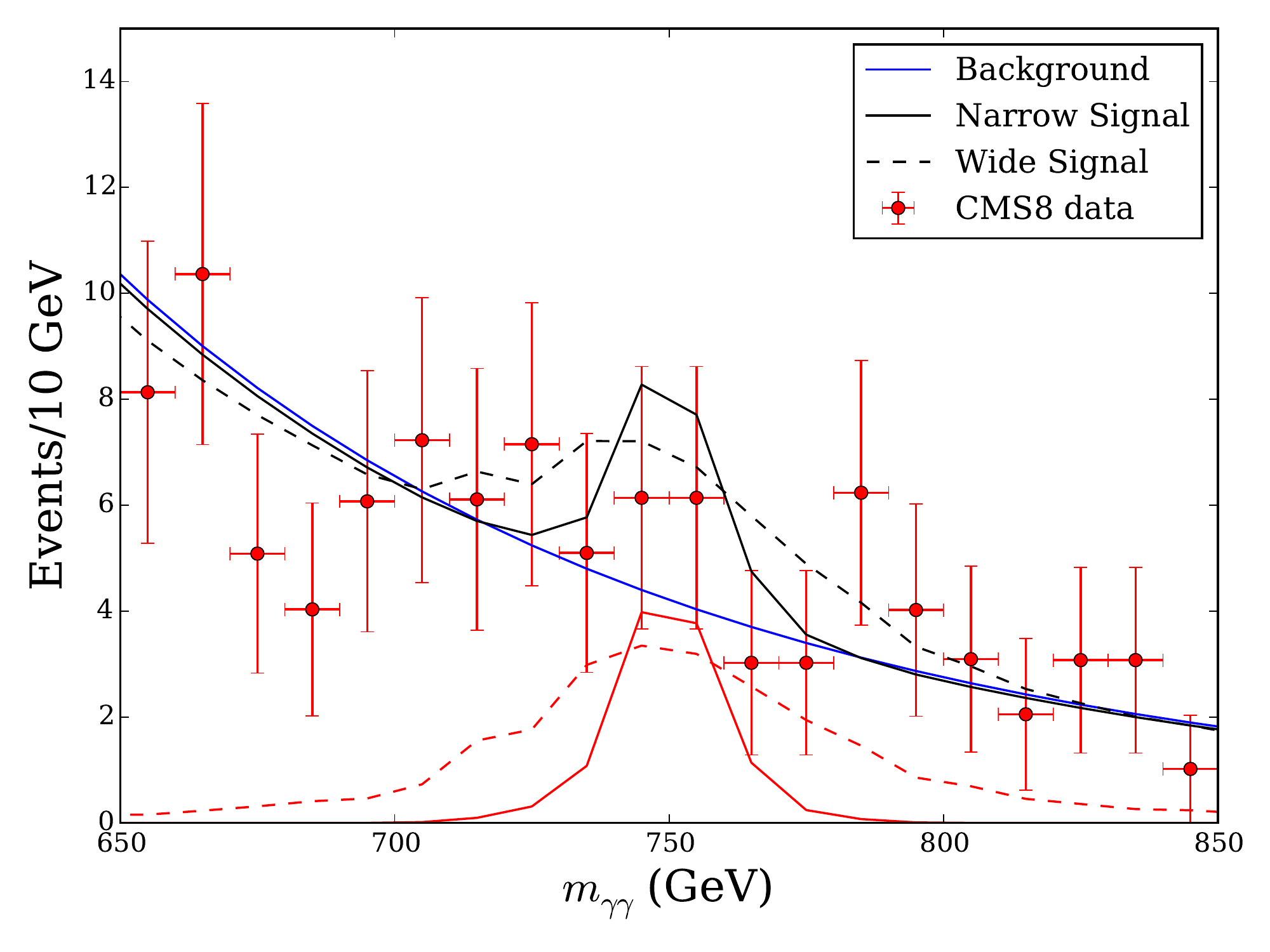}
\caption{Zoomed-in data (red points) for \textsc{Atlas13} (top row), \textsc{Cms13}(\textsc{Cms13/0T}) (barrel-barrel 2$^{\rm nd}$(3$^{\rm rd}$) row left and barrel-endcap 2$^{\rm nd}$(3$^{\rm rd}$) row right), \textsc{Atlas8} (lower left), and \textsc{Cms8} (lower right). Also shown are the best-fit binned differential distributions for a 750~GeV scalar resonance, with a 13~TeV cross section of 4~fb (narrow, solid red line), and 10~fb (wide, dashed red line). The best-fit background is shown in blue, and the sum of background and signal are the solid(dashed) black lines for the narrow(wide) signal. The background is refit to maximize the log-likelihood in the presence of the signal. \label{fig:bestfit13}}
\end{figure}

Now I attempt to address the question of whether there is any preference in the data for a wide resonance over a narrow one. For computational simplicity, I consider the best-fit scenario from Section~\ref{sec:pheno}: a scalar mediator coupling to gluons, with a cross section of 4~fb for the narrow resonance, and 10~fb for the broad resonance. As before, I do not scan over the width, but simply compare the narrow resonance (where the LHC width is controlled entirely by the detector resolution) with a mediator with a 45~GeV width, as suggested by the \textsc{Atlas13} results. In Figure~\ref{fig:bestfit13}, I show the experimental data around the 750~GeV $m_{\gamma\gamma}$ bins, compared to the predicted differential distributions of these best-fit points. It is clear from these examples why the wide resonance is in more tension with the \textsc{Cms13} result, as well as why the 8~TeV data can more easily absorb this type of signal. 

I have already discussed some of the statistical evidence for the question of narrow or wide resonances in the previous section. The combination of \textsc{Atlas13}, \textsc{Cms13}, \textsc{Cms13/0T} increases the overall significance for a narrow resonance, and this significance does not decrease when the 8~TeV data is added, while the combination of 13~TeV data decreases the significance for a wide resonance (though this then increases once the 8~TeV data is included). Thus, one can say that both interpretations have equal statistical significance when combining all the data, though it is  true that the combination of the 13~TeV alone mildly prefers the narrow resonance. 

\begin{figure}[ht]
\includegraphics[width=0.5\columnwidth]{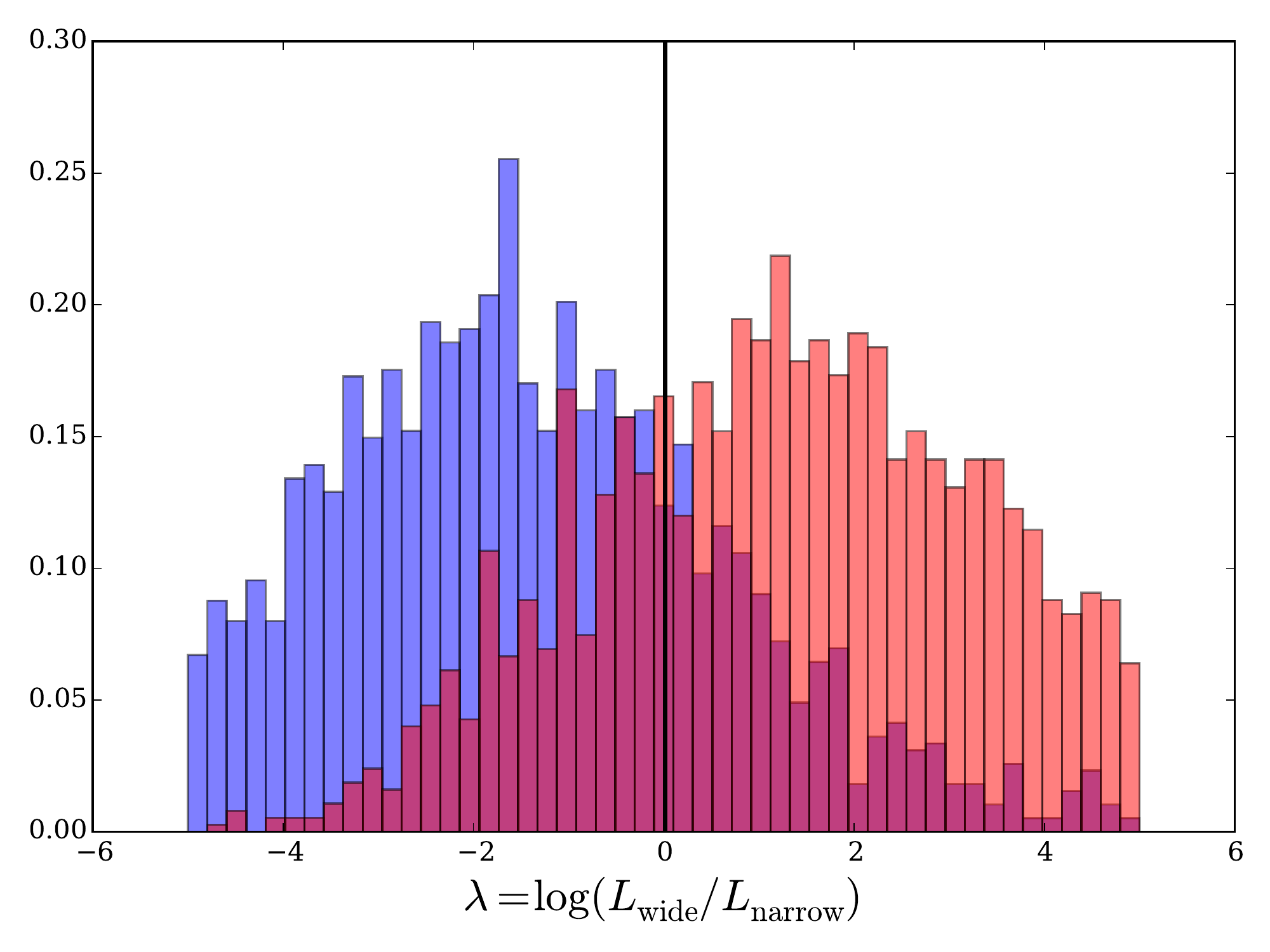}\includegraphics[width=0.5\columnwidth]{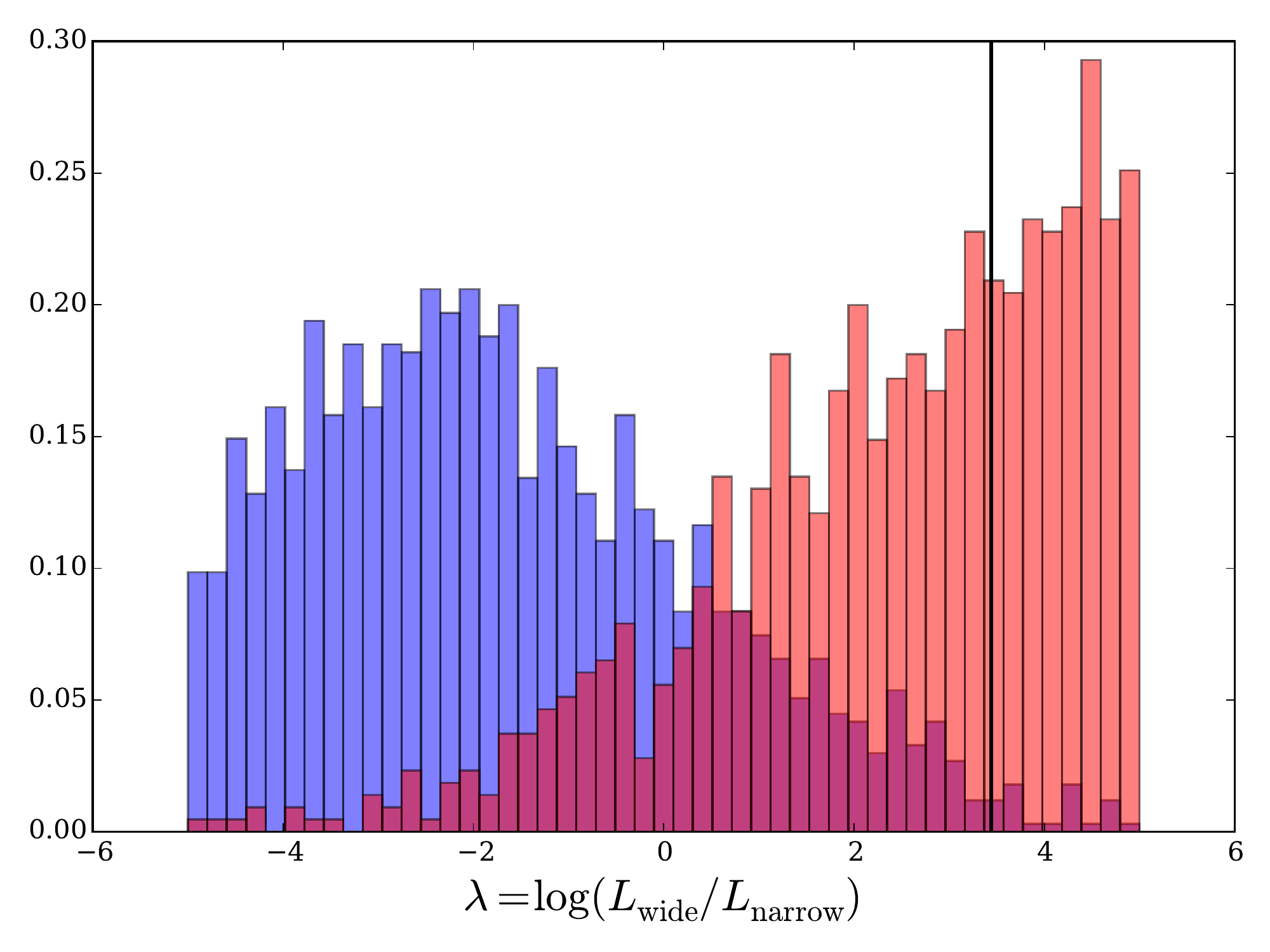}
\caption{Histograms of $\Delta \log$ likelihood of the \textsc{Combo13} (left) and \textsc{Combo} (right) fit to the wide- and narrow-resonance interpretation of the diphoton anomaly. The blue histograms represent 2000 pseudoexperiments where a narrow 750~GeV signal was injected to the ATLAS and CMS data with a cross section of 4~fb and then fit to both a narrow and wide template. The red histogram represent 2000 pseudoexperiments generated where a wide 750~GeV signal was injected with a cross section of 10~fb, then fit to both signal templates. The black vertical line is the value of $\lambda = \log L_{\rm wide}/L_{\rm narrow}$ found in the real data ($\lambda_o$), and the relative bin heights at $\lambda_o$ give the likelihood ratio $R$.  \label{fig:bayes}}
\end{figure}

I perform one additional statistical test, calculating the likelihood ratio for the preference for the narrow resonance over a wide resonance. This is the ratio of the probability of observing some set of data given a narrow resonance over the probability of observing that data given the wide resonance. To calculate this, I define the log-likelihood ratio
\begin{equation}
\lambda = \log(L_{\rm wide}/L_{\rm narrow}),
\end{equation}
where the $L$ values are the maximum likelihoods (given an assumption of the width) marginalized over the background fit parameters. After calculating the {\it observed} $\lambda_o$ from the data, I estimate the probability distribution by generating pseudoexperiments: injecting either a narrow or wide signal (with the best fit cross section) over the best-fit background, and then calculating the $\lambda$ value of that particular pseudoexperiment. The likelihood ratio $R$ is then the ratio at $\lambda_0$ of the normalized probability distribution for narrow signal over the distribution for a wide signal. In Figure~\ref{fig:bayes}, I show these probability distributions along with the $\lambda_0$ fit to the combination of \textsc{Atlas13} and \textsc{Cms13} (left panel), and fit to all data (right panel). As can be seen, when considering only 13~TeV data, there is a very slight preference for a narrow resonance, with a corresponding ratio of $R \sim 1$, which indicates no significance preference for either model using the 13~TeV data only. 

When combining all data, the likelihood ratio for the wide resonance over the narrow width is $R\sim 20$. Thus, if one is considering only the Run-II data, there is no particular preference for a wide or narrow width using this test, while if all data is considered, the 45~GeV width is somewhat more probable. However, it certainly should not be stated that we {\it know} with any degree of confidence that the proposed 750~GeV resonance has a width much larger than one might expect from a perturbatively coupled spin-0 mediator.

\section{Conclusions \label{sec:conclusions}}

When considering the 750~GeV diphoton excess, the theoretical community must balance its natural exuberance with the recognition that the statistical size of the anomalies are very small. As a result, any further slicing of data will yield at best modest statistical preferences for the phenomenological questions that we in the community want answers to. That said, given that this excess is the most significant seen at the LHC since the discovery of the 125~GeV Higgs, and the resulting avalanche of theoretical papers which shows no sign of slowing, it is still a useful exercise to carefully analyze the available data and determine what we do -- and do not -- know at this stage. While there is some useful information to be gleaned from this exercise, we are fortunate that the continuation of Run-II will be upon us shortly.  

From the existing data, we can conclude the following:
\begin{enumerate}
\item Explaining the anomaly through a spin-0 resonance is preferred over a spin-2 mediator, though this preference is less than $1\sigma$ in most cases.

\item Combining the 8 and 13~TeV data from ATLAS and CMS sets yields a $\sim 4.0\sigma$ statistical preference for a signal of $\sim 4(10)$~fb, assuming a narrow (wide) spin-0 resonance. This ignores the look-elsewhere effect, as discussed. Given that the significance of my fits to individual ATLAS and CMS data-sets are underestimates when compared to the full experimental results, it is possible that the actual statistical preferences are larger than these quoted values. However this would require a combined analysis performed by the ATLAS and CMS Collaborations.

\item The cross sections needed for the \textsc{Atlas13}, \textsc{Cms13}, and \textsc{Cms13/0T} data sets are incompatible at the two sigma level, though they agree in mass. The most straightforward reading of this (while maintaining a new physics explanation for the anomalies) is that the larger \textsc{Atlas13} cross section constitutes a modest upward fluctuation from the ``true'' cross section, which is more in line with the \textsc{Cms13} value.\footnote{Of course, in this interpretation, had the \textsc{Atlas13} data not had an upward fluctuation the combined statistical significance would likely not have been large enough to generate the massive response from the theoretical community. Thus, one may be tempted to take an Anthropic Principle view: an upward fluctuation in at least one of the data sets was a necessary condition for this paper to exist.} The reverse is also possible of course, but would bring the diphoton excess in the 13~TeV data in greater tension with the 8~TeV null results.

\item When considering only the 13~TeV data, the \textsc{Cms13} data does not share the \textsc{Atlas13} preference for a 45~GeV width. I find that the ``wide'' interpretation of the resonance has a statistical significance in the \textsc{combo13} data set which is approximately $0.5\sigma$ less likely than the ``narrow'' interpretation. The corresponding likelihood ratio shows no preference for either width. Thus, while the theoretical challenge of a wide resonance may be appealing, the data in no way {\it requires} any new physics explanation to have the unusually large width of $\Gamma \sim 45$~GeV.

\item Combining the 13~TeV data with the 8~TeV, I find that gluon-initiated mediators are preferred, due to having the largest ratio of relevant p.d.f.s. In particular, the combination of all six data sets for a gluon-initiated narrow resonance has the same statistical preference for a signal as the \textsc{Combo13} data alone does, though the best-fit cross section decreases slightly when the 8~TeV data is added ($\sim 4.0\sigma$ for a $\sim 4$~fb signal). In the narrow width assumption, heavy quark-initiated mediators have slightly smaller statistical preference, and a light-quark coupling has a fairly significant decrease in statistical preference, indicating a more serious conflict between the 13 and 8~TeV data.

\item Combining the 13 and 8~TeV data sets under the $\Gamma = 45$~GeV spin-0 model increases the statistical preference for signal as compared to the \textsc{Combo13} result, as the excess can be more easily absorbed by the background model here. Combining all the data sets in this way results in a $\sim 3.5\sigma$ preference for a $\sim 10$~fb signal (a $0.5\sigma$ increase over the \textsc{Combo13} wide-resonance fit), with a likelihood ratio of $\sim 20$ rejecting the narrow interpretation. Again, these statistical preferences are relatively small, thus theorists are free to explore the options, but should keep in mind that the experimental results are inconclusive.

\end{enumerate}

The conclusions of this paper are perhaps not a surprise. There is clear tension between the \textsc{Atlas13} and \textsc{Cms13} results, as well as with the non-observation in 8~TeV data. The question of the width is especially puzzling; but further slicing of the data, as I have demonstrated, leads to somewhat conflicting results which do not have a clear statistical preference towards any one solution. I note that if the ATLAS excess is indeed an upward fluctuation from a signal which is more in line with the \textsc{Cms13} value, then perhaps this could also give a spurious signal of large width. However, the true answers will only come with more data, though I note that, if the signal is indeed real, but on the order of 4~fb, then we may need 10-20~fb$^{-1}$ for a single experiment to have $5\sigma$ discovery. 

\section*{Acknowledgements}
I thank Matthew Baumgart, JP Chou, Yuri Gershtein, Marco Farina, Angelo Monteux,  David Shih, and Scott Thomas for useful feedback and discussions while writing this paper.

\appendix

\section{Additional Analysis \label{sec:appendix}}

\begin{figure}[h]
\includegraphics[width=0.4\columnwidth]{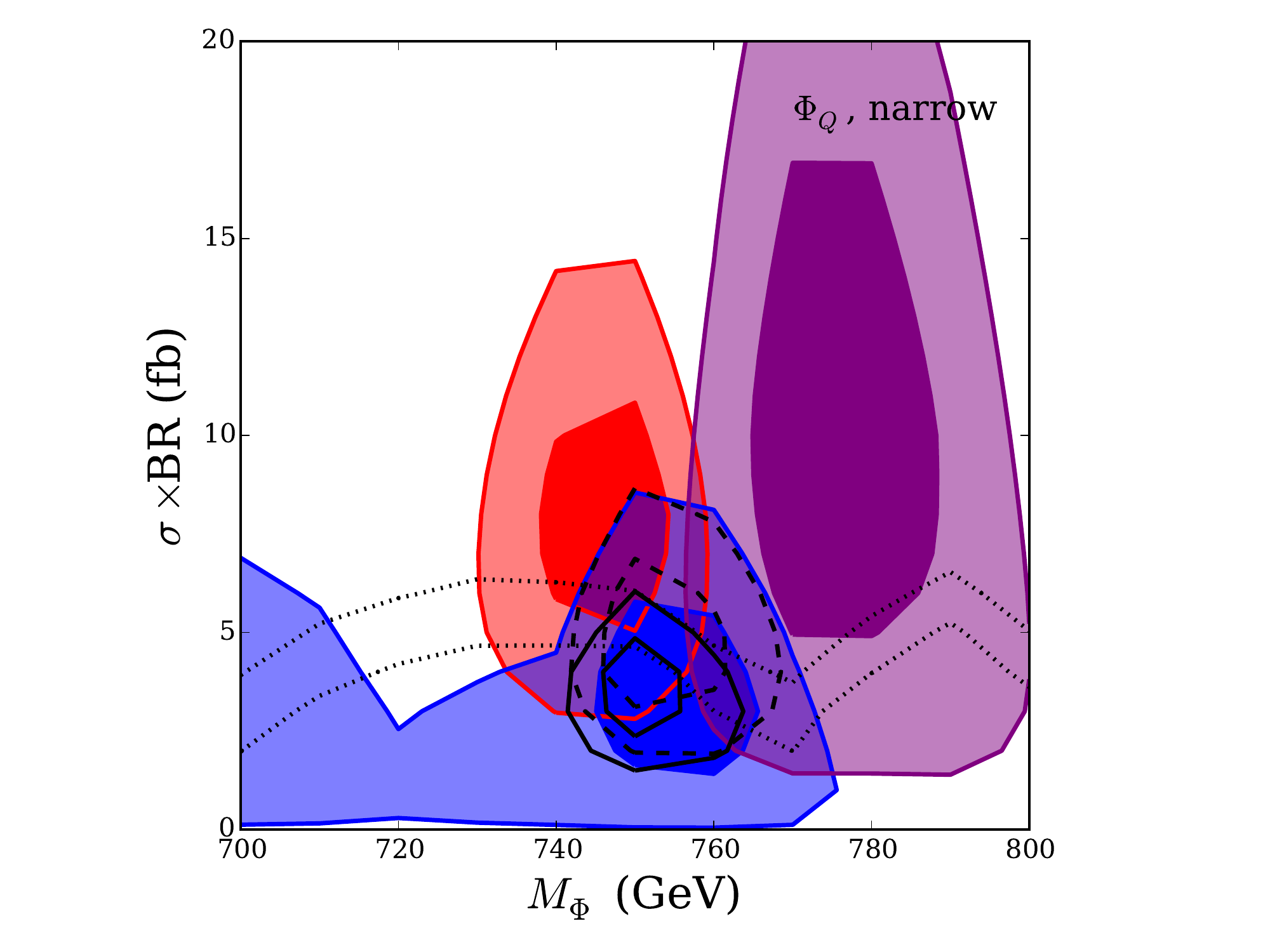}\includegraphics[width=0.4\columnwidth]{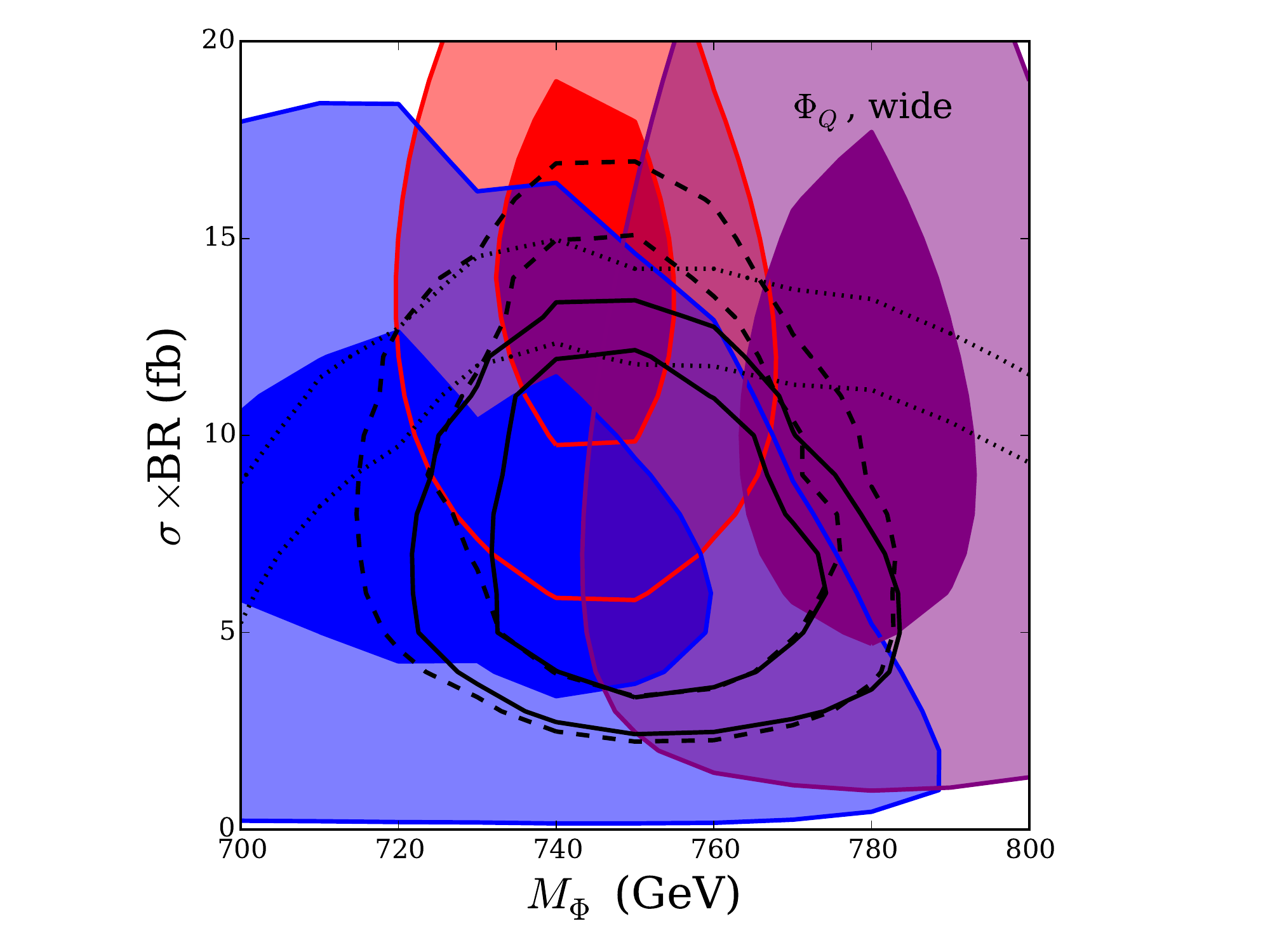}
\includegraphics[width=0.4\columnwidth]{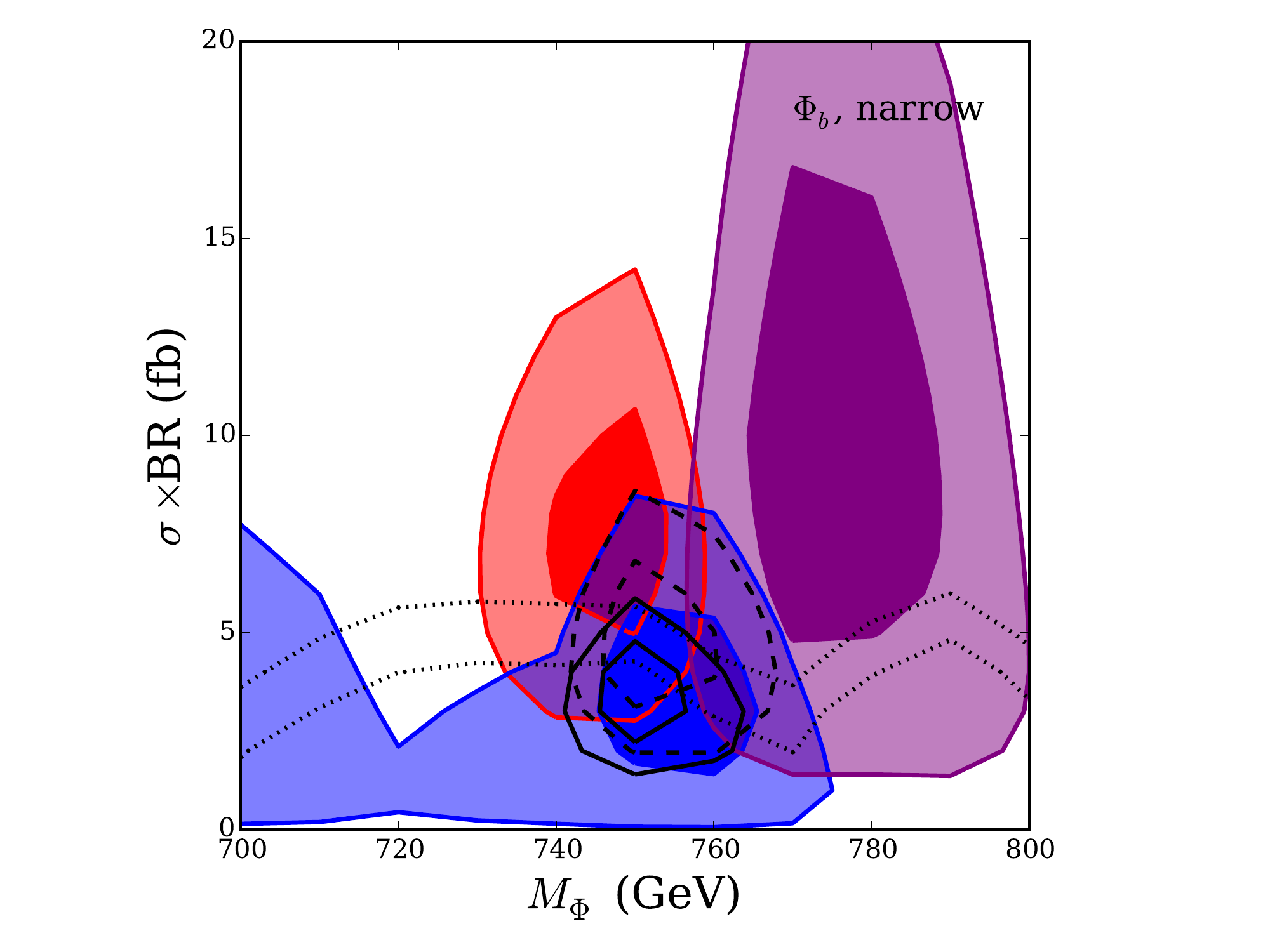}\includegraphics[width=0.4\columnwidth]{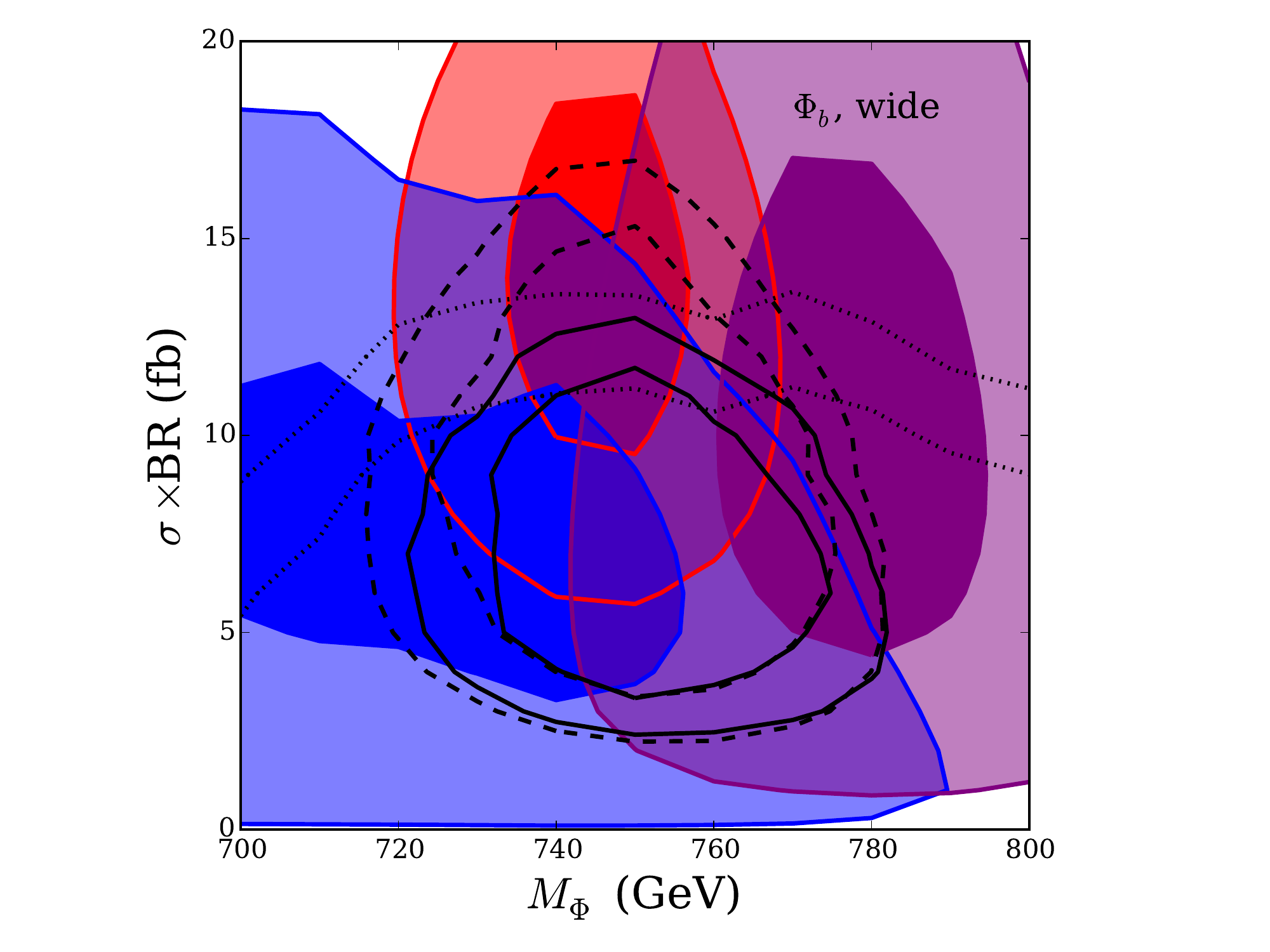}
\caption{Best fit regions (1 and $2\sigma$) of a spin-0 mediator decaying to diphotons, as a function of mediator mass and 13~TeV cross section, assuming the indicated mediator couplings to partons and mediator width. Red regions are the 1 and $2\sigma$ best-fit regions for the \textsc{Atlas13} data, blue is the fit to \textsc{Cms13} data. The combined best fit for both \textsc{Atlas13} and \textsc{Cms13} (\textsc{Combo13}) are the regions outlined in black dashed lines. The 1 and $2\sigma$ upper limits from the combined 8~TeV data (\textsc{Combo8}) are the black dashed regions (with cross sections converted to 13~TeV-equivalents). The best-fit signal combination of all four data sets (\textsc{Combo}) is the black solid regions. \label{fig:spin0_xs_app}}
\end{figure}

\begin{figure}[h]
\includegraphics[width=0.4\columnwidth]{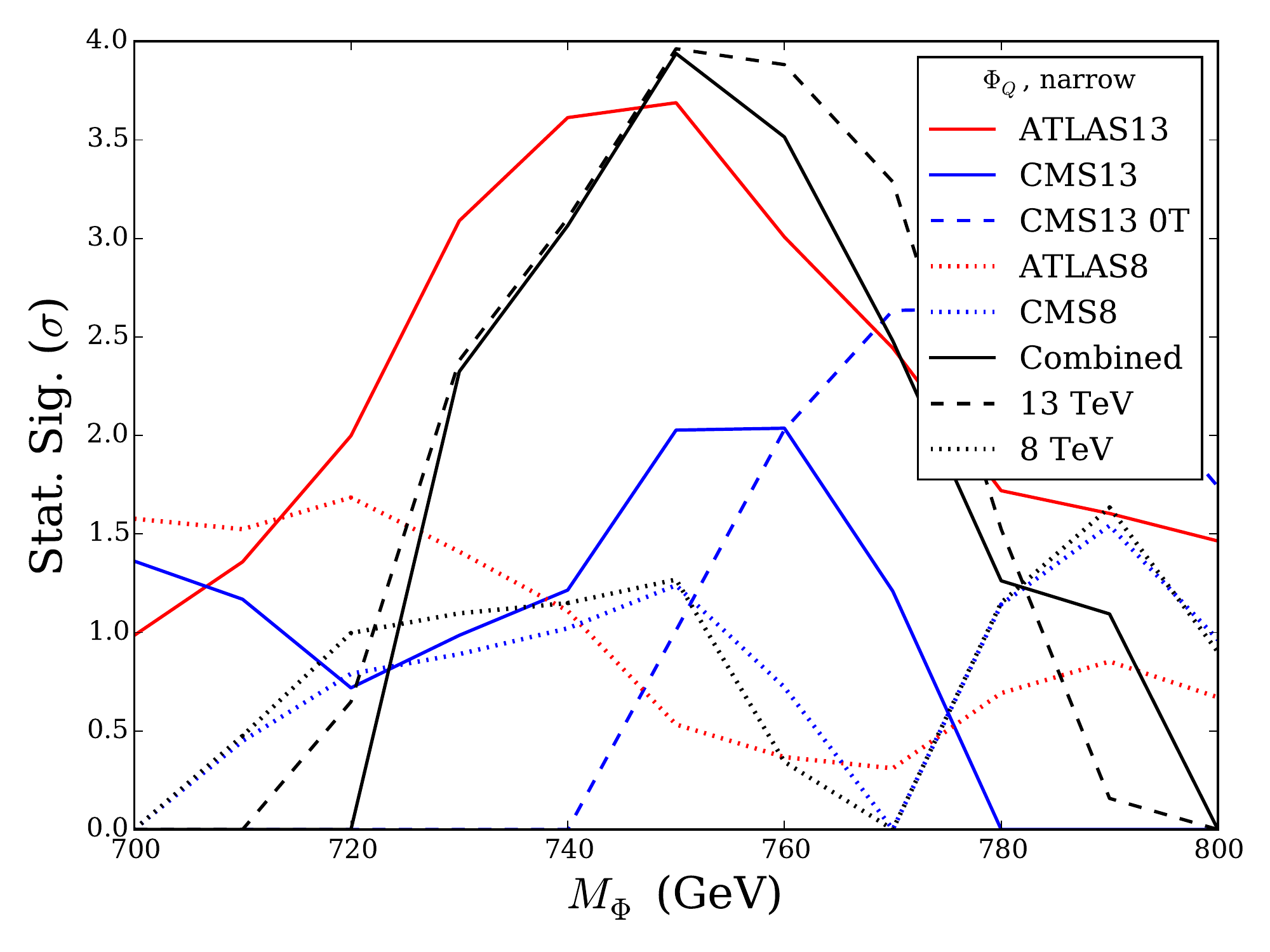}\includegraphics[width=0.4\columnwidth]{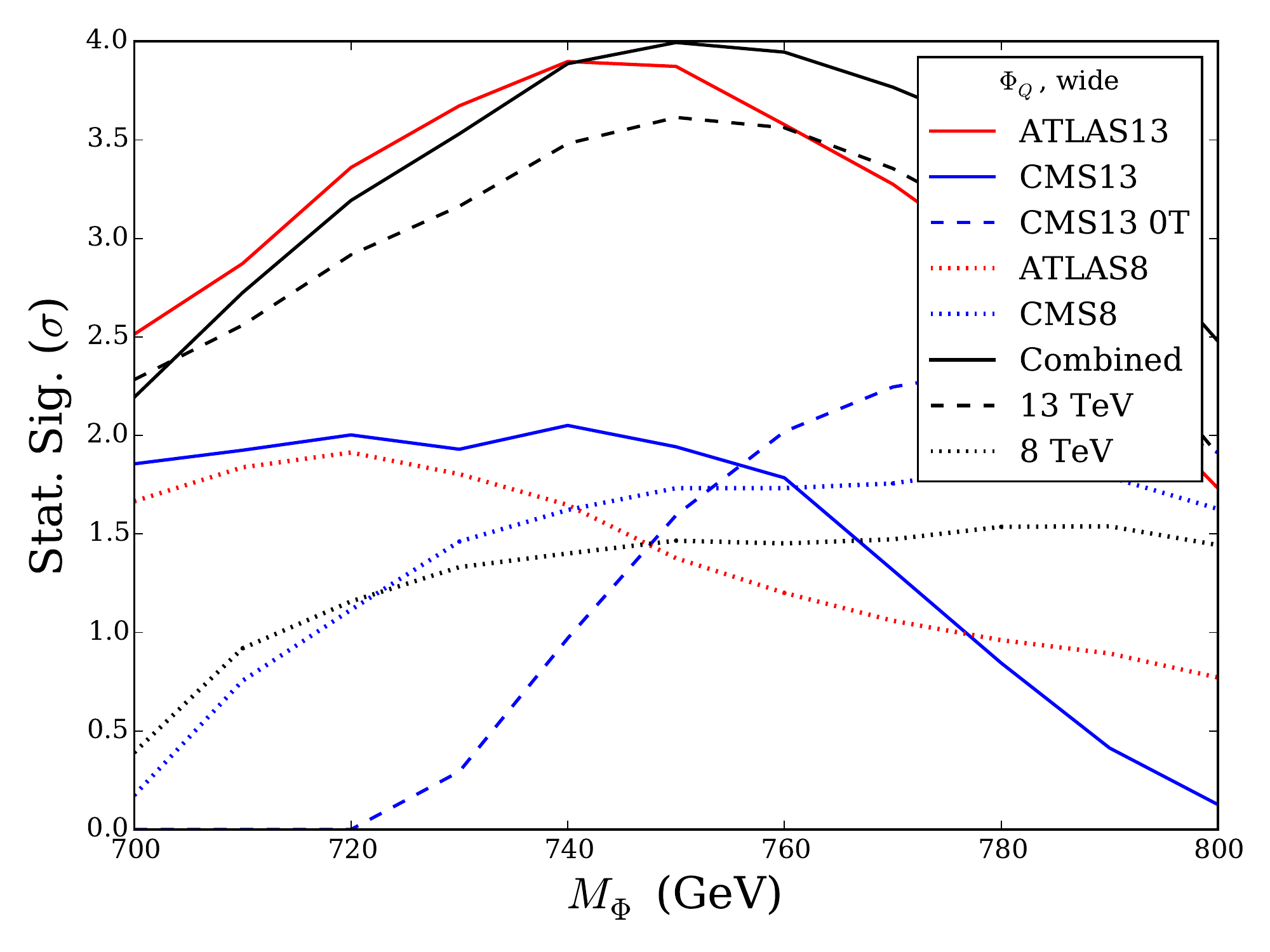}
\includegraphics[width=0.4\columnwidth]{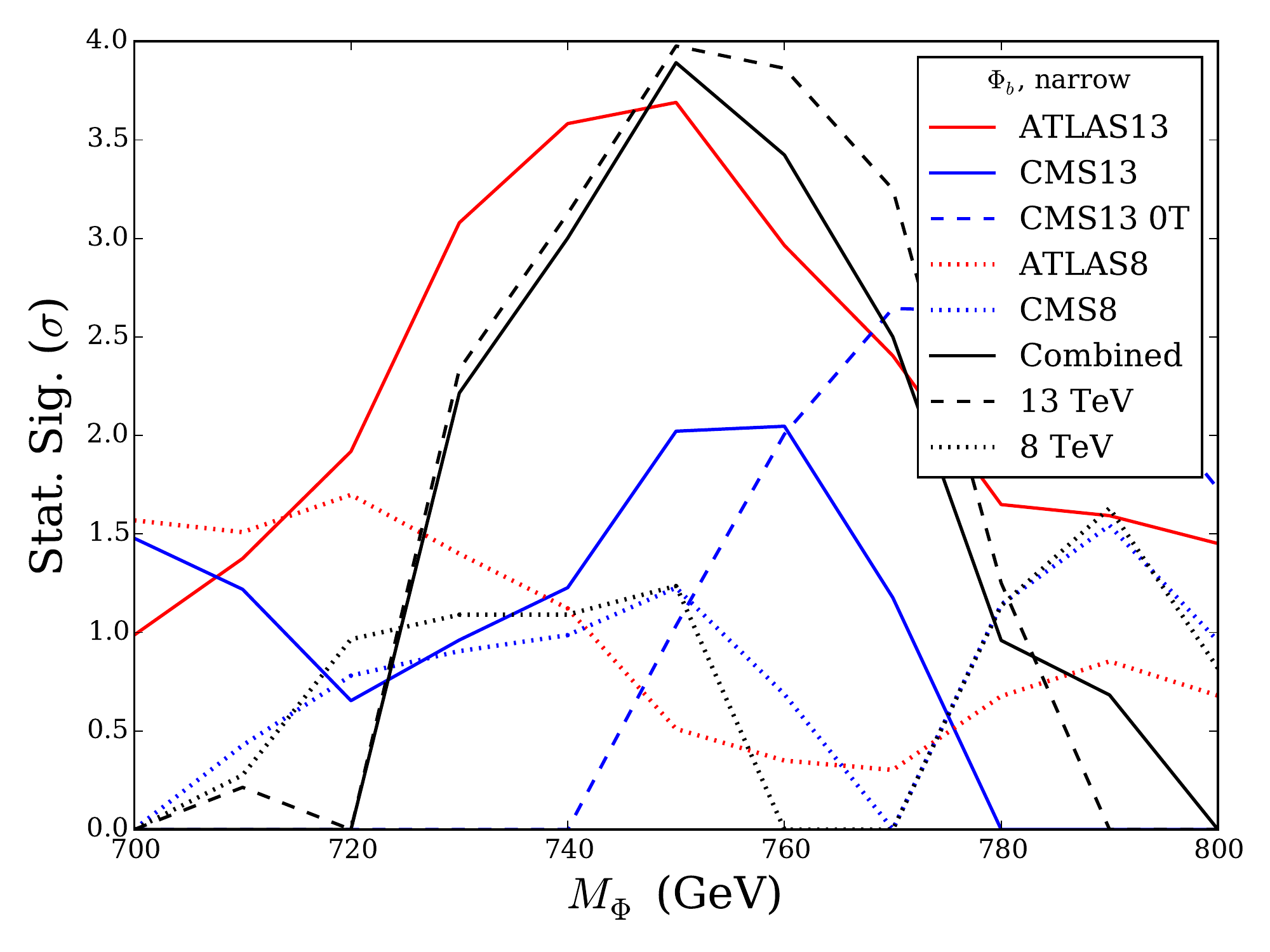}\includegraphics[width=0.4\columnwidth]{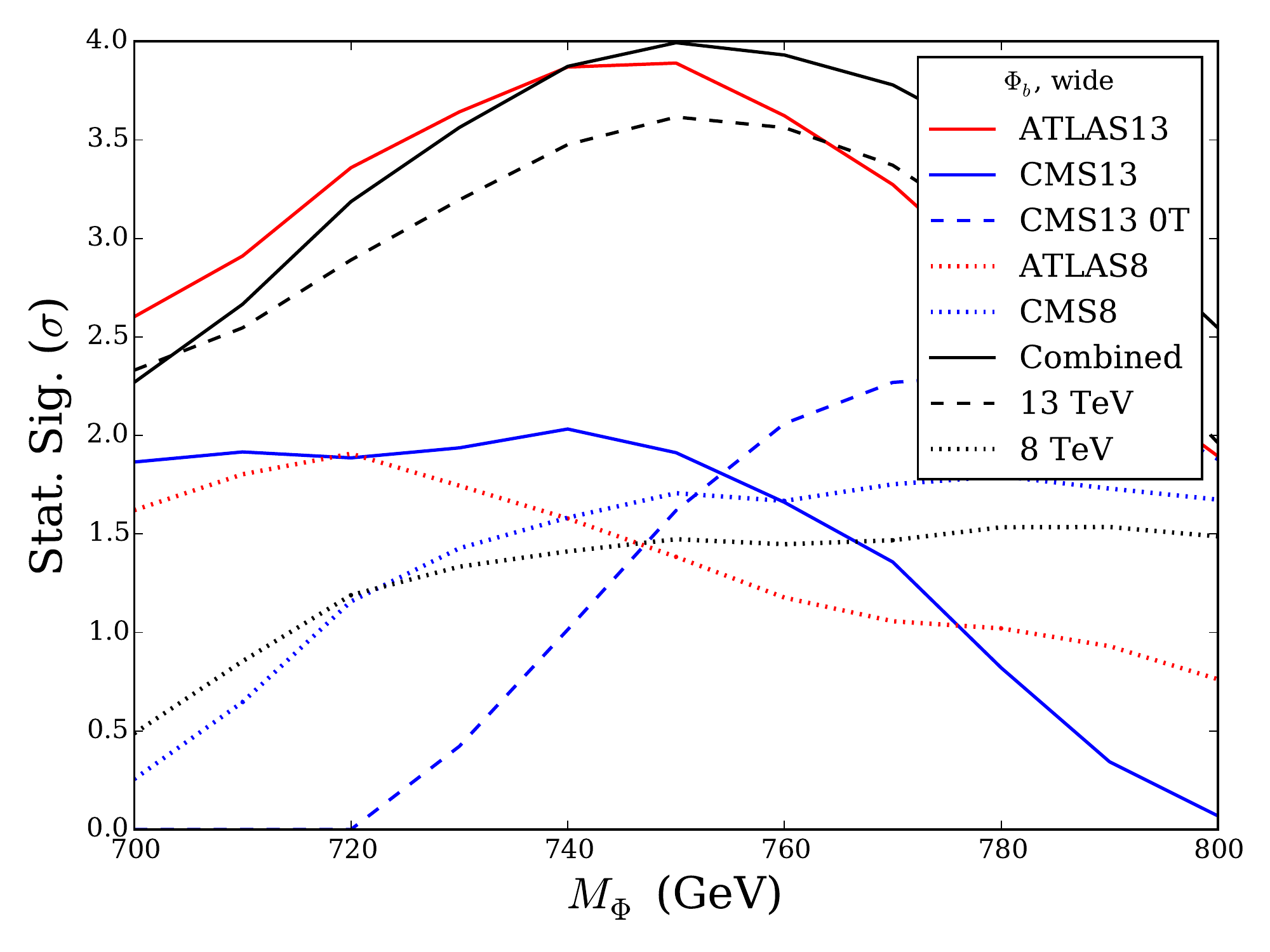}
\caption{Statistical significance for a spin-0 mediator decaying to diphoton, as a function of mediator mass, assuming the indicated mediator couplings to partons and mediator width. At each mass, the cross section is set to the value that maximizes statistical significance for a signal (see Figure~\ref{fig:spin0_xs}). The solid red line is the statistical significance of the \textsc{Atlas13} data alone, solid blue is \textsc{Cms13}, and dotted red and blue lines are \textsc{Atlas8} and \textsc{Cms8}, respectively. When comparing across experiments, note that these significances do not correspond to the same value of the cross section. The dashed (dotted) black line is the combination of 13(8)~TeV data, requiring the same cross section in both ATLAS and CMS. The solid black line is the combined significance of all four data sets.  \label{fig:spin0_stat_app}}
\end{figure}

\begin{figure}[h]
\includegraphics[width=0.4\columnwidth]{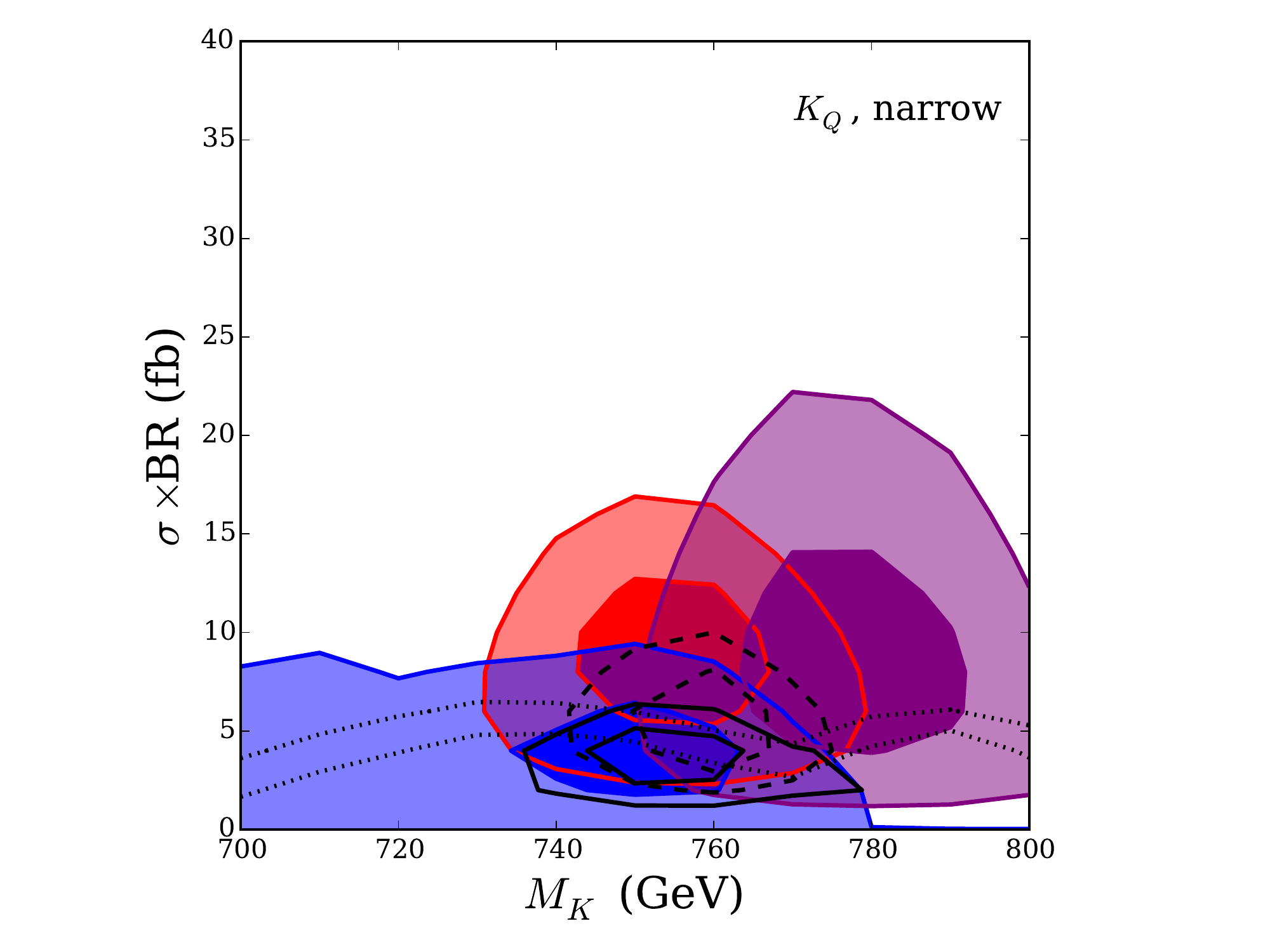}\includegraphics[width=0.4\columnwidth]{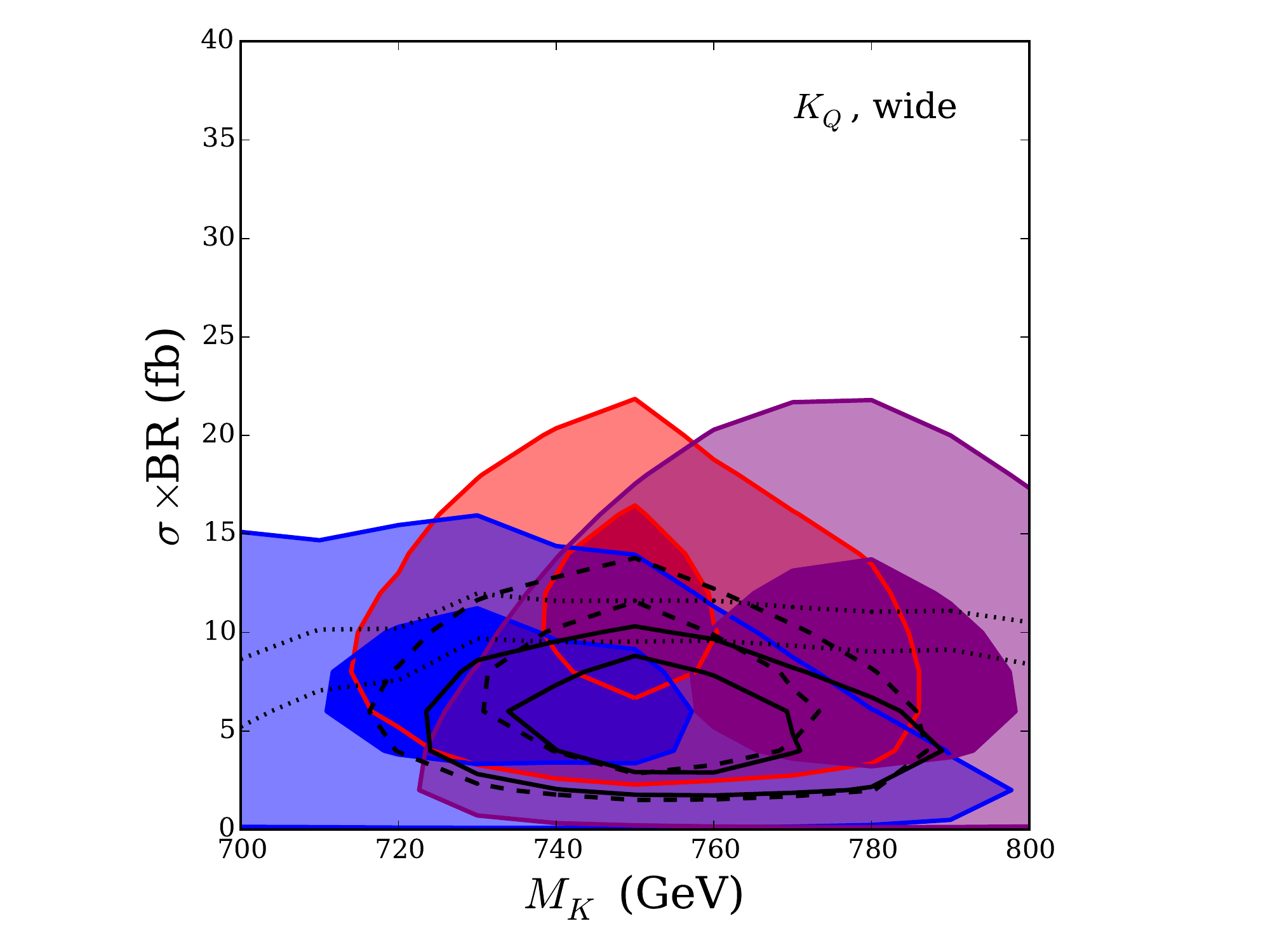}
\includegraphics[width=0.4\columnwidth]{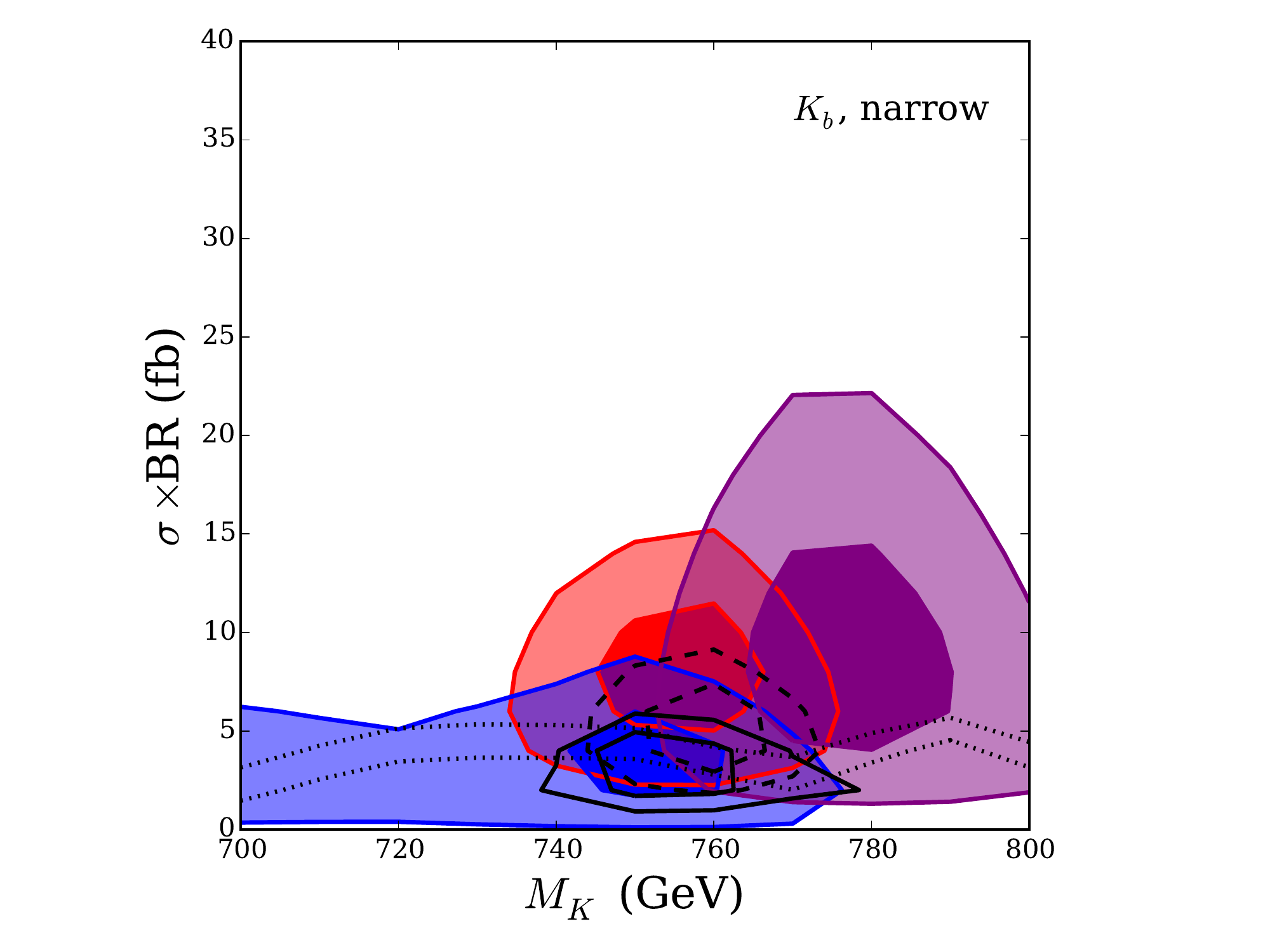}\includegraphics[width=0.4\columnwidth]{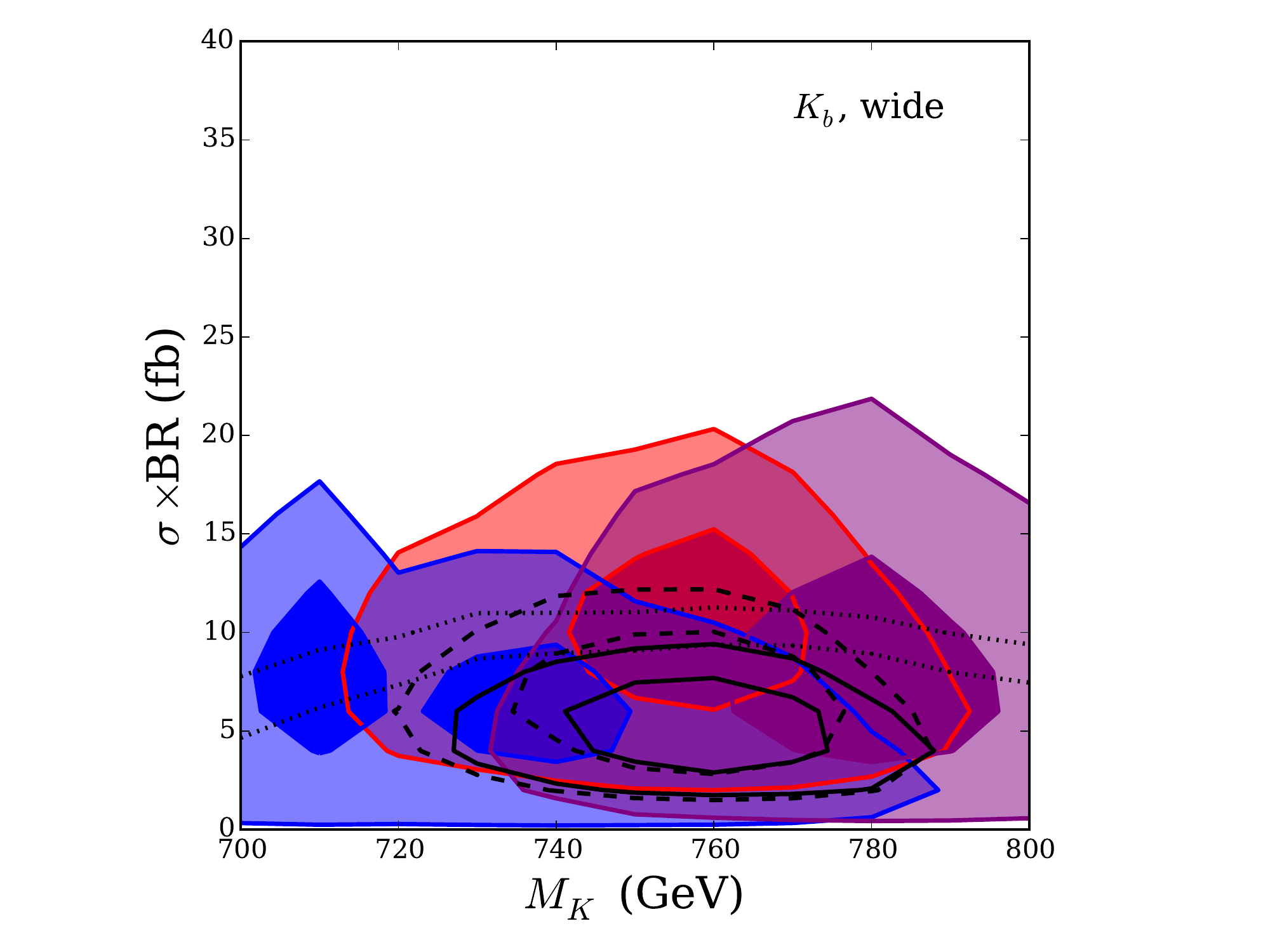}
\caption{Best fit regions (1 and $2\sigma$) of a spin-2 mediator decaying to diphotons, as a function of mediator mass and 13~TeV cross section, assuming the indicated mediator couplings to partons and mediator width. Red regions are the 1 and $2\sigma$ best-fit regions for the \textsc{Atlas13} data, blue is the fit to \textsc{Cms13} data. The combined best fit for both \textsc{Atlas13} and \textsc{Cms13} (\textsc{Combo13}) are the regions outlined in black dashed lines. The 1 and $2\sigma$ upper limits from the combined 8~TeV data (\textsc{Combo8}) are the black dashed regions (with cross sections converted to 13~TeV-equivalents). The best-fit signal combination of all four data sets (\textsc{Combo}) is the black solid regions.  \label{fig:spin2_xs_app}}
\end{figure}

\begin{figure}[h]
\includegraphics[width=0.4\columnwidth]{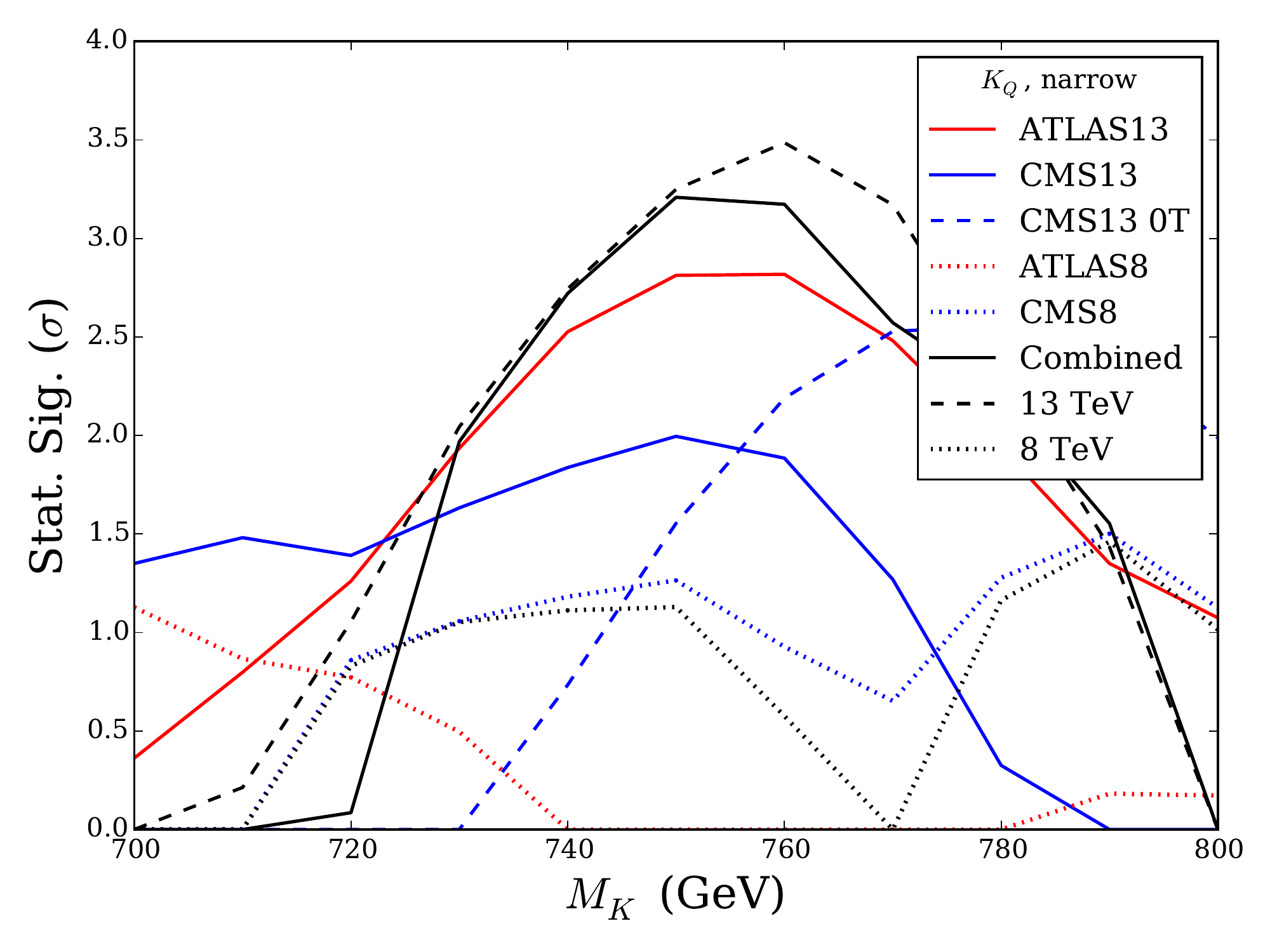}\includegraphics[width=0.4\columnwidth]{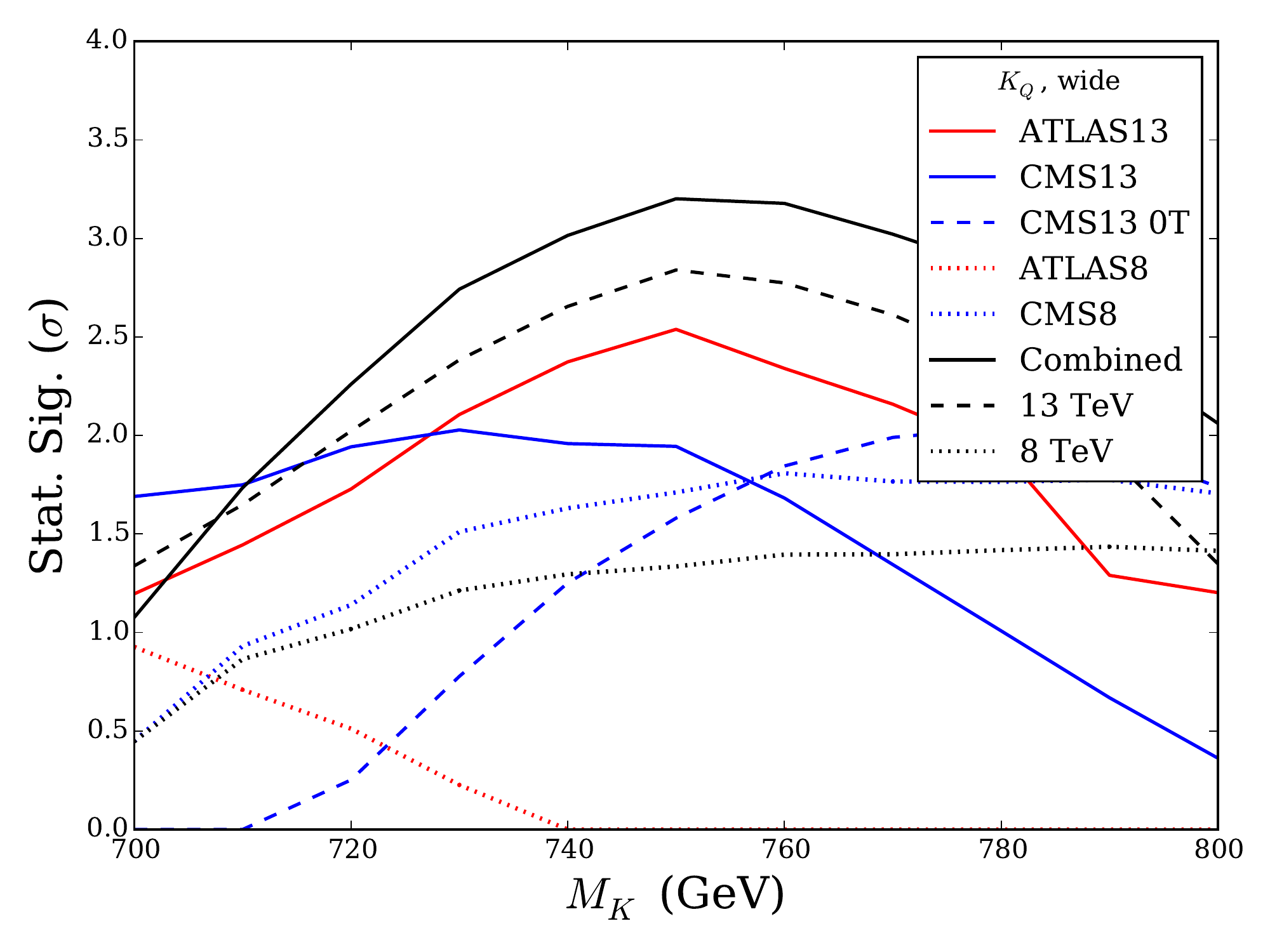}
\includegraphics[width=0.4\columnwidth]{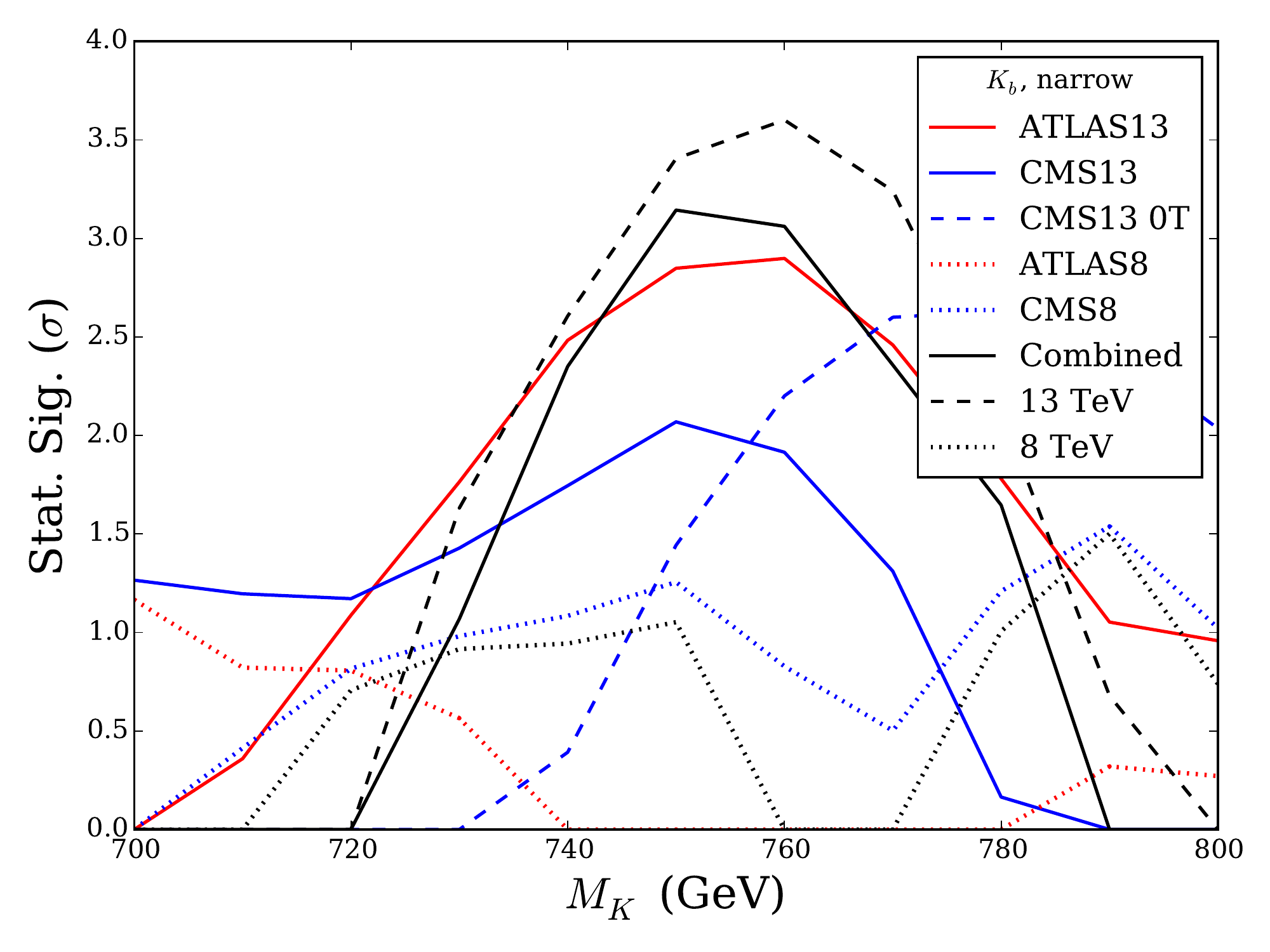}\includegraphics[width=0.4\columnwidth]{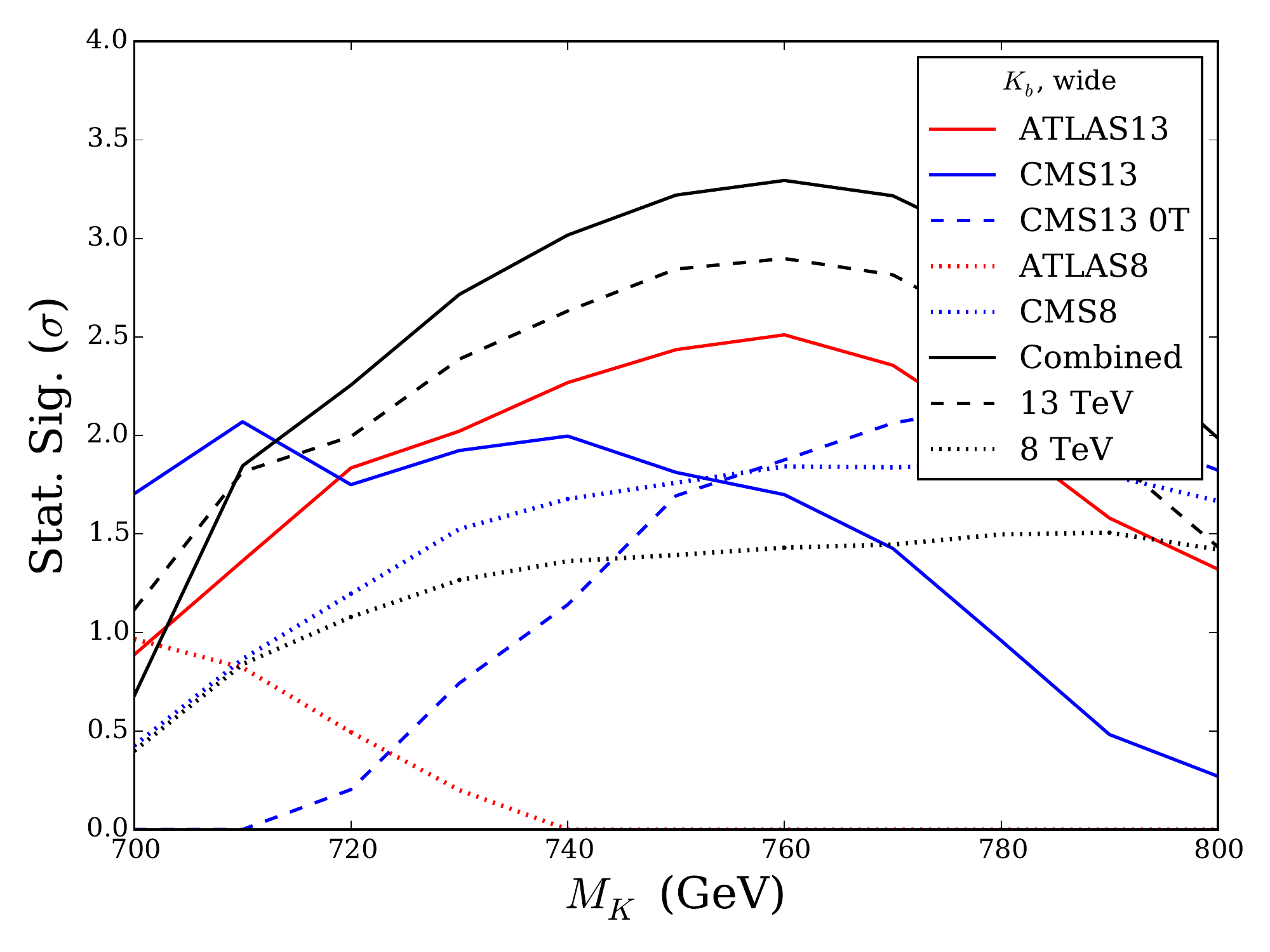}
\caption{Statistical significance for a spin-2 mediator decaying to diphoton, as a function of mediator mass, assuming the indicated mediator couplings to partons and mediator width. At each mass, the cross section is set to the value that maximizes statistical significance for a signal (see Figure~\ref{fig:spin0_xs}). The solid red line is the statistical significance of the \textsc{Atlas13} data alone, solid blue is \textsc{Cms13}, and dotted red and blue lines are \textsc{Atlas8} and \textsc{Cms8}, respectively. When comparing across experiments, note that these significances do not correspond to the same value of the cross section. The dashed (dotted) black line is the combination of 13(8)~TeV data, requiring the same cross section in both ATLAS and CMS. The solid black line is the combined significance of all four data sets.  \label{fig:spin2_stat_app}}
\end{figure}

In the main body of the paper, I presented the best-fit cross sections and statistical significances of spin-0 and spin-2 mediators coupling to gluons and light $u/d$ quarks. Couplings to $c/s$ and $b$ quarks in the proton are also possible, and here I present the equivalent results. As the ratio of p.d.f.s for these partons between 8 and 13~TeV is very close to that found for gluons, these results match the gluon plots in most respects. Figures~\ref{fig:spin0_xs_app} and \ref{fig:spin0_stat_app} show the spin-0 best fit cross sections and statistical significances for $c/s$ and $b$ couplings, while the spin-2 cases are shown in Figures~\ref{fig:spin2_xs_app} and  \ref{fig:spin2_stat_app}.

\end{document}